\documentclass{lmcs}
\pdfoutput=1

\usepackage{lastpage}
\renewcommand{\dOi}{14(4:9)2018}
\lmcsheading{}{1--\pageref{LastPage}}{}{}%
{Jul.~18,~2017}{Oct.~30,~2018}{}

\usepackage{float} % for H option to figure to force "here"

\usepackage{tikz}
\usepackage{amsmath}
\usepackage{amssymb}
\usepackage{xspace}

\usetikzlibrary{matrix,arrows,shapes,positioning,calc,snakes,decorations.markings}

\usetikzlibrary{arrows.meta}
\tikzset{>={Latex[width=3mm,length=3mm]}}

\newcommand{\circuit}{\text{Circuit}}

\pgfdeclarelayer{background}
\pgfsetlayers{background,main}

\DeclareMathOperator{\pp}{P}
\DeclareMathOperator{\OC}{OC}
\DeclareMathOperator{\lsz}{Lsz}

\DeclareMathOperator{\val}{Val}
\DeclareMathOperator{\pri}{pri}
\DeclareMathOperator{\maxdiff}{MaxDiff}
\DeclareMathOperator{\eval}{Eval}
\DeclareMathOperator{\br}{Br}
\DeclareMathOperator{\ornext}{OrNext}
\DeclareMathOperator{\nexbit}{NextBit}

\DeclareMathOperator{\length}{Length}
\DeclareMathOperator{\delay}{Delay}
\DeclareMathOperator{\inputstate}{InputState}
\DeclareMathOperator{\play}{Play}
\DeclareMathOperator{\maxio}{MaxIo}

\newcommand{\nats}{\mathbb N}

\newcommand{\veven}{V_{\text{Even}}}
\newcommand{\vodd}{V_{\text{Odd}}}
\newcommand{\seven}{\Sigma_{\text{Even}}}
\newcommand{\sodd}{\Sigma_{\text{Odd}}}

\newcommand{\inp}{I}
\newcommand{\org}{\ensuremath{\textsc{Or}}\xspace}
\newcommand{\notg}{\ensuremath{\textsc{Not}}\xspace}
\newcommand{\inputg}{\ensuremath{\textsc{Input/Output}}\xspace}
\newcommand{\lale}{\ensuremath{2k + 4n + 6}\xspace}
\newcommand{\laletwo}{\ensuremath{2k + 2n + 6}\xspace}
\newcommand{\C}{K}

\newcommand{\bitswitch}{\textsc{BitSwitch}\xspace}
\newcommand{\circuitvalue}{\textsc{CircuitValue}\xspace}

\usepackage{cite}

\newcommand{\allswitches}{{all-switches strategy improvement}\xspace}
\newcommand{\edgeswitch}{\textsc{EdgeSwitch}\xspace}
\newcommand{\optimalstrategy}{\textsc{OptimalStrategy}\xspace}

\renewcommand{\P}{\ensuremath{\mathtt{P}}\xspace}
\newcommand{\NP}{\ensuremath{\mathtt{NP}}\xspace}
\newcommand{\coNP}{\ensuremath{\mathtt{coNP}}\xspace}
\newcommand{\UP}{\ensuremath{\mathtt{UP}}\xspace}
\newcommand{\coUP}{\ensuremath{\mathtt{coUP}}\xspace}

\newcommand{\PLS}{\ensuremath{\mathtt{PLS}}\xspace}
\newcommand{\PPAD}{\ensuremath{\mathtt{PPAD}}\xspace}
\newcommand{\PSPACE}{\ensuremath{\mathtt{PSPACE}}\xspace}

\theoremstyle{plain}\newtheorem{property}[thm]{Property}

\usetikzlibrary{shapes,positioning,calc,arrows,snakes}

\tikzstyle{even}=[minimum size=1.7cm,inner sep=-0.2cm,draw,regular polygon,
regular polygon sides=4,align=center]
\tikzstyle{odd}=[minimum size=1.2cm,inner sep=0cm,draw,circle]
\tikzstyle{label}=[node distance=1cm,font=\normalsize]

\tikzstyle{evenlarge}=[minimum size=2.0cm,inner sep=-0.2cm,draw,regular polygon,
regular polygon sides=4,align=center]
\tikzstyle{oddlarge}=[minimum size=1.5cm,inner sep=0cm,draw,circle]

\title{The Complexity of All-switches Strategy Improvement}

\author[]{John Fearnley}	%required
\address{University of Liverpool, Liverpool, United Kingdom}	%required
\email{john.fearnley@liverpool.ac.uk}  %optional

\author[]{Rahul Savani}	%optional
\address{\vskip-7pt} % University of Liverpool, Liverpool, United Kingdom}	%required
\email{rahul.savani@liverpool.ac.uk}  %optional


\subjclass{CCS, Theory of computation, Computational complexity and  cryptography, Problems, reductions and completeness}
\titlecomment{A preliminary version of this paper appeared in the proceedings of SODA 2016~\cite{FS16}.}

\begin{document}

\begin{abstract}
Strategy improvement is a widely-used and well-studied class of algorithms for solving graph-based infinite games. These algorithms are parameterized by a switching rule, and one of the most natural rules is ``all switches'' which switches as many edges as possible in each iteration. Continuing a recent line of work, we study all-switches strategy improvement from the perspective of computational complexity. We consider two natural decision problems, both of which have as input a game $G$, a starting strategy $s$, and an edge~$e$. The problems are: 1.) The edge switch problem, namely, is the edge $e$ ever switched by all-switches strategy improvement when it is started from $s$ on game $G$? 2.) The optimal strategy problem, namely, is the edge $e$ used in the final strategy that is found by strategy improvement when it is started from $s$ on game $G$? We show \PSPACE-completeness of the edge switch problem and optimal strategy problem for the following settings: Parity games with the discrete strategy improvement algorithm of V\"oge and Jurdzi\'nski; mean-payoff games with the gain-bias algorithm~\cite{FV97,puterman94}; and discounted-payoff games and simple stochastic games with their standard strategy improvement algorithms. We also show \PSPACE-completeness of an analogous problem to edge switch for the bottom-antipodal algorithm for finding the sink of an Acyclic Unique Sink Orientation on a cube.
\end{abstract}

% \newpage
\maketitle

\section{Introduction}

In this paper we study strategy improvement algorithms for solving two-player
games such as parity games, mean-payoff games, discounted games, and simple
stochastic games~\cite{hoffmankarp66,puterman94,jurdzinski00b}. These games are
interesting both because of their important applications and their unusual
complexity status. Parity games, for example, arise in several areas of
theoretical computer science, for example, in relation to the emptiness problem
for tree automata \cite{GTW02,ej91} and as an algorithmic formulation of model
checking for the modal $\mu$-calculus \cite{EmersonJutla93,stirling95}.
Moreover, all of these problems are in \NP~$\cap$~\coNP, and even
\UP~$\cap$~\coUP~\cite{condon92,jurdzinski98}, so they are unlikely to be
\NP-complete. However, despite much effort from the community, none of these
problems are known to be in~\P, and whether there exists a polynomial-time
algorithm to solve these games is a very important and long-standing open
problem. 

Strategy improvement is a well-studied method for solving these
games~\cite{hoffmankarp66,puterman94,jurdzinski00b}. It is an
extension of the well-known policy iteration algorithms for solving Markov decision
processes. The algorithm selects one of the two players to be the \emph{strategy
improver}. Each strategy of the improver has a set of \emph{switchable} edges,
and switching any subset of these edges produces a strictly better strategy.
So, the algorithm proceeds by first choosing an arbitrary starting strategy, and
then in each iteration, switching some subset of the switchable edges.
Eventually this process will find a strategy with no switchable edges, and it
can be shown that this strategy is an optimal strategy for the improver.

To completely specify the algorithm, a \emph{switching rule} is needed to pick
the subset of switchable edges in each iteration (this is analogous to the pivot
rule used in the simplex method). Many switching rules have been proposed and
studied~\cite{schewe08, fearnley10a, HPZ14, kalai92, ludwig95}. One of the most
natural rules, and the one that we consider in this paper, is the
\emph{all-switches} rule, which always switches a vertex if it has a switchable
edge. In particular, we consider \emph{greedy} all-switches, which chooses
the best edge whenever more than one edge is switchable at a vertex (ties
are broken arbitrarily). For a long time, the all-switches variant of the
discrete strategy improvement algorithm of V\"oge and
Jurdzi\'nski~\cite{jurdzinski00b} was considered the best candidate for a
polynomial-time algorithm to solve parity games. Indeed, no example was known
that required a super-linear number of iterations. However, these hopes were
dashed when Friedmann showed an exponential lower bound for greedy all-switches
strategy improvement for parity games~\cite{F11}. In the same paper, he showed
that his result extends to strategy improvement algorithms for discounted games
and simple stochastic games.

\paragraph{\bf The computational power of pivot algorithms.} 
In this paper, we follow a recent line of work that seeks to explain the poor
theoretical performance of pivoting algorithms using a complexity-theoretic
point of view. The first results in this direction were proved for problems in
the complexity classes \PPAD and \PLS. It is know that, if a problem is \emph{tight-\PLS-complete} 
then computing the solution found by the natural improvement algorithm is
\PSPACE-complete~\cite{JPY88}. Similarly, for the canonical \PPAD-complete
problem End-of-the-Line, computing the solution that is found by the natural
line following algorithm is \PSPACE-complete~\cite{pap94}. This was 
extended to show that is \PSPACE-complete to compute any of the
solutions that can be found by the Lemke-Howson algorithm~\cite{GPS13}, a
pivoting algorithm that solves the \PPAD-complete problem of finding a Nash
equilibrium of a bimatrix game.

Until recently, results of this type were only known for algorithms for
problems that, due to known hardness results, are unlikely to lie in \P. However, a recent series of papers has
shown that similar results hold even for the simplex method for linear
programming.
Adler, Papadimitriou and Rubinstein~\cite{APR14} showed that there
exists an artificially contrived pivot rule for which is \PSPACE-complete to
decide if a given basis is on the path of bases visited by the simplex algorithm
with this pivot rule.
Disser and Skutella~\cite{DS15} studied the natural pivot rule that Dantzig
proposed when he introduced the simplex method, and they showed that
it is \NP-hard to decide whether a given variable enters the basis when the
simplex method is run with this pivot rule.  
Finally, Fearnley and Savani strengthened both these results by showing that the
decision problem that Disser and Skutella considered is actually
\PSPACE-complete~\cite{FS14}, and they also showed that determining if a given
variable is used in the \emph{final optimal solution} found by the algorithm is
\PSPACE-complete.  This result exploited a known connection between
\emph{single-switch} policy iteration for Markov Decision Processes (MDPs) and
the simplex method for a corresponding linear program (LP). The result was first
proved for a greedy variant of single-switch policy iteration, which then
implied the result for the simplex method with Dantzig's pivot rule.

All of the results on linear programming are motivated by the quest to find a
\emph{strongly polynomial} algorithm for linear programming, which was included
in Smale's list of great unsolved problems of the 21st century~\cite{S98}. One
way of resolving this problem would be to design a polynomial-time pivot rule
for the simplex method, and if we are to do this, then it is crucial to
understand why existing pivot rules fail.
The \PSPACE-completeness indicates that they fail because in fact they can do
something far more than is necessary, namely they are capable of solving any
problem that can be computed in polynomial space.

We face a similar quest to find a polynomial-time algorithm for the games 
studied in this paper. Strategy improvement is a prominent algorithm for
solving these games, and indeed it is one of the only algorithms for solving
discounted and simple stochastic games. So, devising a polynomial-time switching
rule is an obvious direction for further study. It may in fact be easier to
devise a polynomial time switching rule, because there is a lot more freedom in
each step of the algorithm: simplex pivot rules correspond to switching rules
\emph{that can only switch a single edge}, whereas strategy improvement rules
can switch \emph{any subset of edges}. Indeed, it may be the case that the
polynomial Hirsch conjecture fails, ruling out a strongly polynomial
simplex method, even though the analogue of the Hirsch conjecture for strategy
improvement is known to be true: one can always reach an optimal strategy
in at most $n$ strategy improvement iterations, where $n$ is the number of
vertices in the game.

\paragraph{\bf Our contribution.}

%In this paper, we provide a complexity-theoretic strengthening of the
%exponential-time results for \allswitches.
%%
%We construct parity games that allow \allswitches to solve any problem in
%$\PSPACE$.  
%%
%Our results will also apply to strategy improvement algorithms for other types
%of games.

Our main results are that, for greedy all-switches strategy improvement,
determining whether the algorithm switches a given edge is \PSPACE-complete, and determining whether
the optimal strategy found by the algorithm uses a
particular edge is \PSPACE-complete. One of the key features that strategy
improvement has is the ability to switch multiple
switchable edges at the same time, rather than just one as in the simplex
method. Our results show that naively using this power does not help to avoid
the \PSPACE-completeness results that now seem to be endemic among pivoting
algorithms. The proof primarily focuses on the strategy improvement algorithm of
V\"oge and Jurdzi\'nski for solving parity games~\cite{jurdzinski00b}. The
following definition formalises the problem that we are interested in.
%
%
%In the following definition, \allswitches refers to the all-switches variant
%of the respective algorithm for the particular type of game.

\begin{defi}
Let $G$ be a game, and let $e$ be an edge and $\sigma$ be a strategy profile of
$G$. 
The problem $\edgeswitch(G,\sigma,e)$ is to decide if the edge $e$ is ever
switched by greedy all-switches strategy improvement when it is applied to $G$
starting from $\sigma$.
\end{defi}
The main technical contribution of the paper is to show the following theorem. 

\begin{thm}
\label{thm:edgeswitch}
\edgeswitch for V\"oge and Jurdzi\'nski's algorithm is
\PSPACE-complete.
\end{thm}

We use this theorem to show similar results for other
games. For mean-payoff games, our results apply to the gain-bias algorithm~\cite{FV97}; and
for discounted and simple stochastic games our results apply to the standard
strategy improvement algorithms~\cite{puri95, condon93}. We utilise the well-known
polynomial-time reductions from parity games to mean-payoff games \cite{puri95},
mean-payoff games to discounted games, and from discounted games
to simple stochastic games~\cite{patersonzwick96}.  The parity games we
construct have the property that when they are reduced to the other games
mentioned above strategy improvement will behave in the same way (for discounted
and simple stochastic games this was already observed by Friedmann~\cite{F11}),
so we get the following corollary of Theorem~\ref{thm:edgeswitch}.

\begin{cor}
\label{cor:edgeswitch}
$\edgeswitch$ for the gain-bias algorithm, and the
standard strategy improvement algorithms for discounted-payoff and simple
stochastic games is \PSPACE-complete.  
\end{cor}

Theorem~\ref{thm:edgeswitch} proves a property about the path taken by strategy
improvement during its computation. Previous results
have also studied the complexity of finding the optimal strategy that is
produced by strategy improvement, which we encode in the following problem.

\begin{defi}
Let $G$ be a game, and let $e$ be an edge and $\sigma$ be a strategy profile of
$G$. The problem $\optimalstrategy(G,\sigma,e)$ is to decide if the edge $e$ is
used in the optimal strategy that is found by greedy all-switches strategy
improvement when applied to $G$ starting from $\sigma$.
\end{defi}

\noindent Augmenting our construction for parity games with an extra gadget 
gives the following theorem.

\begin{thm}
\label{thm:optstrat}
\optimalstrategy for V\"oge and Jurdzi\'nski's algorithm is \PSPACE-complete.
\end{thm}

This result requires that the parity games that we construct have 
multiple optimal solutions because otherwise the \PSPACE
hardness of \optimalstrategy would imply $\NP \cap \coNP = \PSPACE$.
With further modifications, we can again extend this result to
strategy improvement algorithms for other games. 

\begin{cor}
\label{cor:optstrat}
$\optimalstrategy$ for the gain-bias algorithm,  and the
standard strategy improvement algorithms for discounted-payoff and simple
stochastic games is \PSPACE-complete.
\end{cor}

Our results can also be applied to \emph{unique sink orientations} (USOs). An
\emph{orientation} of an $n$-dimensional hypercube is a function that assigns
a direction to each edge of the cube. The \emph{faces} of an $n$-dimensional
cube are the
$k$-dimensional cubes that can be obtained by fixing $n-k$ of the dimensions and
letting the other dimensions be free. An orientation is a USO if every face of
the cube has a unique sink~\cite{SW01}.

Recently, it was shown that recognising a USO is \coNP-complete, and that
recognising an \emph{acyclic} USO (AUSO) is \PSPACE-complete~\cite{GT15}.
As we will see, the games that we consider will be guaranteed to induce AUSOs.
The fundamental algorithmic problem for USOs is to find the global sink 
assuming oracle access to the edge orientation.
The design and analysis of algorithms for this problem is
an active research area~\cite{SS05,GS06,MS04,FHZ11,FHZ11b,HPZ14}, in particular
for AUSOs.
The \textsc{BottomAntipodal} algorithm~\cite{SS05} for AUSOs on cubes starts at
an arbitrary vertex and in each iteration jumps to the antipodal vertex in the
sub-cube spanned by the outgoing edges at the current vertex.

For binary games, where vertices have outdegree at most two, the valuation functions used by strategy
improvement induce an AUSO on a cube, and all-switches strategy improvement corresponds to
\textsc{BottomAntipodal} on this AUSO.
For non-binary games we instead get an AUSO on a grid~\cite{gmr08}, so
our results immediately give a \PSPACE-hardness result for grid USOs for 
a problem analogous to \edgeswitch.
To get a similar result for AUSOs on cubes we turn our construction into a
binary parity game, and we get the following.

\begin{cor}
\label{cor:uso}
Let $C$ be a $d$-dimensional cube AUSO, specified by a $poly(d)$-sized
circuit that computes the edge orientations for each vertex of $C$.
Given a dimension $k \in \{1,\ldots, d\}$, and a vertex $v$, it is
\PSPACE-complete to decide if \textsc{BottomAntipodal}
started at $v$, ever switches the $k$th coordinate.
\end{cor}

Since a USO has a unique solution, by definition, we cannot hope to get a
result for AUSOs that is analogous to \optimalstrategy, since, as noted above,
\PSPACE-hardness of \optimalstrategy requires multiple optimal solutions under
standard complexity-theoretic assumptions.

\paragraph{\bf Related work.} 

The best known algorithms for  mean-payoff, discounted, and simple stochastic
games have subexponential running time: the
\emph{random facet strategy improvement} algorithms combine strategy improvement
with the random-facet algorithm for LPs~\cite{ludwig95,kalai92,msw96}.
Following the work of Friedmann~\cite{F11} that we build on heavily in this 
paper,
Friedmann, Hansen, and Zwick showed a sub-exponential lower bound for the
random facet strategy improvement algorithm%
%for parity games
~\cite{FHZ11b}.
They also used a construction of Fearnley~\cite{F10} 
%can be simulated by a probabilistic action, to show an exponential lower bound for the
%all-switches variant of policy iteration of average-reward MDPs~\cite{F10},
to extend the
bound to the random facet pivot rule for the simplex method~\cite{FHZ11}.

For parity games, there are several algorithms that perform better than random
facet strategy improvement. First, a \emph{deterministic} subexponential-time
algorithm was found~\cite{jpz06}. Very recent work has improved this 
even further by
producing an algorithm that uses quasi-polynomial time and space~\cite{CJKLS17},
and it has subsequently been shown that there are algorithms that use
quasi-polynomial time and polynomial space~\cite{JL17,FJSSW17}.

%Friedmann also gave a sub-exponential lower bound for Zadeh's pivot rule for the
%simplex method~\cite{F11b}.

%This connection is well known, and has been applied in other contexts.
%Friedmann, Hansen, and Zwick used this connection in the expected total-reward
%setting, to show sub-exponential lower bounds for some randomized pivot
%rules~\cite{FHZ11}.
%Post and Ye have shown that Dantzig's pivot rule is strongly
%polynomial for \emph{deterministic discounted} MDPs~\cite{PY13}, while
%Hansen, Kaplan, and Zwick went on to prove various further bounds
%for this setting~\cite{HKZ14}.

%%% Local Variables:
%%% mode: latex
%%% TeX-master: "note.tex"
%%% End:

\paragraph{\bf Roadmap.}
In Section~\ref{sec:prelim}, we give a formal definition of parity games, and
more specifically the \emph{one-sink} games used by Friedmann that we also
use for our construction.
We then give a high-level overview of how \allswitches works.
Our main reduction starts with an iterated circuit evaluation problem.
In Section~\ref{sec:construct}, we describe our main construction of a parity 
game that will implement iterated circuit iteration when strategy improvement
is run on it.
In Section \ref{sec:strategies}, we describe the sequence of strategies that \allswitches
will go through as it implements the iterated circuit evaluation.
In Section~\ref{sec:proof}, we show that the construction works as claimed 
and thus prove that \edgeswitch is \PSPACE-hard for parity games.
In Section~\ref{sec:other}, we show how this result for \edgeswitch extends
to strategy improvement algorithms for other games.
In Section~\ref{sec:optstrat}, we show how to augment our construction with an
extra gadget to give \PSPACE-hardness results for \optimalstrategy.
In Section~\ref{sec:conc}, we state some open problems.

\section{Preliminaries}
\label{sec:prelim}

\subsection{Parity games}
A parity game is defined by a tuple $G = (V, \veven, \vodd, E, \pri)$, where
$(V, E)$ is a directed graph. The sets $\veven$ and $\vodd$ partition $V$ into
the vertices belonging to player Even, and the vertices belonging to player Odd,
respectively. The \emph{priority} function $\pri : V \rightarrow \{1, 2,
\dots\}$ assigns a positive natural number to each vertex. We make the standard
assumption that there
are no terminal vertices, which means that every vertex is required to have at
least one outgoing edge. The strategy improvement algorithm of 
V\"oge and Jurdzi\'nski also requires that we assume, without loss of
generality, that every priority is assigned to at most one vertex. 

A strategy for player Even is a function that picks one outgoing edge for each
Even vertex. More formally, a \emph{deterministic positional strategy} for Even
is a function $\sigma : \veven \rightarrow V$ such that, for each $v \in \veven$
we have that $(v, \sigma(v)) \in E$. Deterministic positional strategies for
player Odd are defined analogously. Throughout this paper, we will only consider
deterministic positional strategies, and from this point onwards, we will refer
to them simply as \emph{strategies}. We use $\seven$ and $\sodd$ to denote the
set of strategies for players Even and Odd, respectively.

A \emph{play} of the game is an infinite path through the game. More precisely,
a play is a sequence $v_0, v_1, \dots $ such that for all $i\in
\nats$ we have $v_i \in V$ and $(v_i, v_{i+1}) \in E$.
Given a pair of strategies $\sigma \in \seven$ and $\tau \in \sodd$, and a
starting vertex $v_0$, there is a unique play that occurs when the
game starts at $v_0$ and both players follow their respective strategies.
So, we define $\play(v_0, \sigma, \tau) = v_0, v_1, \dots$,
where for each $i \in \nats$ we have
$v_{i+1} = \sigma(v_i)$ if $v_i \in \veven$, and 
$v_{i+1} = \tau(v_i)$ if $v_i \in \vodd$.

Given a play $\pi = v_0, v_1, \dots$ we define 
\begin{equation*}
\maxio(\pi) = \max \{ p \; : \; \exists \text{ infinitely many } i \in \nats \text{ s.t. }
\pri(v_i) = p\},
\end{equation*}
to be the maximum priority that occurs \emph{infinitely often} along $\pi$. We
say that a play $\pi$ is \emph{winning} for player Even if $\maxio(\pi)$ is
even, and we say that $\pi$ is winning for Odd if $\maxio(\pi)$ is odd.
A strategy $\sigma \in \seven$ is a \emph{winning strategy} for a vertex $v \in
V$ if, for every strategy $\tau \in \sodd$, we have that $\play(v, \sigma,
\tau)$ is winning for player Even. Likewise, a strategy $\tau \in \sodd$ is a
winning strategy for $v$ if, for every strategy $\sigma \in \seven$, we have
that $\play(v, \sigma, \tau)$ is winning for player Odd. The following
fundamental theorem states that parity games are \emph{positionally determined}.
\begin{thm}[\cite{ej91,mostowski91}]
\label{thm:posdet}
In every parity game, the set of vertices~$V$ can be partitioned into
\emph{winning sets} $(W_\text{0}, W_\text{1})$, where Even has a positional
winning strategy for all $v \in W_\text{0}$, and Odd has a positional winning
strategy for all $v \in W_\text{1}$.
\end{thm}

\noindent The computational problem that we are interested in is, given a parity game,
to determine the partition $(W_\text{0}, W_\text{1})$.

\subsection{Strategy improvement}

We now describe the strategy improvement algorithm of V\"oge and
Jurdzi\'nski~\cite{jurdzinski00b} for solving parity games, which will be the
primary focus of this paper. 

\paragraph{\bf Valuations.} 

The algorithm assigns a \emph{valuation} to each vertex $v$ under every pair of
strategies $\sigma \in \seven$ and $\tau \in \sodd$.
Since both of these strategies are positional, we know that $\play(v, \sigma,
\tau)$ consists of a finite
initial path followed by an infinitely repeated simple cycle.
 Let $p$ be the largest
priority that is seen infinitely often along $\play(v, \sigma, \tau)$. Since
every priority is assigned to at most one vertex, there is a unique vertex $u$
with $\pri(u) = p$. We use this vertex to decompose the play: let $P(v, \sigma,
\tau)$ be the finite simple path that starts at $v$ and ends at $u$, and let
$C(v, \sigma, \tau)$ be the infinitely-repeated cycle that starts at $u$ and
ends at $u$.
We can now define the valuation function $\val_{\text{VJ}}^{\sigma, \tau}(v) = (p, S, d)$
where $p$ is as above and:
\begin{itemize}
\item $S$ is the set of priorities on the finite path that are strictly greater
than $p$:
\begin{equation*}
S = \{\pri(u) \; : \; u \in P(v, \sigma, \tau) \text{ and } \pri(u) > p\}.
\end{equation*}
\item $d$ is the length of the finite path: $d = |P(v, \sigma, \tau)|$.
\end{itemize}

We now define an order over valuations. First we define an order $\preceq$ over
priorities: we have that $p \prec q$ if one of the following holds:
\begin{itemize}
\item $p$ is odd and $q$ is even.
\item $p$ and $q$ are both even and $p < q$.
\item $p$ and $q$ are both odd and $p > q$.
\end{itemize}
Furthermore, we have that $p \preceq q$ if either $p \prec q$ or $p = q$.

Next we define an order of the sets of priorities that are used in the second
component of the valuation. Let $P, Q \subset \nats$. We first define:
\begin{equation*}
\maxdiff(P, Q) = \max\bigl((P \setminus Q) \cup (Q \setminus P)\bigr).
\end{equation*}
If $d = \maxdiff(P, Q)$ then we define $P \sqsubset Q$ to hold if one of the
following conditions holds:
\begin{itemize}
\item $d$ is even and $d \in Q$.
\item $d$ is odd and $d \in P$.
\end{itemize}
Furthermore, we have that $P \sqsubseteq Q$ if either $P = Q$ or $P \sqsubset
Q$.

Finally, we can provide an order over valuations. We have that $(p, S, d)
\prec (p', S', d')$ if one of the following conditions holds:
\begin{itemize}
\item $p \prec p'$.
\item $p = p'$ and $S \sqsubset S'$.
\item $p = p'$ and $S = S'$ and $p$ is odd and $d < d'$.
\item $p = p'$ and $S = S'$ and $p$ is even and $d > d'$.
\end{itemize}
Furthermore, we have that $(p, S, d) \preceq (p', S', d')$ if either $(p, S, d)
\prec (p', S', d')$ or $(p, S, d) = (p', S', d')$.

\paragraph{\bf Best responses.}

Given a strategy $\sigma \in \seven$, a \emph{best response} against $\sigma$
is a strategy $\tau^{*} \in \sodd$ such that, for every $\tau \in \sodd$ and
every vertex $v$ we have $\val_{\text{VJ}}^{\sigma, \tau}(v) \preceq \val_{\text{VJ}}^{\sigma,
\tau^{*}}(v)$. 
V\"oge and
Jurdzi\'nski proved the following properties.
\begin{lem}[\cite{jurdzinski00b}]
For every $\sigma \in \seven$ a best response $\tau^*$ can be computed in
polynomial time.
\end{lem}

\noindent We define $\br(\sigma)$ to be an arbitrarily chosen best response
strategy against $\sigma$. Furthermore, we define $\val_{\text{VJ}}^{\sigma}(v) =
\val_{\text{VJ}}^{\sigma, \br(\sigma)}(v)$.

\paragraph{\bf Switchable edges.}

Let $\sigma$ be a strategy and $(v, u) \in E$ be an edge such that $\sigma(v)
\ne u$. We say that $(v, u)$ is \emph{switchable} in $\sigma$ if
$\val_{\text{VJ}}^{\sigma}(\sigma(v)) \prec \val_{\text{VJ}}^{\sigma}(u)$. Furthermore, we define a
\emph{most appealing} outgoing edge at a vertex $v$ to be an edge $(v, u)$ such
that, for all edges $(v, u')$ we have $\val_{\text{VJ}}^{\sigma}(u') \preceq
\val_{\text{VJ}}^{\sigma}(u)$.

There are two fundamental properties of switchable edges that underlie the
strategy improvement technique. The first property is that switching any subset
of the switchable edges will produce an improved strategy. Let $\sigma$ be a
strategy, and let $W \subseteq E$ be a set of switchable edges in $\sigma$ such
that, for each vertex $v$, there is at most one edge of the form $(v, u) \in W$.
\emph{Switching} $W$ in $\sigma$ creates a new strategy $\sigma[W]$ where for
all $v$ we have:
\begin{equation*}
\sigma[W](v) = \begin{cases}
u & \text{ if $(v, u) \in W$,} \\
\sigma(v) & \text{otherwise.}
\end{cases}
\end{equation*}
We can now formally state the first property.

\begin{lem}[\cite{jurdzinski00b}] 
Let $\sigma$ be a strategy and let $W \subseteq E$ be a set of switchable edges
in $\sigma$ such that, for each vertex $v$, there is at most one edge of the
form $(v, u) \in W$. We have:
\begin{itemize}
\item For every vertex $v$ we have $\val_{\text{VJ}}^{\sigma}(v) \preceq \val_{\text{VJ}}^{\sigma[W]}(v)$.
\item There exists a vertex $v$ for which 
$\val_{\text{VJ}}^{\sigma}(v) \prec \val_{\text{VJ}}^{\sigma[W]}(v)$.
\end{itemize}
\end{lem}

The second property concerns strategies with no switchable edges. A
strategy $\sigma \in \seven$ is \emph{optimal} if for every vertex $v$ and every
strategy $\sigma' \in \seven$ we have $\val_{\text{VJ}}^{\sigma'}(v) \preceq
\val_{\text{VJ}}^{\sigma}(v)$. 
\begin{lem}[\cite{jurdzinski00b}] 
A strategy with no switchable edges is optimal.
\end{lem}

\noindent V\"oge and Jurdzi\'nski also showed that winning sets for both players
can be extracted from an optimal strategy.  If $\sigma$ is an optimal strategy,
then $W_\text{0}$ contains every vertex $v$ for which the first component of
$\val_{\text{VJ}}^{\sigma}(v)$ is even, and $W_\text{1}$ contains every vertex $v$ for which
the first component of $\val_{\text{VJ}}^{\sigma}(v)$ is odd. Hence, to solve the parity
game problem, it is sufficient to find an optimal strategy.

\paragraph{\bf The algorithm.} 
The two properties that we have just described give rise to an obvious
\emph{strategy improvement} algorithm that finds an optimal strategy.
The algorithm begins by selecting an arbitrary strategy
$\sigma \in \seven$. In each iteration, the algorithm
performs the following steps:
\begin{enumerate}
\item If there are no switchable edges, then terminate.
\item Otherwise, select a set $W \subseteq E$ of switchable edges in $\sigma$
such that, for each vertex~$v$, there is at most one edge of the form $(v, u)
\in W$. 
\item Set $\sigma := \sigma[W]$ and go to step 1.
\end{enumerate}
By the first property, each iteration of this algorithm produces a strictly
better strategy according to the $\prec$ ordering, and therefore the algorithm
must eventually terminate. However, the algorithm can only terminate when there
are no switchable edges, and therefore the second property implies that the
algorithm will always find an optimal strategy.

The algorithm given above does not specify a complete algorithm, because it does
not specify \emph{which} subset of switchable edges should be chosen in each
iteration. Indeed, there are many variants of the algorithm that use a variety
of different \emph{switching rules}. In this paper, we focus on the \emph{greedy
all-switches} switching rule. This rule switches every vertex that has a
switchable edge, and if there is more than one switchable edge, it arbitrarily
picks one of the most appealing edges. 

\paragraph{\bf One-sink games.}

Friedmann observed that, for the purposes of showing lower bounds, it is
possible to simplify the V\"oge-Jurdzi\'nski algorithm by restricting the input
to be a \emph{one-sink game}~\cite{F11}.

A one-sink parity game contains a sink vertex $s$ such that $\pri(s) = 1$. An
even strategy $\sigma \in \seven$ is called a \emph{terminating strategy} if,
for every vertex $v$, the first component of $\val_{\text{VJ}}^{\sigma}(v)$ is
$1$. This means that, when the opponent plays optimally against $\sigma$, the
play will terminate in the sink $s$.
Formally, a parity game is a one-sink parity game if:
\begin{itemize}
\item There is a vertex $s \in V$ such that $\pri(s) = 1$, and $(s, s)$ is the
only outgoing edge from $s$. Furthermore, there is no vertex $v$ with $\pri(v) =
0$.
\item All optimal strategies are terminating. 
\end{itemize}

Now, suppose that we apply the V\"oge-Jurdzi\'nski algorithm, and furthermore
suppose that the initial strategy is terminating. Since the initial and optimal
strategies are both terminating, we have that, for every strategy $\sigma$
visited by the algorithm and every vertex $v$, the first component of
$\val_{\text{VJ}}^{\sigma}(v) = 1$, and so it can be ignored. Furthermore, since there is no
vertex with priority $0$, the second component of $\val_{\text{VJ}}^{\sigma}(v)$ must be
different from the second component of $\val_{\text{VJ}}^{\sigma}(u)$, for every $v, u \in
V$. Therefore, the third component of the valuation can be ignored.

Thus, for a one-sink game, we can define a simplified version of the
V\"oge-Jurdzi\'nski algorithm that only uses the second component. So, we define
$\val^{\sigma}(v)$ to be equal to the second component of
$\val_\text{VJ}^\sigma(v)$, and we carry out strategy improvement using the
definitions given above, but with $\val^{\sigma}(v)$ substituted for
$\val_\text{VJ}^\sigma(v)$. Note, in particular, that in this strategy
improvement algorithm, and edge $(v, u)$ is switchable in $\sigma$ if
$\val^{\sigma}(\sigma(v)) \sqsubset \val^{\sigma}(u)$.

In our proofs, we will frequently want to determine the maximum difference
between two valuations. For this reason, we introduce the following notation.
For every strategy $\sigma$, and every pair of vertices $v, u \in V$, we define
$\maxdiff^\sigma(v, u) = \maxdiff(\val^{\sigma}(v), \val^{\sigma}(u))$.

\subsection{Circuit iteration problems}

To prove our \PSPACE-completeness results, we will reduce from two \emph{circuit
iteration problems}, which we now define. 

\paragraph{\bf The problems.}

A \emph{circuit iteration} instance is
a triple $(F, B, z)$, where:
\begin{itemize}
\item $F : \{0, 1\}^n \rightarrow \{0, 1\}^n$ is a function represented as a
boolean circuit $C$,
\item $B \in \{0, 1\}^n$ is an initial bit-string, and
\item $z$ is an integer such that $1 \le z \le n$. 
\end{itemize}
We use standard notation for function iteration: given a bit-string $B \in
\{0,1\}^n$, we recursively define $F^{1}(B) = F(B)$, and $F^{i}(B) =
F(F^{i-1}(B))$ for all $i > 1$. 

We now define two problems that will be used as the starting point for our
reduction. Both are decision problems that take as input a circuit iteration
instance $(F, B, z)$.
\begin{itemize}
\itemsep2mm
\item $\bitswitch(F, B, z)$: decide whether there exists an even $i \le 2^n$
such that the $z$-th bit of $F^{i}(B)$ is $1$.
\item $\circuitvalue(F, B, z)$: decide whether the $z$-th bit of $F^{2^n}(B)$ is $1$.
\end{itemize}
The requirement for $i$ to be even in $\bitswitch$ is a technical requirement
that is necessary in order to make our reduction to strategy improvement work.

The fact that these problems are \PSPACE-complete should not be too surprising,
because~$F$ can simulate a single step of a space-bounded Turing machine, so
when $F$ is iterated, it simulates a run of the space-bounded Turing machine.
The following result was shown in~\cite{FS14}.

\begin{lem}{\cite[Lemma 7]{FS14}}
\label{lem:pspace}
$\bitswitch$ and $\circuitvalue$ are \PSPACE-complete.
\end{lem}

\textbf{Circuits.}
For the purposes of our reduction, we must make some assumptions about the
format of the circuits that represent $F$. Let $C$ be a boolean circuit with $n$
input bits, $n$ output bits, and $k$ gates. We assume, w.l.o.g., that
all gates are or-gates or not-gates. The circuit will be represented as
a list of gates indexed~$1$ through $n+k$. The indices $1$ through~$n$
represent the $n$ \emph{inputs} to the circuit. Then, for each $i > n$, we have:
\begin{itemize}
\item If gate $i$ is an or-gate, then we define $\inp_1(i)$ and $\inp_2(i)$ to
give the indices of its two inputs.
\item If gate $i$ is a not-gate, then we define $\inp(i)$ to give the index of
its input.
\end{itemize}
The gates $k+1$ through $k + n$ correspond to the $n$ \emph{output bits} of the
circuit, respectively. For the sake of convenience, for each input bit $i$, we
define $\inp(i) = k+i$, which indicates that, if the circuit is applied to its
own output, input bit $i$ should copy from output bit $\inp(i)$. Moreover, we
assume that the gate ordering is topological. That is, for each or-gate~$i$ we
assume that~$i > \inp_1(i)$ and~$i > \inp_2(i)$, and we assume that for each
not-gate $i$ we have $i > \inp(i)$.

For each gate $i$, let $d(i)$ denote the \emph{depth} of gate $i$, which is the
length of the longest path from~$i$ to an input bit. So, in particular, the
input bits are at depth $0$. Observe that we can increase the depth of a gate by
inserting dummy or-gates: given a gate~$i$, we can add an or-gate $j$ with
$\inp_1(j) = i$ and $\inp_2(j) = i$, so that $d(j) = d(i)+1$. We use this fact
in order to make the following assumptions about our circuits:
\begin{itemize}
\item For each or-gate $i$, we have $d(\inp_1(i)) = d(\inp_2(i))$.
\item There is a constant $c$ such that, for every output bit $i \in \{k+1,
k+n\}$, we have $d(i) = c$.
\end{itemize}
From now on, we assume that all circuits that we consider satisfy these
properties. Note that, since all outputs gates have the same depth, we can
define $d(C) = d(k+1)$, which is the depth of all the output bits of the
circuit.

Given an input bit-string $B \in \{0, 1\}^n$, the output of each gate in $C$ can
be determined.
We define $\eval(B, i) = 1$ if gate $i$ is true for
input $B$, and $\eval(B, i) = 0$ if gate $i$ is false for input $B$.

Given a circuit $C$, we define the \emph{negated form} of $C$ to be a
transformation of $C$ in which each output bit is negated. More formally, we
transform $C$ into a circuit $C'$ using the following operation: for each output
bit $n+i$ in $C$, we add a \notg gate $n+k+i$ with $\inp(n+k+i) = n+i$.  

\section{The Construction}
\label{sec:construct}

%%%%%%%%%%%%%%%%%%%%%%%%%%%%%%%%%%%%%%%%%%%%%%%%%%%%%%%%%%%%%%%%%%%%%%%%%%%%%%%
% OVERVIEW FIGURES
%%%%%%%%%%%%%%%%%%%%%%%%%%%%%%%%%%%%%%%%%%%%%%%%%%%%%%%%%%%%%%%%%%%%%%%%%%%%%%%
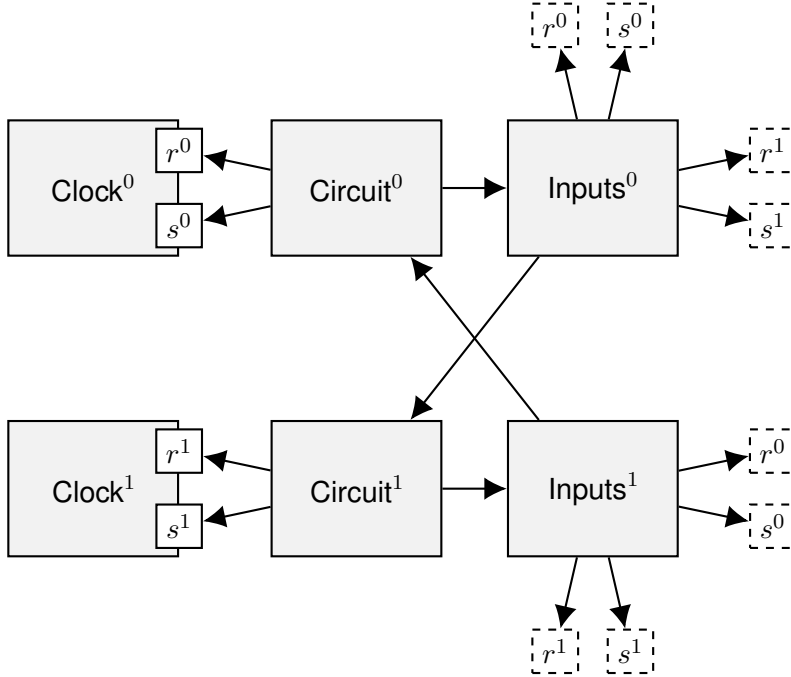
\begin{figure}
% fill white since these will be used in the foreground
\tikzset{box/.style={font=\sffamily,draw, thick, minimum height=1.8cm,minimum width=0.5cm}}
\tikzset{main/.style={box,fill=gray!10}} %,label={below:{\text{Circuit} 1}}}}
\tikzstyle{component}=[fill=white, font=\sffamily,draw, thick, minimum size=0cm,align=center]

\begin{tikzpicture}[
	decoration={markings,mark=at position 0.6 with {\arrow{triangle 60}}},
	path/.style={thick},
	clip=false
	%path/.style={ultra thick,>=stealth,postaction=decorate}
	%every node/.style={
]

% BOXES IN THE BACKGROUND
\begin{pgfonlayer}{background}
\node[main, xshift=0cm, yshift=2cm, align=center, text width=2cm] (circuit0) {$\circuit^0$};
\node[main, xshift=0cm, yshift=-2cm, align=center,text width=2cm] (circuit1) {$\circuit^1$};
\node[main, xshift=-3.5cm, yshift=2cm, align=center, text width=2cm] (clock0) {${\text{Clock}}^0$};
\node[main, xshift=-3.5cm, yshift=-2cm, align=center,text width=2cm] (clock1) {${\text{Clock}}^1$};
\node[main, xshift=2cm, yshift=0cm, align=center, text width=2cm] (inputs0) at (circuit0.east) {${\text{Inputs}}^0$};
\node[main, xshift=2cm, yshift=0cm, align=center,text width=2cm] (inputs1) at (circuit1.east) {${\text{Inputs}}^1$};
\end{pgfonlayer}{background}

%%%%%%%%%%%%%%%%%%%%%%%%%%%%%%%%%%%%%%%%%%%%%%%%%%%%%%%%%%%%%%%%%%%%%%%%%%%%%%
% INPUTS/OUTPUTS
%%%%%%%%%%%%%%%%%%%%%%%%%%%%%%%%%%%%%%%%%%%%%%%%%%%%%%%%%%%%%%%%%%%%%%%%%%%%%%

% circuit 0
\node[component,xshift=0cm,yshift=0.5cm]  (r0) at (clock0.east) {$r^0$};
\node[component,xshift=0cm,yshift=-0.5cm]  (s0) at (clock0.east) {$s^0$};

% circuit 1
\node[component,xshift=0cm,yshift=0.5cm]  (r1) at (clock1.east) {$r^1$};
\node[component,xshift=0cm,yshift=-0.5cm]  (s1) at (clock1.east) {$s^1$};

% DASHED INPUTS

% s0 r0 dashed for inputs0
\node[component,dashed, xshift=-0.5cm, yshift=1.25cm]  (r0-dashed-inputs0) at (inputs0.north) {$r^0$};
\node[component,dashed, xshift=0.5cm, yshift=1.25cm]  (s0-dashed-inputs0) at (inputs0.north) {$s^0$};

% s1 r1 dashed for inputs0
\node[component,dashed, xshift=1.25cm,yshift=0.5cm]  (r1-dashed) at (inputs0.east) {$r^1$};
\node[component,dashed, xshift=1.25cm,yshift=-0.5cm]  (s1-dashed) at (inputs0.east) {$s^1$};

% s0 r0 dashed for inputs1
\node[component,dashed, xshift=1.25cm,yshift=0.5cm]  (r0-dashed) at (inputs1.east) {$r^0$};
\node[component,dashed, xshift=1.25cm,yshift=-0.5cm]  (s0-dashed) at (inputs1.east) {$s^0$};

% s1 r1 dashed for inputs1
\node[component,dashed, xshift=0.5cm,yshift=-1.25cm]  (s1-dashed-inputs1) at (inputs1.south) {$s^1$};
\node[component,dashed, xshift=-0.5cm,yshift=-1.25cm]  (r1-dashed-inputs1) at (inputs1.south) {$r^1$};

% EDGES
%%%%%%%%%%%%%%%%%%%%%%%%%%%%%%%%%%%%%%%%%%%%%%%%%%%%%%%%%%%%%%%%%%%%%%%%%%%%%%

% Circuits to Inputs 
\draw[path,->] (circuit0) -- node[right] {} (inputs0);
\draw[path,->] (circuit1) -- node[right] {} (inputs1);

% Inputs to Circuits
\draw[path,->] (inputs0) -- node[right] {} (circuit1);
\draw[path,->] (inputs1) -- node[right] {} (circuit0);

% Circuits to Clock
\draw[path,->] (circuit0) -- node[right] {} (s0) ;
\draw[path,->] (circuit0) -- node[right] {} (r0) ;

\draw[path,->] (circuit1) -- node[right] {} (s1) ;
\draw[path,->] (circuit1) -- node[right] {} (r1) ;

% Inputs to Clock
\draw[path,->] (inputs0) -- node[right] {} (s1-dashed) ;
\draw[path,->] (inputs0) -- node[right] {} (r1-dashed) ;
\draw[path,->] (inputs0) -- node[right] {} (s0-dashed-inputs0) ;
\draw[path,->] (inputs0) -- node[right] {} (r0-dashed-inputs0) ;

\draw[path,->] (inputs1) -- node[right] {} (s0-dashed) ;
\draw[path,->] (inputs1) -- node[right] {} (r0-dashed) ;
\draw[path,->] (inputs1) -- node[right] {} (s1-dashed-inputs1) ;
\draw[path,->] (inputs1) -- node[right] {} (r1-dashed-inputs1) ;

\end{tikzpicture}

%%% Local Variables:
%%% mode: latex
%%% TeX-master: "../note.tex"
%%% End:
\caption{\label{fig:overview1} High-level overview of our construction. There are
two copies of the underlying circuit, and two clocks. The two are synchronized
via the nodes $r^0$, $s^0$, $r^1$, and $s^1$. In this diagram the directions of arrows
are consistent with the directed edges in the corresponding parity game.}
\end{figure}

\paragraph{\bf Overview.}

We will show that $\edgeswitch$ is \PSPACE-complete by reducing from the circuit
iteration problem $\bitswitch$. Figure~\ref{fig:overview1} gives a high level
picture of the construction. Given a circuit $F$ that is to be iterated, we
create a gadget that is capable of computing~$F$ on a given input. Our
construction will contain two copies of this gadget, which will be numbered $0$
and $1$. The two circuits alternate, with the output of one circuit being passed
to the input of the other circuit. So, given an initial bit-string $B$,
circuit~$0$ computes $F(B)$, then circuit~$1$ computes $F(F(B))$, then
circuit~$0$ computes $F(F(F(B)))$, and so on. The technical reason for having
two copies of the circuit is that our circuit gadget cannot handle the
input bits being changed before the output bits are read, and so a single
circuit gadget cannot feed its own outputs back into its inputs.

Figure~\ref{fig:overview1} also shows the \emph{clocks} which play a fundamental
role in driving the construction. Each copy of the circuit is equipped with its
own clock, which controls the timing of that circuit. In particular, each clock
has two states $r$ and $s$. Ordinarily, the valuation of $r$ is larger than the
valuation of $s$. Every so often, the clock produces a signal, which is
transmitted by the valuation of $s$ being larger than the valuation of $r$. This
signal causes the associated circuit to begin computing based on its current
input. Thus, the clocks plays an important role in synchronising the two
circuits, and ensuring that each circuit starts computing only after its partner
has finished computing the previous iteration.

Each clock is implemented by a modified version of Friedmann's exponential-time
example. Friedmann's examples are designed to force greedy all-switches strategy
improvement to mimic a binary counter. The signal sent by the clock occurs when
Friedmann's example increments the binary counter to the next number. Our
modifications serve only to increase the number of strategy improvement
iterations that take place between each increment.

Finally, Figure~\ref{fig:overview1} shows the \inputg gadgets. These gadgets are
responsible for transmitting bit-strings between the two circuits, and so they
are the most complex part of the construction. These gadgets have two modes.
When they are in \emph{output mode}, they are connected at the outputs of a
circuit, where they read and store the outputs. When they are in \emph{input
mode} they are connected at the inputs of the other circuit, where they allow
the stored bit-string to be read. For this reason, the \inputg gadgets must be
connected to both clocks, so that they are able to transition between the
circuits at the correct time.

\paragraph{\bf The reduction.}

Formally, let $(F, B, z)$ be the input to the
circuit iteration problem, and let $C$ be the negated form of the circuit that
computes $F$. Throughout this section, we will use $n$ as the bit-length of $B$,
and $k = |C|$ as the number of gates used in $C$. We will use $\org$, $\notg$,
and $\inputg$ to denote the set of or-gates, not-gates, and input/output-gates,
respectively.

In the rest of this section, we describe the construction. We begin by giving an
overview of Friedmann's example, both because it plays a key role in our
construction, and because the \notg gate gadgets in our circuits are a
modification of the bit-gadget used by Friedmann. We then move on to describe
the gate gadgets, and how they compute the function $F$.

\paragraph{\bf Priorities.}

As we have mentioned, the strategy improvement algorithm that we consider
requires that every priority is assigned to at most one vertex. This is
unfortunately a rather cumbersome requirement when designing more complex
constructions. To help with this, we define a shorthand for specifying
priorities. Let $c \in \nats$, let $i,l \in \{1, \dots, |V|\}$, let $j \in \{0,
1, 2\}$, and let $e \in \{0,1\}$. We define 

$$\pp(c, i, l, j, e) = 6 \cdot |V|^2 \cdot c + 6 \cdot |V| \cdot i + 6 \cdot l + 2 \cdot j + e.$$

\medskip

The first four parameters should be thought of as a lexicographic ordering,
which determines how large the priority is. The final number $e$ determines
whether the priority is odd or even. Note that $\pp(c, i, l, j, e)$ is an
injective function, so if we ensure that the same set of arguments are never
used twice, then we will never assign the same priority to two different
vertices. One thing to note is that, since this priority notation is rather
cumbersome, it is not possible to use it in our diagrams. Instead, when we draw
parts of the construction, we will use \emph{representative} priorities, which
preserve the order and parity of the priorities used in the gadgets, but not
their actual values.

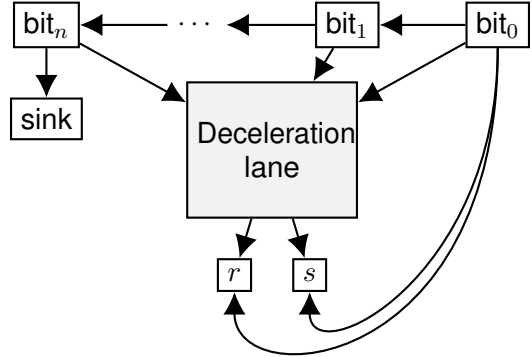
\begin{figure}
% fill white since these will be used in the foreground
\tikzset{box/.style={font=\sffamily,draw, thick, minimum height=1.8cm,minimum width=0.5cm}}
\tikzset{main/.style={box,fill=gray!10}} %,label={below:{\text{Circuit} 1}}}}
\tikzstyle{component}=[fill=white, font=\sffamily,draw, thick, minimum size=0cm,align=center]

\begin{tikzpicture}[
	decoration={markings,mark=at position 0.6 with {\arrow{triangle 60}}},
	path/.style={thick},
	clip=false
	%path/.style={ultra thick,>=stealth,postaction=decorate}
	%every node/.style={
]

% BOXES IN THE BACKGROUND
\begin{pgfonlayer}{background}
\node[main, xshift=0cm, yshift=2cm, align=center, text width=2cm] (dlane) {Deceleration lane};
\end{pgfonlayer}{background}

%%%%%%%%%%%%%%%%%%%%%%%%%%%%%%%%%%%%%%%%%%%%%%%%%%%%%%%%%%%%%%%%%%%%%%%%%%%%%%
% BITS
%%%%%%%%%%%%%%%%%%%%%%%%%%%%%%%%%%%%%%%%%%%%%%%%%%%%%%%%%%%%%%%%%%%%%%%%%%%%%%

\node[component,xshift=3cm,yshift=0.75cm]  (b0) at (dlane.north) {${\text{bit}}_0$};
\node[component,xshift=1cm,yshift=0.75cm]  (b1) at (dlane.north) {${\text{bit}}_1$};
\node[xshift=-1cm,yshift=0.75cm]  (bphantom) at (dlane.north) {$\cdots$};
\node[component,xshift=-3cm,yshift=0.75cm]  (bn) at (dlane.north) {${\text{bit}}_n$};
\node[component,xshift=-3cm,yshift=-0.5cm]  (sink) at (dlane.north) {sink};

%%%%%%%%%%%%%%%%%%%%%%%%%%%%%%%%%%%%%%%%%%%%%%%%%%%%%%%%%%%%%%%%%%%%%%%%%%%%%%
% s and r
%%%%%%%%%%%%%%%%%%%%%%%%%%%%%%%%%%%%%%%%%%%%%%%%%%%%%%%%%%%%%%%%%%%%%%%%%%%%%%

\node[component,xshift=0.5cm,yshift=-0.75cm]  (r) at (dlane.south) {$s$};
\node[component,xshift=-0.5cm,yshift=-0.75cm]  (s) at (dlane.south) {$r$};

% EDGES
%%%%%%%%%%%%%%%%%%%%%%%%%%%%%%%%%%%%%%%%%%%%%%%%%%%%%%%%%%%%%%%%%%%%%%%%%%%%%%

% Bit 0 to r and s
\draw[path,->] (b0.south) to[out=270,in=270,yshift=-2cm] (r);
\draw[path,->] (b0.south) to[out=270,in=270,yshift=-3cm] (s);

% Bits to Dlane
\draw[path,->] (b0) -- node[right] {} (dlane);
\draw[path,->] (b1) -- node[right] {} (dlane);
\draw[path,->] (bn) -- node[right] {} (dlane);

% Bits to Bits
\draw[path,->] (b0) -- node[right] {} (b1);
\draw[path,->] (b1) -- node[right] {} (bphantom);
\draw[path,->] (bphantom) -- node[right] {} (bn);
\draw[path,->] (bn) -- node[right] {} (sink);

% Dlane to r and s
\draw[path,->] (dlane) -- node[right] {} (r);
\draw[path,->] (dlane) -- node[right] {} (s);

\end{tikzpicture}

%%% Local Variables:
%%% mode: latex
%%% TeX-master: "../note.tex"
%%% End:
\caption{\label{fig:overview2} High-level overview of a clock.}
\end{figure}

\subsection{Friedmann's exponential-time example}

In this section, we give an overview of some important properties of Friedmann's
exponential-time examples. In particular, we focus on the properties that will
be important for our construction. A more detailed description of the example
can be found in Friedmann's original paper~\cite{F11}.

A high level view of Friedmann's construction is shown in
Figure~\ref{fig:overview2}. It works by forcing greedy all-switches strategy
improvement to simulate an~$n$ bit binary counter. It consists of two
components: a \emph{bit gadget} that is used to store one of the bits of the
counter, and a \emph{deceleration lane} that is used to ensure that the counter
correctly moves from one bit-string to the next.

\begin{figure}
\begin{center}
%\resizebox{0.9\textwidth}{!}{
\begin{tikzpicture}[node distance=1.8cm]
\node[even] (c) {$t_0$ \\ $16$};
\node [even,left of=c] (t1) {$t_1$ \\ $7$};
\node [even,left of=t1] (t2) {$t_2$ \\ $9$};
\node [even,left of=t2] (t3) {$t_3$ \\ $11$};
%\node [left of=t3] (t4) {$\dots$};
\node [even,left of=t3] (t5) {$t_4$ \\ $13$};

\node[even,above of=t1] (a1) {$a_1$ \\ $8$};
\node[even,above of=t2] (a2) {$a_2$ \\ $10$};
\node[even,above of=t3] (a3) {$a_3$ \\ $12$};
\node[even,above of=t5] (a5) {$a_4$ \\ $14$};

\path[->]
	(t5) edge (t3)
	%(t4) edge (t3)
	(t3) edge (t2)
	(t2) edge (t1)
	(t1) edge (c)
	;

\path[->]
	(a5) edge (t5)
	(a3) edge (t3)
	(a2) edge (t2)
	(a1) edge (t1)
	;

\foreach \x in {c, t1, t2, t3, t5}
{
	\node [even,draw,minimum size=0.8cm] (\x1) at ($(\x) - (0.5,1.8)$) {$r$};
	\node [even,draw,minimum size=0.8cm] (\x2) at ($(\x) - (-0.5,1.8)$) {$s$};

	\path[->]
		(\x) edge (\x1)
		(\x) edge (\x2)
		;
}
\end{tikzpicture}
%}
\end{center}
\caption{Friedmann's deceleration lane of length $4$. Each $t_i$ vertex with $i
> 0$ has an
odd priority, while each $a_i$ vertex has an even priority that is equal to
$\pri(t_i) + 1$. The vertex $t_0$ has an even priority that is larger than any
of the
priorities assigned to the $a_i$ vertices. }
\label{fig:decel}
\end{figure}
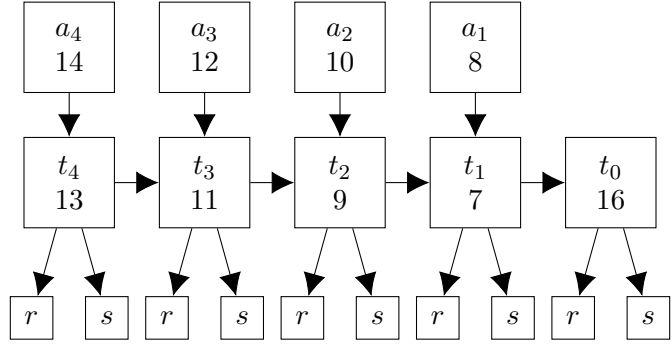

\paragraph{\bf The deceleration lane.} Friedmann's example contains one copy of
the deceleration lane. The deceleration lane has a specified length $m$, and
Figure~\ref{fig:decel} shows an example of a deceleration lane of length~$4$.
Friedmann's construction contains one copy of the deceleration lane of
length~$2n$. 
Remember that our diagrams use representative priorities, which preserve the
order and parity of the priorities used, but not their values.

A key property of the deceleration lane is that greedy all-switches strategy
improvement requires $m$ iterations to find the optimal strategy. Consider an
initial strategy in which each vertex $t_i$ uses the edge to $r$, and that the
valuation of $r$ is always larger than the valuation of $s$. First note that,
since there is a large even priority on $t_0$, the optimal strategy is for every
vertex $t_i$, with $i \ge 1$, to use the edge to $t_{i-1}$. However, since the
vertices $t_i$ with $i \ge 1$ are all assigned odd priorities, in the initial
strategy only the edge from $t_1$ to $t_0$ is switchable. Furthermore, once this
edge has been switched, only the edge from $t_2$ to $t_1$ is switchable. In this
way, the gadget ensures that $m$ iterations are required to move from the
initial strategy to the optimal strategy for this gadget.

Another important property is that the gadget can be \emph{reset}. This is
achieved by having a single iteration in which the valuation of $s$ is much
larger than the valuation of $r$, followed by another iteration in which the
valuation of $r$ is much larger than the valuation of $s$. In the first
iteration all vertices $t_i$ switch to $s$, and in the second iteration all
vertices switch back to $r$. Note that after the second iteration, we have
arrived back at the initial strategy described above.

\paragraph{\bf The bit gadget.}

\begin{figure}
\begin{center}
\begin{tikzpicture}[node distance=2.8cm,font=\normalsize]
\node[even,align=center] (d) {$d_i$ \\ $3$};
\node[odd,right of=d,align=center] (e) {$e_i$ \\ $4$};
\node[even,right of=e,align=center] (f) {$f_i$ \\ $15$};
\node[even,above of=e,align=center] (h) {$h_i$ \\ $16$};
\node[even,above of=f,align=center] (g) {$g_i$ \\ $5$};
\node[even,above of=h,align=center] (k) {$k_i$ \\ $13$};

\node[even,below right=2cm and 2cm of d] (a_1) {$a_1$};
\node[even,left=1cm of a_1] (a_2) {$a_2$};
\node[left=1cm of a_2] (a_3) {$\dots$};
\node[even,left=1cm of a_3] (a_4) [align=center] {$a_{2i}$};

\node[even,above left=0.5cm and 2cm of k] (g_1) {$g_{i+1}$};
\node[even,below left=0.5cm and 2cm of k] (g_2) {$g_n$};
\node[even,below left=2.5cm and 2cm of k] (g_n) {$x$};

\node [left=2.45cm of k] (dots2) {$\vdots$};

\node[even,left=2cm of d] (r) {$r$};
\node[even,above left=0.45cm and 2cm of d] (s) {$s$};

\path[->]
	(g) edge (f)
	(g) edge (k)
	(f) edge (e)
	(e) edge (h)
	(h) edge (k)
	(e) edge [bend left] (d)
	(d) edge [bend left] (e)

	(d) edge (a_4)
	(d) edge (a_2)
	(d) edge (a_1)

	(k) edge (g_1)
	(k) edge (g_2)
	(k) edge (g_n)

	(d) edge (r)
	(d) edge (s)
	;
\end{tikzpicture}
\end{center}
\caption{An example of Friedman's bit gadget. The vertex $d_i$ has a small odd
priority, while the vertex $e_i$ has an even priority that is equal to
$\pri(d_i)~+~1$. The vertex $f_i$ has a large odd priority, while the vertex
$h_i$ has a large even priority that is equal to $\pri(h_i) + 1$. The
priorities assigned to $k_i$ and $g_i$ are not relevant to the operation of
Friedmann's construction.}
\label{fig:bit}
\end{figure}
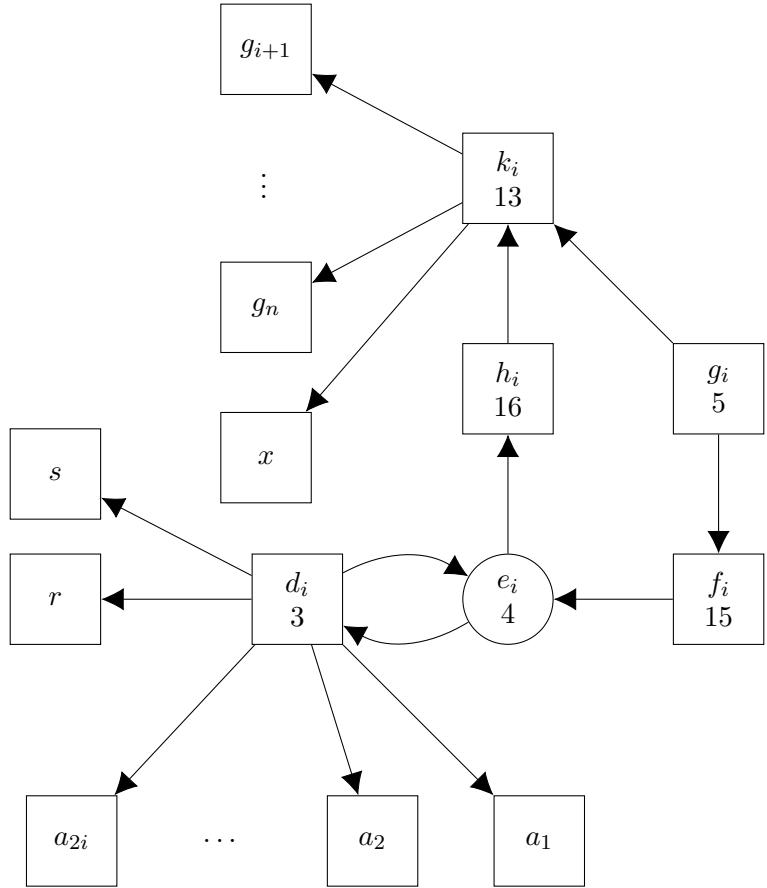

The bit gadget is designed to store one bit of a binary counter. The
clock construction will contain $n$ copies of this gadget, which will be indexed $1$
through $n$. Figure~\ref{fig:bit} gives a depiction of a bit gadget with index
$i$.

The current value of the bit for index $i$ is represented by the choice that the current
strategy makes at~$d_i$. More precisely, for every strategy $\sigma$ we have:
\begin{itemize}
\item If $\sigma(d_i) = e_i$, then bit $i$ is $1$ in $\sigma$.
\item If $\sigma(d_i) \ne e_i$, then bit $i$ is $0$ in $\sigma$.
\end{itemize}
\noindent The Odd vertex $e_i$ plays a crucial role in this gadget. If
$\sigma(d_i) = e_i$, then Odd's best response is to use edge $(e_i,h_i)$, to
avoid creating the even cycle between $d_i$ and $e_i$. On the other hand, if
$\sigma(d_i) \ne e_i$, then Odd's best response is to use $(e_i,d_i)$, to avoid
seeing the large even priority at $h_i$.

One thing to note is that, in the case where $\sigma(d_i) \ne e_i$, the edge to
$e_i$ is always switchable. To prevent $d_i$ from immediately switching to
$e_i$, we must ensure that there is always a more appealing outgoing edge from
$e_i$, so that the greedy all-switches rule will switch that edge instead. The
edges from $d_i$ to the deceleration lane provide this. Once~$t_1$ has switched
to $t_0$, the edge from $d_i$ to $a_{1}$ becomes more appealing than the edge to
$e_i$, once~$t_2$ has switched to $t_1$, the edge from $d_i$ to $a_{2}$ becomes
more appealing than the edge to $e_i$, and so on. In this way, we are able to
prevent $d_i$ from switching to $e_i$ for $2i$ iterations by providing outgoing
edges to the first $2i$ vertices of the deceleration lane.

\paragraph{\bf The vertices $s$ and $r$.}

The vertex $s$ has outgoing edges to every vertex $f_i$ in the bit gadgets, and
the vertex $r$ has outgoing edges to every vertex $g_i$ in the bit gadgets. If
$i$ is the index of the least significant $1$ bit, then $s$ chooses the edge to
$f_i$ and $r$ chooses the edge to $g_i$. The priority assigned to $r$ is larger
than the priority assigned to $s$, which ensures that the valuation of $r$ is
usually larger than the valuation of $s$, as required to make the deceleration
lane work.

When the counter moves from one bit-string to the next, the index of the least
significant~$1$ changes to some $i' \ne i$. The vertex $s$ switches to $f_{i'}$
one iteration before the vertex~$r$ switches to $g_{i'}$. This creates the
single iteration in which the valuation of~$s$ is larger than the valuation
of~$r$, which resets the deceleration lane.

\paragraph{\bf Simulating a binary counter.}

To simulate a binary counter, we must do two things. Firstly, we must ensure
that if the counter is currently at some bit-string $\C \in \{0, 1\}^n$, then
the \emph{least significant zero} in $\C$ must be flipped to a one. Secondly,
once this has been done, all bits whose index is smaller than the least
significant zero must be set to $0$. If these two operations are always
performed, then strategy improvement will indeed count through all binary
strings.

The least significant zero is always flipped because each bit $i$ has $2i$ edges
to the deceleration lane. Since the purpose of the deceleration lane is to
prevent the vertex $d_i$ switching to $e_i$, the vertex $d_{i'}$ where $i'$ is
the index of the least significant zero, is the first to run out of edges, and
subsequently switch to $e_{i'}$.

Once this has occurred, all bits with index smaller than the least significant
zero are set to $0$ due to the following chain of events. The vertex $s$
switches $f_{i'}$,  and then the vertex~$d_{i''}$ in all bits with index $i'' <
i'$ will be switched to $s$. Since~$d_{i''}$ no longer uses the edge to~$e_{i''}$, 
the bit has now been set to $0$.

\paragraph{\bf Our modifications to Friedmann's example.}

%We use two copies of Friedmann's example, which will be numbered $0$ and $1$. T
In order to use Friedmann's example as a clock, we make a few minor adjustments
to it. Firstly, we make the deceleration lane longer. Friedmann's example uses a
deceleration lane of length $2n$, but we use a deceleration lane of length
$\lale$. Furthermore, while the vertex $d_i$ has outgoing edges to each $a_j$
with $j \le 2i$ in Friedmann's version, in our modified version the vertex $d_i$
has outgoing edges to each $a_j$ with $j \le 2i + 2k + 2n + 6$. 

The reason for this is that Friedmann's example can move from one bit-string to
the next in as little as four iterations, but we need more time in order to
compute the circuit~$F$. By making the deceleration lane longer, we slow down
the construction, and ensure that there are at least $2k + 2n + 6$ iterations
before the clock moves from one bit-string to the next. 

The second change that we make is to change the priorities, because we need to
make room for the gadgets that we add later. However, we have not made any
fundamental changes to the priorities: the ordering of priorities between the
vertices and their parity is maintained. We have simply added larger gaps
between them.

The following table specifies the version of the construction that we use.
Observe that two copies are specified: one for $j = 0$ and the other
for $j = 1$. Furthermore, observe that the vertex $x$ will be the sink in our
one-sink game.

\setlength{\tabcolsep}{6pt}
\begin{small}
\begin{center}
\begin{tabular}{l|l|l|l|l}
Vertex & Conditions  & Edges  & Priority & Player \\
\hline
$t_0^j$ & $j \in \{0, 1\}$ & $r^j$, $s^j$ & $\pp(2, 0, 2k+4n+4, j, 0)$ & Even \\
$t_l^j$ & $j \in \{0, 1\}$, & $r^j$, $s^j$, $t^j_{l-1}$ &
$\pp(2, 0, l, j, 1)$ & Even \\
	& $1 \le l \le \lale$ &&& \\ 
$a_l^j$ & $j \in \{0, 1\}$, & $t^j_l$ & $\pp(2, 0, l
+1, j,
0)$ & Even \\
	& $1 \le l \le \lale$ &&& \\
\hline
$d^j_i$ & $j \in \{0, 1\}$, $1 \le i \le n$ & $e^j_i$, $s^j$, $r^j$, $a^j_l $ for & $\pp(1, i, 0, j, 1)$ & Even  \\
& & $1 \le l \le \laletwo + 2i$  & &\\
$e^j_i$ & $j \in \{0, 1\}$, $1 \le i \le n$ & $h^j_i$, $d_i$ & $\pp(1, i, 1, j, 0)$ &
Odd \\
$g^j_i$ & $j \in \{0, 1\}$, $1 \le i \le n$ & $f^j_i$ & $\pp(1, i, 2, j, 1)$ &
Even \\
$k^j_i$ & $j \in \{0, 1\}$, $1 \le i \le n$ & $x$, $g^j_l$, for $i < l \le
n$ & $\pp(8, i, 0, j, 1)$ & Even \\
$f^j_i$ & $j \in \{0, 1\}$, $1 \le i \le n$ & $e^j_i$ & $\pp(8, i, 1, j, 1)$ &
Even \\
$h^j_i$ & $j \in \{0, 1\}$, $1 \le i \le n$ & $k^j_i$ & $\pp(8, i, 2, j, 0)$ &
Even \\
\hline
$s^j$ & $j \in \{0, 1\}$ & $x$, $f^j_l$ for $1 \le l \le n$ & $\pp(7, 0, 0,
j, 0)$ & Even \\
$r^j$ & $j \in \{0, 1\}$ & $x$, $g^j_l$ for $1 \le l \le n$ & $\pp(7, 0, 1,
j, 0)$ & Even \\
$x$ &  & $x$ & $\pp(0, 0, 0, 0, 1)$ & Even \\
\end{tabular}
\end{center}
\end{small}

\subsection{Our construction}

\begin{figure}
% fill white since these will be used in the foreground
\tikzset{box/.style={font=\sffamily,draw, thick, minimum height=1.8cm,minimum width=2cm}}
\tikzset{main/.style={box,fill=gray!10}} %,label={below:{\text{Circuit} 1}}}}
\tikzstyle{component}=[fill=white, font=\sffamily,draw, thick, minimum size=0cm,align=center]

\begin{tikzpicture}[
	decoration={markings,mark=at position 0.6 with {\arrow{triangle 60}}},
	path/.style={thick},
	clip=false
	%path/.style={ultra thick,>=stealth,postaction=decorate}
	%every node/.style={
]

% BOXES IN THE BACKGROUND
\begin{pgfonlayer}{background}
\node[main, xshift=-2cm, yshift=0cm, align=center, text width=1cm] (input1) {${\text{Input}}_1^0$};
\node[main, xshift=2cm, yshift=0cm, align=center, text width=1cm] (input2) {${\text{Input}}_2^0$};
\node[main, xshift=-2cm, yshift=3cm, align=center, text width=1cm] (gate3) {${\text{Gate}}_3^0$ (NOT)};
\node[main, xshift=2cm, yshift=3cm, align=center, text width=1cm] (gate4) {${\text{Gate}}_4^0$ (OR)};
\node[main, xshift=-2cm, yshift=6cm, align=center, text width=1cm] (gate5) {${\text{Gate}}_5^0$ (OR)};
\node[main, xshift=2cm, yshift=6cm, align=center, text width=1cm] (gate6) {${\text{Gate}}_6^0$ (NOT)};
\node[main, xshift=-2cm, yshift=9cm, align=center, text width=1cm] (input7) {${\text{Input}}_1^1$};
\node[main, xshift=2cm, yshift=9cm, align=center, text width=1cm] (input8) {${\text{Input}}_2^1$};
\end{pgfonlayer}{background}

% circuit 
\node[component,dashed, xshift=-1cm,yshift=0.5cm]  (r-dashed1) at (input1.west) {$r^0$};
\node[component,dashed, xshift=-1cm,yshift=-0.5cm]  (s-dashed1) at (input1.west) {$s^0$};
%\node[component,dashed, xshift=0.5cm,yshift=-1cm]  (s1-dashed1) at (input1.south) {$s^1$};
%\node[component,dashed, xshift=-0.5cm,yshift=-1cm]  (r1-dashed1) at (input1.south) {$r^1$};
\node[component,dashed, xshift=1cm,yshift=0.5cm]  (r-dashed2)  at (input2.east) {$r^0$};
\node[component,dashed, xshift=1cm,yshift=-0.5cm]  (s-dashed2)  at (input2.east) {$s^0$};
\node[component,dashed, xshift=-1cm,yshift=0.5cm]  (s1-dashed2)  at (input2.west) {$s^1$};
\node[component,dashed, xshift=-1cm,yshift=-0.5cm]  (r1-dashed2)  at (input2.west) {$r^1$};
\node[component,dashed, xshift=-1cm,yshift=0.5cm]  (r-dashed3) at (gate3.west) {$r^0$};
\node[component,dashed, xshift=-1cm,yshift=-0.5cm]  (s-dashed3) at (gate3.west) {$s^0$};
\node[component,dashed, xshift=1cm,yshift=0.5cm]  (r-dashed4)  at (gate4.east) {$r^0$};
\node[component,dashed, xshift=1cm,yshift=-0.5cm]  (s-dashed4)  at (gate4.east) {$s^0$};
\node[component,dashed, xshift=-1cm,yshift=0.5cm]  (r-dashed5)  at (gate5.west) {$r^0$};
\node[component,dashed, xshift=-1cm,yshift=-0.5cm]  (s-dashed5)  at (gate5.west) {$s^0$};
\node[component,dashed, xshift=1cm,yshift=0.5cm]  (r-dashed6)  at (gate6.east) {$r^0$};
\node[component,dashed, xshift=1cm,yshift=-0.5cm]  (s-dashed6)  at (gate6.east) {$s^0$};
\node[component,dashed, xshift=-1cm,yshift=0.5cm]  (r-dashed7)  at (input7.west) {$r^1$};
\node[component,dashed, xshift=-1cm,yshift=-0.5cm]  (s-dashed7)  at (input7.west) {$s^1$};
\node[component,dashed, xshift=1cm,yshift=0.5cm]  (r-dashed8)  at (input8.east) {$r^1$};
\node[component,dashed, xshift=1cm,yshift=-0.5cm]  (s-dashed8)  at (input8.east) {$s^1$};
\node[component,dashed, xshift=-1cm,yshift=0.5cm]  (r0-dashed8)  at (input8.west) {$r^0$};
\node[component,dashed, xshift=-1cm,yshift=-0.5cm]  (s0-dashed8)  at (input8.west) {$s^0$};

% gate outputs
\node[component,xshift=0cm,yshift=0cm]  (o1) at (input1.north) {$o_1^0$};
\node[component,xshift=0cm,yshift=0cm]  (o2) at (input2.north) {$o_2^0$};
\node[component,xshift=0cm,yshift=0cm]  (o3) at (gate3.north) {$o_3^0$};
\node[component,xshift=0cm,yshift=0cm]  (o4) at (gate4.north) {$o_4^0$};
\node[component,xshift=0cm,yshift=0cm]  (o5) at (gate5.north) {$o_5^0$};
\node[component,xshift=0cm,yshift=0cm]  (o6) at (gate6.north) {$o_6^0$};
\node[component,xshift=0cm,yshift=0cm]  (o7) at (input7.north) {$o_1^1$};
\node[component,xshift=0cm,yshift=0cm]  (o8) at (input8.north) {$o_2^1$};

% depth labels
\node[xshift=-3cm] () at (input1.west) {Depth $0$};
\node[xshift=-3cm] () at (gate3.west) {Depth $1$};
\node[xshift=-3cm] () at (gate5.west) {Depth $2$};

% EDGES
%%%%%%%%%%%%%%%%%%%%%%%%%%%%%%%%%%%%%%%%%%%%%%%%%%%%%%%%%%%%%%%%%%%%%%%%%%%%%%

% level 1 to inputs
\draw[path,->] (gate3.south) -- (o1);
\draw[path,->] (gate4.south) -- (o1);
\draw[path,->] (gate4.south) -- (o2);

% level 2 to level 1
\draw[path,->] (gate5.south) -- (o3);
\draw[path,->] (gate5.south) -- (o4);
\draw[path,->] (gate6.south) -- (o4);

% inputs 1 to outputs 0
\draw[path,->] (input7.south) -- (o5);
\draw[path,->] (input8.south) -- (o6);

% gates to s and r
\draw[path,->] (input1.west) -- (r-dashed1);
\draw[path,->] (input1.west) -- (s-dashed1);
\draw[path,->] (input1.east) -- (r1-dashed2);
\draw[path,->] (input1.east) -- (s1-dashed2);
\draw[path,->] (input2.east) -- (r-dashed2);
\draw[path,->] (input2.east) -- (s-dashed2);
\draw[path,->] (input2.west) -- (r1-dashed2);
\draw[path,->] (input2.west) -- (s1-dashed2);
\draw[path,->] (gate3.west) -- (r-dashed3);
\draw[path,->] (gate3.west) -- (s-dashed3);
\draw[path,->] (gate4.east) -- (r-dashed4);
\draw[path,->] (gate4.east) -- (s-dashed4);
\draw[path,->] (gate5.west) -- (r-dashed5);
\draw[path,->] (gate5.west) -- (s-dashed5);
\draw[path,->] (gate6.east) -- (r-dashed6);
\draw[path,->] (gate6.east) -- (s-dashed6);
\draw[path,->] (input7.west) -- (r-dashed7);
\draw[path,->] (input7.west) -- (s-dashed7);
\draw[path,->] (input8.east) -- (r-dashed8);
\draw[path,->] (input8.east) -- (s-dashed8);
\draw[path,->] (input7.east) -- (r0-dashed8);
\draw[path,->] (input7.east) -- (s0-dashed8);
\draw[path,->] (input8.west) -- (r0-dashed8);
\draw[path,->] (input8.west) -- (s0-dashed8);

\end{tikzpicture}

%%% Local Variables:
%%% mode: latex
%%% TeX-master: "../note.tex"
%%% End:
\caption{\label{fig:overview3} Example of how we implement a specific circuit with three gates.}
\end{figure}
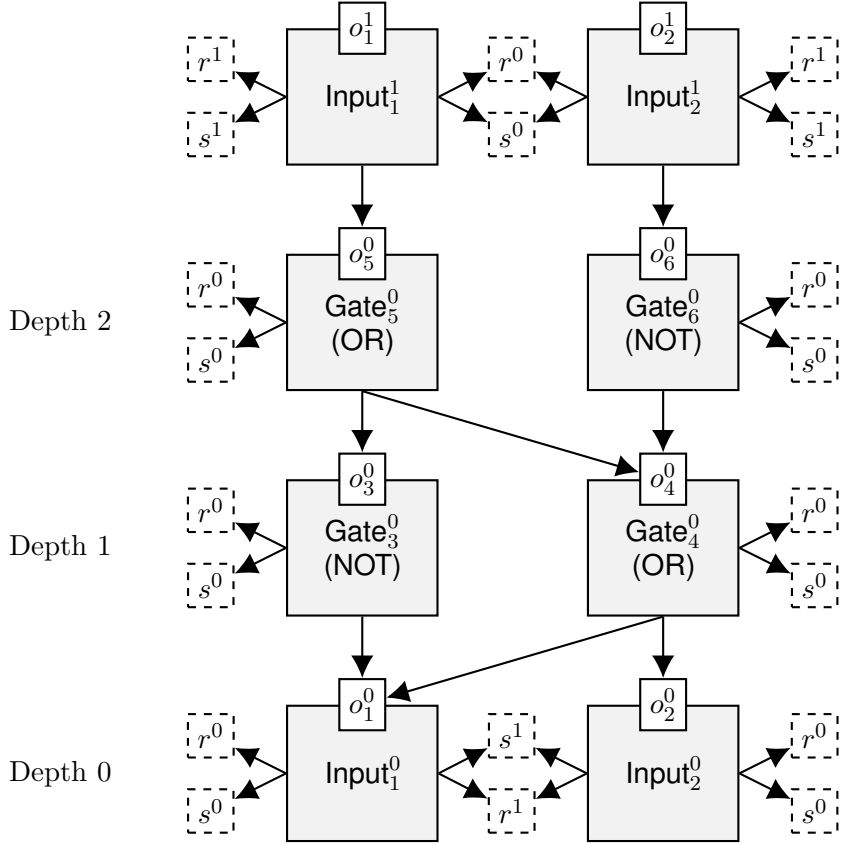

\paragraph{\bf Circuits.}
Given a circuit, we will produce a gadget that simulates that circuit. An
example is given in Figure~\ref{fig:overview3}.
For each gate in the circuit, we design a gadget that computes the output
of that gate. The idea is that greedy all-switches strategy improvement will
compute these gates in depth order. Starting from an initial strategy, the first
iteration will compute the outputs for all gates of depth $1$, the next
iteration will use these outputs to compute the outputs for all gates of depth
$2$, and so on. In this way, after $k$ iterations of strategy improvement, the
outputs of the circuit will have been computed. We then use one additional
iteration to store these outputs in an input/output gadget.

Strategy improvement valuations will be used to represent the output of each
gate. Each gate $i$ has a state $o^j_i$, and the valuation of this state will
indicate whether the gate evaluates to true or false. In particular the
following rules will be followed. 
\begin{property}
\label{prop:rules}
In every strategy $\sigma$ we have the following properties.
\begin{enumerate}
\item Before the gate has been evaluated, we will have $\val^\sigma(o^j_i)
\sqsubset \val^{\sigma}(r^j)$.
\item If the gate has been evaluated to false, we will continue to have 
$\val^\sigma(o^j_i) \sqsubset \val^{\sigma}(r^j)$.
\item If the gate has been evaluated to true, then we will instead have
$\val^\sigma(r^j) \sqsubset \val^{\sigma}(o^j_i)$, and $\maxdiff^{\sigma}(r^j,
o^j_i)$ will be a large even priority.
\end{enumerate}
\end{property}

\noindent The input/output gadgets are connected to both circuits, and these
gadgets have two modes.
\begin{enumerate}
\item When circuit $j$ is computing, the gadget is in \emph{output mode}, where
it reads the output of circuit $j$ and stores it.
\item When circuit $1-j$ is computing, the gadget is in \emph{input mode}, where
it outputs the value that was stored from the previous computation into circuit
$1-j$.
\end{enumerate}
Therefore, the gates of depth $1$ in circuit $j$ read their input from the
input/output gadgets in circuit $1-j$, while the input/output gadgets in circuit
$j$ read their input from the outputs of circuit $j$. To formalise this, we
introduce the following notation.
For every $\notg$-gate, we define $\inputstate(i,j)$ as follows:
\begin{equation*}
\inputstate(i,j) = \begin{cases}
o^{1-j}_{\inp(i)} & \text{if $d(i) = 1$,} \\
o^j_{\inp(i)} & \text{if $d(i) > 1$.} 
\end{cases}
\end{equation*}
\noindent For every $\org$-gate, and every $l \in \{1, 2\}$, we define
$\inputstate(i,j,l)$ as follows:
\begin{equation*}
\inputstate(i,j,l) = \begin{cases}
o^{1-j}_{\inp_l(i)} & \text{if $d(i) = 1$,} \\
o^j_{\inp_l(i)} & \text{if $d(i) > 1$.} 
\end{cases}
\end{equation*}

\paragraph{\bf The clocks.}
As we have mentioned, we use two copies of Friedmann's example to act as clocks
in our construction. These clocks will be used to drive the computation. In
particular, the vertices $r^j$ and $s^j$ will play a crucial role in
synchronising the two circuits. As described in the previous section, when the
clock \emph{advances}, i.e., when it moves from one bit-string to the next, there
is a single iteration in which the valuation of $s^j$ is much larger than the
valuation of $r^j$. This event will trigger the computation.
\begin{itemize}
\item The iteration in which the valuation of $s^0$ is much larger than the
valuation of $r^0$ will trigger the start of computation in circuit $0$.
\item The iteration in which the valuation of $s^1$ is much larger than the
valuation of $r^1$ will trigger the start of computation in circuit $1$.
\end{itemize}

In order for this approach to work, we must ensure that the two clocks are
properly synchronised. In particular, the gap between computation starting in
circuit $j$ and computation starting in circuit $1-j$ must be at least $k+3$, to
give enough time for circuit $j$ to compute the output values, and for these
values to be stored. We now define notation for this purpose. First we define
the number of iterations that it takes for a clock to move from bit-string $\C$
to $\C + 1$. For every bit-string $\C \in \{0, 1\}^n$, we define $\lsz(\C)$ to
be the index of the \emph{least significant zero} in $\C$: that is, the smallest
index $i$ such that $\C_i = 0$. For each $\C \in \{0, 1\}^n$, we define:
\begin{equation*}
\length(\C) = \Bigl( \laletwo \Bigr) + 2\lsz(\C) + 5.
\end{equation*}
This term can be understood as the length of the deceleration lane to which all
bits in the clock have edges, plus the number of extra iterations it takes to
flip the least-significant zero, plus five extra iterations needed to transition
between the two bit-strings.

Next we introduce the following \emph{delay} function, which gives the amount of
time each circuit spends computing. For each $j \in \{0, 1\}$ and each $\C \in
\{0, 1\}^n$, we define:
\begin{equation*}
\delay(j, \C) = \begin{cases}
\Bigl( d(C) + 3 \Bigr) + 2n & \text{if $j = 0$,} \\
\Bigl( d(C) + 3 \Bigr) + 2 \cdot \lsz(\C) + 5 & \text{if $j = 1$.} 
\end{cases}
\end{equation*}
Circuit $1$ starts computing $\delay(0, \C)$ iterations after Circuit $0$
started computing, and Circuit $0$ starts computing $\delay(1, \C)$ iterations
after circuit $1$ started computing. Observe that $\delay(0, \C) + \delay(1, \C)
= \length(\C)$, which ensures that the two circuits do not drift relative to
each other. The term $d(C) + 3$ in each of the delays ensures that there is
always enough time to compute the circuit, before the next circuit begins the
subsequent computation.

\begin{figure}[H]
%%%%%%%%%%%%%%%%%%%%%%%%%%%%%%%%%%%%%%%%%%%%%%%%%%%%%%%%%%%%%%%%%%%%%%%%%%%%
% OR GATE 
%%%%%%%%%%%%%%%%%%%%%%%%%%%%%%%%%%%%%%%%%%%%%%%%%%%%%%%%%%%%%%%%%%%%%%%%%%%%
\begin{minipage}{0.28\textwidth}
% \begin{center}
\begin{tikzpicture}[node distance=1cm,font=\scriptsize]
\node [even] (o) {$o^j_i$ \\ $1$};
\node [even] (r) [left=1cm of o] {$r^j$};
\node [even] (s) [above left=1cm and 1cm of o] {$s^j$};
\node [even] (i1) [below left=1cm and 0.5cm of o] {$o^j_{\inp_1(i)}$};
\node [even] (i2) [below right=1cm and 0.5cm of o] {$o^j_{\inp_2(i)}$};

\path[->]
	(o) edge (r)
	(o) edge (s)
	(o) edge (i1)
	(o) edge (i2)
	;
\end{tikzpicture}
% \end{center}
%\caption{\org gate.}
\end{minipage}
\begin{minipage}{0.71\textwidth}
%%%%%%%%%%%%%%%%%%%%%%%%%%%%%%%%%%%%%%%%%%%%%%%%%%%%%%%%%%%%%%%%%%%%%%%%%%%%
% OR GATE TABLE
%%%%%%%%%%%%%%%%%%%%%%%%%%%%%%%%%%%%%%%%%%%%%%%%%%%%%%%%%%%%%%%%%%%%%%%%%%%%
\begin{center}
\begin{tabular}{l|l|l|l|l}
Vertex & Conditions  & Edges  & Priority & Player\\
\hline
$o^j_i$ & $j \in \{0, 1\}$, & $s^j$, $r^j$, 
& $\pp(4, i, 0, j, 1)$ & Even \\
& $i \in \org$& $\inputstate(i,j,1)$ & & \\
& & $\inputstate(i,j,2)$ & & 
\end{tabular}
\end{center}
%\caption{\org gate priorities.}
\end{minipage}
\caption{The \org gate.}
\label{fig:or}
\end{figure}
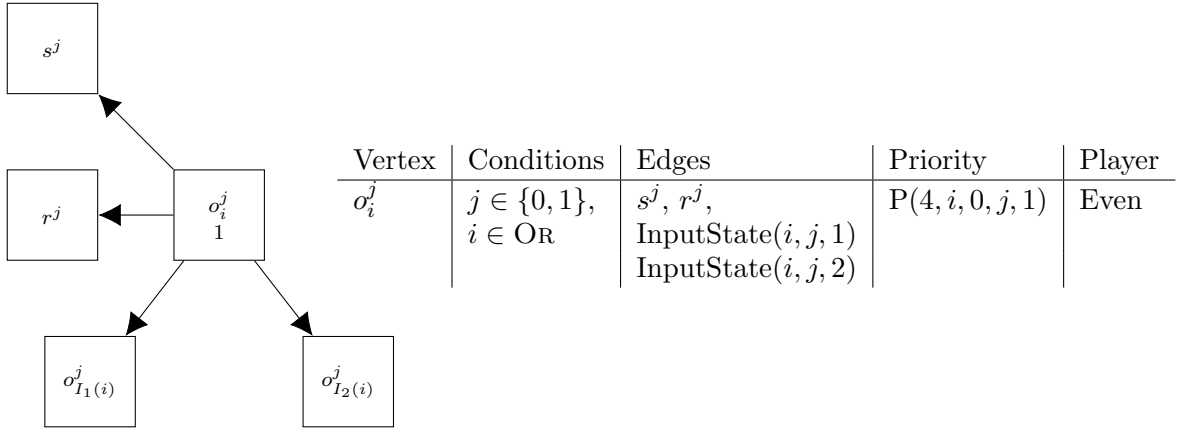

\paragraph{\bf Or gates.}
The gadget for a gate $i \in \org$ is quite simple, and is shown in
Figure~\ref{fig:or}. It is not difficult to verify that the three rules given in
Property~\ref{prop:rules} hold for this gate. Before both inputs have been
evaluated, the best strategy at $o^j_i$ is to move directly to $r^j$, since the
valuation of both inputs is lower than the valuation of $r^j$. Note that in this
configuration the valuation of $o^j_i$ is smaller than the valuation of $r^j$,
since $o^j_i$ has been assigned an odd priority.

Since, by assumption, both inputs have the same depth, they will both be
evaluated at the same time. If they both evaluate to false, then nothing changes
and the optimal strategy at $o^j_i$ will still be $r^j$. This satisfies the
second rule. On the other hand, if at least one input evaluates to true, then
the optimal strategy at $o^j_i$ is to switch to the corresponding input states.
Since the valuation of this input state is now bigger than $r^j$, the valuation
of $o^j_i$ will also be bigger than $r^j$, so the third rule is also satisfied.

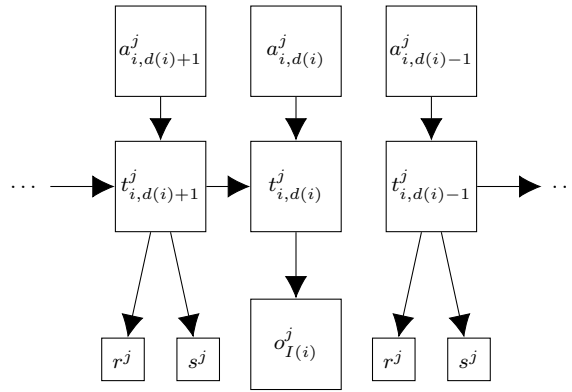
\begin{figure}[b]
\begin{center}
\begin{tikzpicture}[node distance=1.8cm,font=\scriptsize]
\node [] (t1) {$\dots$};
\node [even,left of=t1] (t2) {$t^j_{i,d(i)-1}$};
\node [even,left of=t2] (t3) {$t^j_{i,d(i)}$};
\node [even,left of=t3] (t4) {$t^j_{i,d(i)+1}$};
\node [left of=t4] (t5) {$\dots$};

\node[even,above of=t2] (a2) {$a^j_{i,d(i)-1}$};
\node[even,above of=t3] (a3) {$a^j_{i,d(i)}$};
\node[even,above of=t4] (a4) {$a^j_{i,d(i)+1}$};

\node[even,below of=t3,node distance=2.1cm] (v) {$o^j_{\inp(i)}$};

\path[->]
	(t5) edge (t4)
	(t4) edge (t3)
	(t2) edge (t1)
	;

\path[->]
	%(a5) edge (t5)
	(a4) edge (t4)
	(a3) edge (t3)
	(a2) edge (t2)
	%(a1) edge (t1)
	;

\path[->]
	(t3) edge (v)
	;

\foreach \x in {t2, t4}
{
	\node [even,draw,minimum size=0.8cm] (\x1) at ($(\x) - (0.5,2.3)$) {$r^j$};
	\node [even,draw,minimum size=0.8cm] (\x2) at ($(\x) - (-0.5,2.3)$) {$s^j$};

	\path[->]
		(\x) edge (\x1)
		(\x) edge (\x2)
		;
}

\end{tikzpicture}
\end{center}
\caption{\label{fig:modified} Modified deceleration lane for a \notg gate $i$ in circuit $j$.}
\end{figure}

\paragraph{\bf Not gates.} The construction for a gate $i \in \notg$ is more
involved. The gadget is quite similar to a bit-gadget from Friedmann's
construction. However, we use a special \emph{modified deceleration lane}, which
is shown in Figure~\ref{fig:modified}.

The modified deceleration lane is almost identical to Friedmann's deceleration
lane, except that state $t^j_{i, d(i)}$ is connected to the output state of the
input gate. The idea is that, for the first $d(i) - 1$ iterations the
deceleration lane behaves as normal. Then, in iteration $d(i)$, the input gate
is evaluated. If it evaluates to true then the valuation of $t^j_{i, d(i)}$ will
be large, and the deceleration lane continues switching as normal. If it evaluates to
false, then the valuation of $t^j_{i, d(i)}$ will be low, and the deceleration
lane will stop switching.

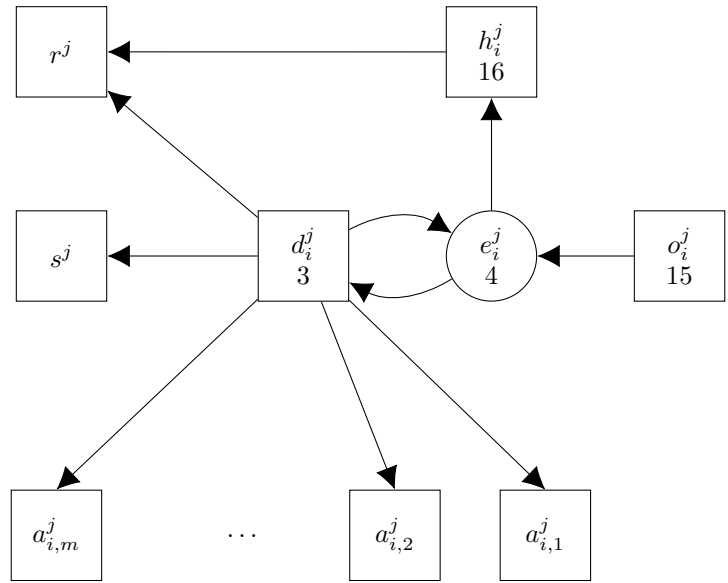
\begin{figure}
\begin{center}
\begin{tikzpicture}[node distance=2.5cm,font=\small]%\scriptsize]
\node[even,align=center] (d) {$d^j_i$ \\ $3$};
\node[odd,right of=d,align=center] (e) {$e^j_i$ \\ $4$};
\node[even,right of=e,align=center] (f) {$o^j_i$ \\ $15$};
\node[even,above=1.5cm of e,align=center] (k) {$h^j_i$ \\ $16$};

\node[even,below right=2.5cm and 2cm of d] (a_1) {$a^j_{i,1}$};
\node[even,left of=a_1,node distance=2cm] (a_2) {$a^j_{i,2}$};
\node[left of=a_2,node distance=2cm] (a_3) {$\dots$};
\node[even,left of=a_3,node distance=2.5cm] (a_4) [align=center] {$a^j_{i,m}$};

%\node[even,above left=0.00cm and 2cm of k] (g_1) {$g^j_{k+1}$};
%\node[even,below left=0.00cm and 2cm of k] (g_2) {$g^j_{k+n}$};
%\node[even,below left=1.5cm and 2cm of k] (g_n) {$x^j$};

%\node [left=2.45cm of k] (dots2) {$\vdots$};

\node[even,left=2cm of d] (s) {$s^j$};
\node[even,above left=1.5cm and 2cm of d] (r) {$r^j$};

\path[->]
	(f) edge (e)
	(e) edge (k)
	(e) edge [bend left] (d)
	(d) edge [bend left] (e)

	(d) edge (a_4.north)
	(d) edge (a_2.north)
	(d) edge (a_1.north)

	%(k) edge (g_1)
	%(k) edge (g_2)
	%(k) edge (g_n)
	(k) edge (r)

	(d) edge (r)
	(d) edge (s)
	;
\end{tikzpicture}
\end{center}
\caption{\notg gate with index $i$ in circuit $j$.}
\label{fig:not}
\end{figure}

The \notg gate gadget, which is shown in Figure~\ref{fig:not} is a simplified
bit gadget that is connected to the modified deceleration lane. As in
Friedmann's construction, the strategy chosen at $d^j_i$ will represent the
output of the gate. In a strategy $\sigma$, the gate outputs $1$ if
$\sigma(d^j_i) = e^j_i$, and it outputs $0$ otherwise. As we know, Friedmann's
bit gadget is distracted from switching $d^j_i$ to $e^j_i$ by the deceleration
lane. By using the modified deceleration lane, we instead obtain a \notg gate.
Since the deceleration lane keeps on switching if and only if the input gate
evaluates to true, the state $d^j_i$ will switch to $e^j_i$ in iteration $d(i)$
if and only if the input gate evaluates to false. This is the key property that
makes the \notg gate work.

To see that the three rules specified in Property~\ref{prop:rules} are
respected, observe that there is a large odd priority on the state $o^j_i$, and
an even larger even priority on the state $h^j_i$. This causes the valuation of
$o^j_i$ to only be larger than the valuation of $r^j$ if and only if $d^j_i$
chooses the edge to $e^j_i$, which only happens when the gate evaluates to true.

Finally, when the computation in circuit $j$ begins again, the \notg-gate is
reset. This is ensured by giving the vertex $d^j_i$ edges to both $s^j$ and
$r^j$. So, when the clock for circuit $j$ advances, no matter what strategy is
currently chosen, the vertex $d^j_i$ first switches to $s^j$, and then to $r^j$,
and then begins switching to the deceleration lane.

The following table formally specifies the \notg gate gadgets that we use in the
construction.

\begin{small}
\begin{center}
\begin{tabular}{l|l|l|l|l}
Vertex & Conditions  & Edges  & Priority & Player \\
\hline
$t^j_{i, 0}$ & $j \in \{0, 1\}$, $i\in \notg$ & $r^j$, $s^j$ & $\pp(5, i,
2k+4n+4, j, 0)$ & Even \\
$t_{i,l}^j$ & $j \in \{0, 1\}$, $i\in \notg$, & $r^j$, $s^j$, $t^j_{i, l-1}$ & $\pp(5, i, l, j, 1)$ & Even \\
	& $1 \le l \le \lale$, & & & \\
& \hfill and $l \ne d(i)$ & & & \\
$t_{i,d(i)}^j$ & $j \in \{0, 1\}$, $i\in \notg$ & $\inputstate(i,j)$ & $\pp(5, i,
d(i), j, 1)$ & Even
\\
$a_{i,l}^j$ & $j \in \{0, 1\}$, $i \in \notg$,  & $t_{i,l}$ & $\pp(5, i, l+1, j, 0)$ & Even \\
	& $1 \le l \le \lale$, & & & \\
& \hfill and $l \ne d(i)$ & & & \\
$a_{i,d(i)}^j$ & $j \in \{0, 1\}$, $i \in \notg$ &
$t_{i,d(i)}$ & $\pp(4, i, 0, j, 0)$ & Even \\ \hline
$d^j_i$ & $j \in \{0, 1\}$, $i \in \notg$ & $s^j$, $r^j$, $e^j_i$, $a^j_{i,l}$
 & $\pp(4,i,0,j,1)$ & Even \\
& &  for $1 \le l \le \lale$ & & \\
$e^j_i$ & $j \in \{0, 1\}$, $i \in \notg$ & $h^j_i$, $d^j_i$ & $\pp(4,i,1,j,0)$
& Odd
\\
$o^j_i$ & $j \in \{0, 1\}$, $i \in \notg$ & $e^j_i$ & $\pp(6,i,0,j,1)$ & Even \\
$h^j_i$ & $j \in \{0, 1\}$, $i \in \notg$ & $r^j$ & $\pp(6,i,1,j,0)$ & Even
\end{tabular}
\end{center}
\end{small}

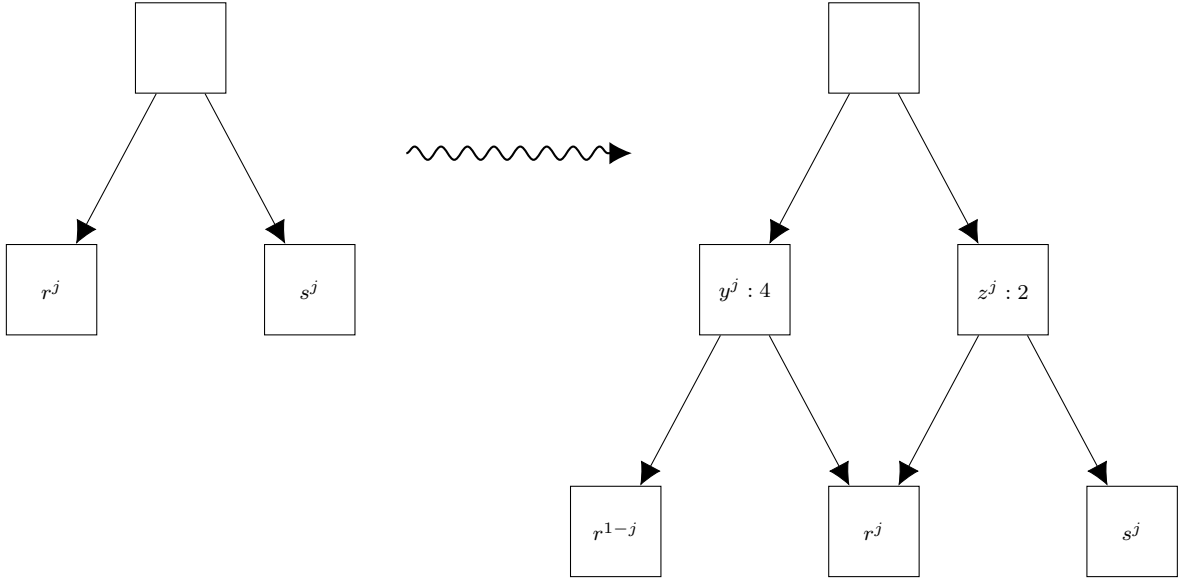
\begin{figure}
\begin{center}
\begin{tikzpicture}[node distance=1cm,font=\scriptsize]
\node [even] (orig) {};
\node [even] (origr) [below left=2cm and 0.5cm of orig] {$r^j$};
\node [even] (origs) [below right=2cm and 0.5cm of orig] {$s^j$};

\path[->]
	(orig) edge (origr)
	(orig) edge (origs)
	;

\path[snake=snake,draw,thick,line after snake=0.2cm,->]
    ($(orig) + (3, -1.4)$) -- ($(orig) + (6, -1.4)$) 
    ;

\node[even] (new) [right=8cm of orig] {};
\node[even] (newy) [below left= 2cm and 0.5cm of new] {$y^j: 4$};
\node[even] (newz) [below right= 2cm and 0.5cm of new] {$z^j: 2$};
\node[even] (news) [below right= 2cm and 0.5cm of newz] {$s^j$};
\node[even] (newr) [below left= 2cm and 0.5cm of newz] {$r^j$};
\node[even] (newrtwo) [below left= 2cm and 0.5cm of newy] {$r^{1-j}$};

\path[->]
	(new) edge (newy)
	(new) edge (newz)
	(newy) edge (newrtwo)
	(newy) edge (newr)
	(newz) edge (news)
	(newz) edge (newr)
	;

\end{tikzpicture}
\end{center}
\caption{Circuit mover gadget for circuit $j$. Left: the edges to $r^j$ and
$s^j$ in a vertex from a \notg-gate. Right: the replacement for these outgoing
edges.}
\label{fig:circuitmover}
\end{figure}

\paragraph{\bf Input/output gates.}
For each input-bit in each copy of the circuit, have an input/output gate.
Recall that these gadgets have two modes. When Circuit~$j$ is computing, the
input/output gadgets in circuit $j$ are in output mode, in which they store the
output of the circuit, and the input/output gadgets in circuit $1-j$ are in
input mode, in which they output the value that was stored in the previous
computation.

\begin{figure}
\begin{center}
\begin{tikzpicture}[node distance=2.5cm,font=\scriptsize]
\node[even,align=center] (d) {$d^j_i$ \\ $3$};
\node[odd,right of=d,align=center] (e) {$e^j_i$ \\ $4$};
\node[even,right=3cm of e,align=center] (f) {$o^j_i$ \\ $15$};
\node[odd,right=1.0cm of e,align=center] (q) {$q^j_{i,0}$ \\ $6$};
\node[even,above=1.5cm of q,align=center] (q_1) {$q^j_{i,1}$ \\ $32$};
\node[even,above=1cm of q_1,align=center] (romj) {$r^{1-j}$ \\ $4$};
\node[even,above=1.5cm of e,align=center] (k) {$h^j_{i,0}$ \\ $2$};

\node[even,above left=1.0cm and 1.5cm of k,align=center] (h_1) {$h^j_{i,1}$ \\ $30$};
\node[even,left=1.5cm of k,align=center] (h_2) {$h^j_{i,2}$ \\ $12$};

\node[even,below right=2.5cm and 2.5cm of d] (a_1) {$a^j_{i,1}$};
\node[even,left of=a_1,node distance=2cm] (a_2) {$a^j_{i,2}$};
\node[left of=a_2,node distance=2cm] (a_3) {$\dots$};
\node[even,left of=a_3,node distance=2cm] (a_4) [align=center] {$a^j_{i,m}$};

\node[even,left=2cm of d] (s) {$z^j$};
%\node[even,above left=1.5cm and 2cm of d] (r) {$y^j$};
\node[even,above left=.5cm and 2cm of d] (r) {$y^j$};
\node[even,above=.5cm of r] (rone) {$r^{1-j}$};
\node[even,above=.5cm of rone] (rtwo) {$r^{j}$};

\path[->]
	(f) edge (q)
    (q) edge (e)
    (q) edge (q_1)
    (q_1) edge (romj)
	(e) edge (k)
	(e) edge [bend left] (d)
	(d) edge [bend left] (e)

	(d) edge (a_4.north)
	(d) edge (a_2.north)
	(d) edge (a_1.north)

	(k) edge (h_1)
	(k) edge (h_2)
    (h_1) edge (rtwo)
    (h_2) edge (rone)

	(d) edge (r)
	(d) edge (s)
	;
\end{tikzpicture}
\end{center}
\caption{\inputg gate with index $i$ in circuit $j$.}
\label{fig:input}
\end{figure}
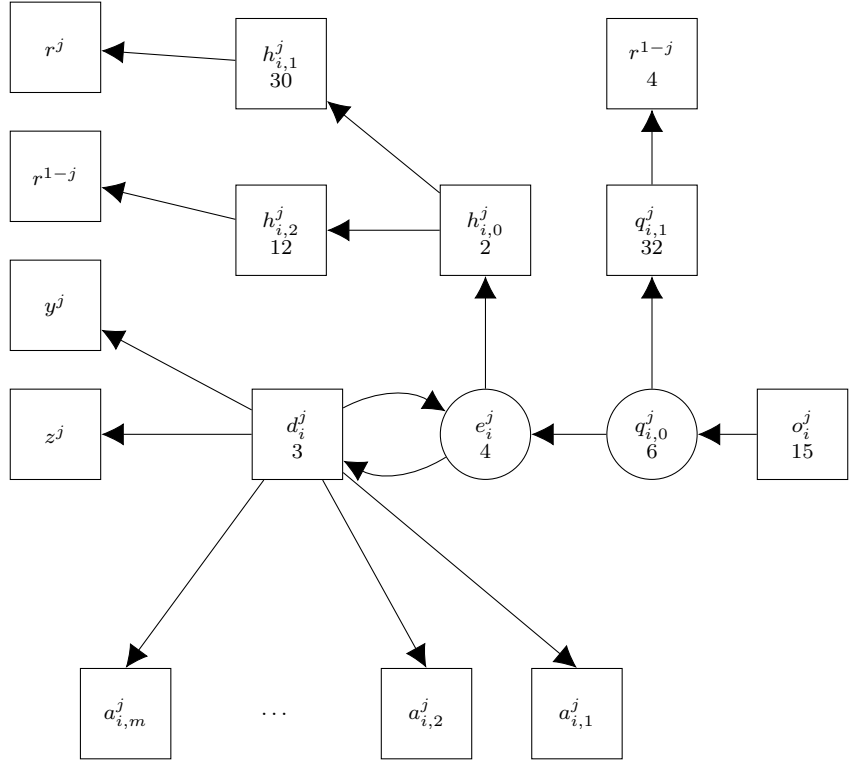

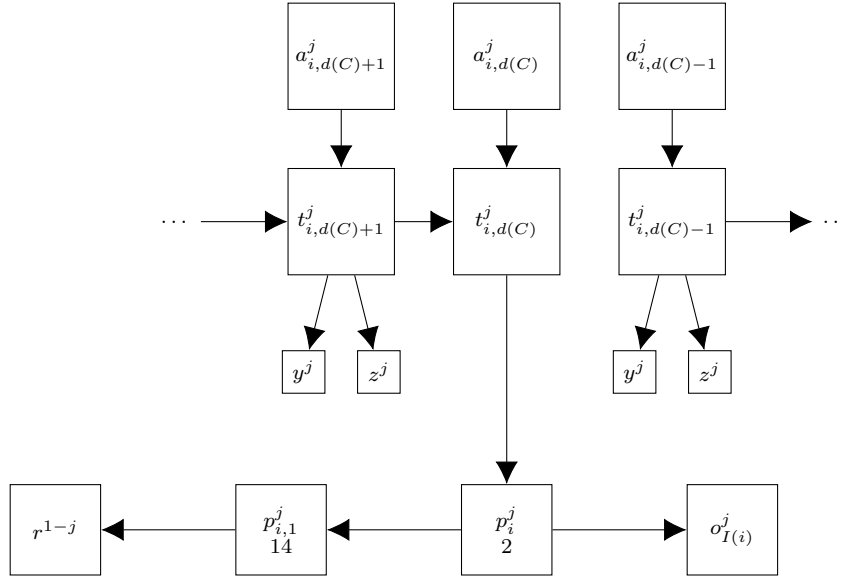
\begin{figure}
\begin{center}
\begin{tikzpicture}[node distance=2.2cm,font=\scriptsize]
\node [] (t1) {$\dots$};
\node [evenlarge,left of=t1] (t2) {$t^j_{i,d(C)-1}$};
\node [evenlarge,left of=t2] (t3) {$t^j_{i,d(C)}$};
\node [evenlarge,left of=t3] (t4) {$t^j_{i,d(C)+1}$};
\node [left of=t4] (t5) {$\dots$};

\node[evenlarge,above of=t2] (a2) {$a^j_{i,d(C)-1}$};
\node[evenlarge,above of=t3] (a3) {$a^j_{i,d(C)}$};
\node[evenlarge,above of=t4] (a4) {$a^j_{i,d(C)+1}$};

\node[even,below of=t3,node distance=4.1cm,align=center] (v) {$p^j_{i}$\\ $2$};
\node[even,right of=v,node distance=3cm] (o) {$o^j_{\inp(i)}$};
\node[even,left of=v,node distance=3cm,align=center] (p) {$p^j_{i,1}$ \\ $14$};
\node[even,left of=p,node distance=3cm] (romj) {$r^{1-j}$};

\path[->]
	(t5) edge (t4)
	(t4) edge (t3)
	(t2) edge (t1)
	;

\path[->]
	%(a5) edge (t5)
	(a4) edge (t4)
	(a3) edge (t3)
	(a2) edge (t2)
	%(a1) edge (t1)
	;

\path[->]
	(t3) edge (v)
    (v) edge (o)
    (v) edge (p)
    (p) edge (romj)
	;

\foreach \x in {t2, t4}
{
	\node [even,draw,minimum size=0.8cm] (\x1) at ($(\x) - (0.5,2)$) {$y^j$};
	\node [even,draw,minimum size=0.8cm] (\x2) at ($(\x) - (-0.5,2)$) {$z^j$};

	\path[->]
		(\x) edge (\x1)
		(\x) edge (\x2)
		;
}

\end{tikzpicture}
\end{center}
\caption{Modified deceleration lane for an \inputg gate $i$ in circuit $j$.}
\label{fig:modifiedinput}
\end{figure}

At its core, the input/output gadget is simply another copy of the \notg-gate
gadget that is connected to the $i$th output bit of circuit $j$. However, we
modify the \notg-gate gadget by adding in extra vertices that allow it to be
\emph{moved} between the two circuits. The most important part of this circuit
mover apparatus is shown in Figure~\ref{fig:circuitmover}: all of the vertices
in the \notg-gate gadget that have edges to $r^j$ and $s^j$ are modified so that
they instead have edges to $y^j$ and $z^j$. Figures~\ref{fig:input}
and~\ref{fig:modifiedinput} show the \inputg gadget and its associated modified
deceleration lane, respectively. There are three differences between this gadget
and the \notg gate, which are the inclusion of the vertices $h^j_{i, \star}$ and
$q^j_{i, \star}$ (shown in Figure~\ref{fig:input}), and 
the vertices $p^j_i$ and $p^j_{i, 1}$ (shown in Figure~\ref{fig:modifiedinput}).
All of these vertices are involved in the operation of moving the gadget between
the two circuits.

When the gadget is in output mode, the vertex $y^j$ chooses the edge to $r^j$,
the vertex $h^j_{i, 0}$ chooses the edge to $h^j_{i, 1}$, and the vertex $p^j_i$
chooses the edge to $o^j_{\inp(i)}$. When these edges are chosen, the gadget is
essentially the same as a \notg-gate at the top of the circuit. So, once the
circuit has finished computing, the vertex $d_i^j$ chooses the edge to $e_i^j$
(i.e., the stored bit is $1$) if and only if the $i$th output from the circuit was
a $0$. Since the circuit was given in negated form, the gadget has therefore
correctly stored the $i$th bit of $F(B)$.

Throughout the computation in circuit $j$, the valuation of $r^j$ is much larger
than the valuation of $r^{1-j}$. The computation in circuit $1-j$ begins when
the clock in circuit $1-j$ advances, which causes the valuation of $r^{1-j}$ to
become much larger than the valuation of $r^j$. When this occurs, the
input/output gate then transitions to input mode. The transition involves the
vertex $y^j$ switching to $r^{1-j}$, the vertex $h^j_{i, 0}$ switching to
$h^j_{i, 2}$, and the vertex $p^j_i$ switching to $p^j_{i, 1}$. Moreover, the
player Odd vertex $q^j_{i, 0}$ switches $e^j_i$. This vertex acts as a circuit
breaker, which makes sure that the output of the gadget is only transmitted to
circuit $1-j$ when the gadget is in input mode.

The key thing is that all of these switches occur \emph{simultaneously} in the
same iteration. Since strategy improvement only cares about the \emph{relative}
difference between the outgoing edges from the vertex, and since all edges
leaving the gadget switch at the same time, the operation of the \notg-gate is
not interrupted. So, the strategy chosen at $d^j_i$ will continue to hold the
$i$th bit of $F(B)$, and the gadget has transitioned to input mode.

When the gadget is in input mode, it can be viewed as a \notg at the bottom of
circuit $1-j$ that has already been computed. In particular, the switch from
$h^j_{i, 1}$ to $h^j_{i, 2}$ ensures that, if the output is $1$, then the gadget
has the correct output priority. Moreover, the deceleration lane has enough
states to ensure that, if the output is $0$, then output of the gadget will not
flip from $0$ to $1$ while circuit $1-j$ is computing.

Finally, once circuit $1-j$ has finished computing, the clock for circuit $j$
advances, and the input/output gadget moves back to output mode. This involves
resetting the \notg gate gadget back to its initial state. This occurs because,
when the clock in circuit $j$ advances, there is a single iteration in which the
valuation of $s^j$ is higher than the valuation of $r^j$. This causes $z^j$ to
switch to $s^j$ which in turn causes a single iteration in which the valuation
of $z^j$ is higher than the valuation of $y^j$. Then, in the next iteration the
vertex $y^j$ switches to $r^j$, and so the valuation of $y^j$ is then larger
than the valuation of $z^j$. So, the valuations of $y^j$ and $z^j$ give exactly
the same sequence of events as $r^j$ and $s^j$, which allows the \notg-gate to
reset.

The following table specifies the input/output gadget.

\begin{small}
\begin{center}
\begin{tabular}{l|l|l|l|l}
Vertex & Conditions  & Edges  & Priority & Player \\ \hline
$y^j$ & $j \in \{0, 1\}$ & $r^{1-j}$, $r^j$ & $\pp(3, 0, 1, j, 0)$ & Even \\
$z^j$ & $j \in \{0, 1\}$ & $r^j$, $s^j$ & $\pp(3, 0, 0, j, 0)$ & Even \\ 
\hline
$t^j_{i,0}$ & $j \in \{0, 1\}$, $i\in \inputg$ & $y^j$, $z^j$ & $\pp(5, i,
2k+4n+4, j, 0)$ & Even \\
$t_{i,l}^j$ & $j \in \{0, 1\}$, $i\in \inputg$, & $y^j$, $z^j$, $t^j_{i, l-1}$ & $\pp(5, i, l, j, 1)$ & Even \\
	& \hfill $1 < l \le \lale$, &&& \\
& \hfill and $l \ne d(C)$ &&& \\
$t_{i,d(C)}^j$ & $j \in \{0, 1\}$, $i\in \inputg$ & $p^j_{i}$ & $\pp(5, i,
d(C), j, 1)$ & Even \\
$p_{i}^j$ & $j \in \{0, 1\}$, $i \in \inputg$ & $o^j_{\inp(i)}$, $p^j_{i,1}$ &
$\pp(3, i, 2, j, 0)$ & Even \\
$p_{i,1}^j$ & $j \in \{0, 1\}$, $i \in \inputg$ & $r^{1-j}$ & $\pp(5, i, 2k+4n+5, j, 0)$ & Even \\
$a_{i,l}^j$ & $j \in \{0, 1\}$ $i \in \inputg$, & $t_{i,l}$ & $\pp(5, i, l+1, j, 0)$ & Even \\
& $1 \le l \le \lale$  &&& \\
\hline
$d^j_i$ & $j \in \{0, 1\}$, $i \in \inputg$ & $y^j$, $z^j$, $e^j_i$, $a^j_{i,l}$ for
 & $\pp(4,i,0,j,1)$ & Even \\
& & \hfill $1 \le l \le \lale $  && \\
$e^j_i$ & $j \in \{0, 1\}$, $i \in \inputg$ & $h^j_{i,0}$, $d^j_i$ & $\pp(4,i,1,j,0)$
& Odd
\\
$q^j_{i,0}$ & $j \in \{0, 1\}$, $i \in \inputg$ & $e^j_i$, $q^j_{i,1}$ & $\pp(4,
i, 2, j, 0)$ & Odd \\
$q^j_{i,1}$ & $j \in \{0, 1\}$, $i \in \inputg$ & $r^{1-j}_i$ & $\pp(6, d(C)+2,
0, j, 0)$ & Even \\
$o^j_i$ & $j \in \{0, 1\}$, $i \in \inputg$ & $q^j_{i,0}$ & $\pp(6,i,0,j,1)$ & Even \\
\hline
$h_{i,0}^j$ & $j \in \{0, 1\}$, $i \in \inputg$ & $h_{i,1}^j$, $h_{i,2}^j$ & $\pp(3, i, 3, j, 0)$
& Even \\
$h_{i,1}^j$ & $j \in \{0, 1\}$, $i \in \inputg$ & $r^j$ & $\pp(6,d(C)+1,1,j,0)$ & Even \\
$h_{i,2}^j$ & $j \in \{0, 1\}$, $i \in \inputg$ & $r^{1-j}$ & $\pp(6,0,1,j,0)$ & Even \\
%$h^j_i$ & $j \in \{0, 1\}$, $i \in \inputg$ & $y^j$ & $\pp(6,i,1,j,0)$ & Even
\end{tabular}
\end{center}
\end{small}

\paragraph{\bf One-sink game.} If we are to use the simplified strategy
improvement algorithm, we must first show that this construction is a one-sink
game. We do so in the following lemma.

\begin{lem}
The construction is a one-sink game.
\end{lem}
\begin{proof}
In order to show that the construction is a one-sink game, we must show that
the two required properties hold. Firstly, we must show that there is a
vertex the satisfies the required properties of a sink vertex. It is not
difficult to verify that vertex $x$ does indeed satisfy these properties: the
only outgoing edge from $x$ is the edge $(x, x)$, and we have $\pri(x) = \pp(0,
0, 0, 0, 0) = 1$. Furthermore, no vertex is assigned priority $0$.

Secondly, we must argue that all optimal strategies are terminating. Recall that
a terminating strategy has the property that the first component of the
V\"oge-Jurdzi\'nski valuation is $1$, which implies that all paths starting at
all vertices eventually arrive at the sink $x$. So, consider a strategy $\sigma$
that is not terminating, and let $v$ be a vertex such that the first component
of  $\val^{\sigma}_{VJ}$ is strictly greater than 1. Let $C$ be the cycle that
is eventually reached by following $\sigma$ and $\br(\sigma)$ from $v$. There
are two cases to consider:
\begin{itemize}
\item If $C$ contains at least one vertex from a clock, then
$C$ must be entirely contained within that clock, because there are no edges
that leave either of the two clocks. In this case we have that $\sigma$ is not
optimal because Friedmann has shown that his construction is a one-sink game.
\item If $C$ does not contain a vertex from a clock, then it is entirely
contained within the circuits. First observe that $C$ cannot be a two-vertex
cycle using the vertices $d^j_i$ and $e^j_i$, because it is not a best-response
for Odd allow a cycle with an even priority to be formed, since he can always
move to $r^j$, and from there eventually reach a cycle with priority $p \preceq
1$ (because the clock is a one-sink game). But, the only other way to form a
cycle in the circuits is to pass through both of the circuits. In this case, the
highest priority on the cycle will be an odd priority assigned to the state
$o^j_i$ in either a \notg-gate (if there is one on the path), or an input/output
gate (otherwise). Since this odd priority is strictly greater than $1$, and
since player Even can always assure a priority of $1$ by, for example, moving to
$r^j$ in every input/output state $d^j_i$, we have that $\sigma$ is not an
optimal strategy.
\end{itemize}
Therefore, we have shown that the construction is a one-sink game. 
\end{proof}

\section{Strategies} 
\label{sec:strategies}

In this section, we define an initial strategy, and describe the sequence of
strategies that greedy all-switches strategy improvement switches through when
it is applied to this initial strategy. We will define strategies for each of
the gadgets in turn, and then combine these into a full strategy for the entire
construction.

It should be noted that we will only define \emph{partial strategies} in this
section, which means that some states will have no strategy specified. This is
because our construction will work no matter which strategy is chosen at these
states.

To deal with this, we must define what it means to apply strategy improvement to
a partial strategy. If $\chi$ is a partial strategy and $\sigma \in \seven$ is a
strategy, then we say that $\sigma$ \emph{agrees} with $\chi$ if $\sigma(v) =
\chi(v)$ for every vertex $v \in V$ for which $\chi$ is defined. So, if $\chi_1$
and $\chi_2$ are partial strategies, then we say that greedy all-switches
strategy improvement switches $\chi_1$ to $\chi_2$ if, for every strategy
$\sigma_1 \in \seven$ that agrees with $\chi_1$, greedy all-switches strategy
improvement switches $\sigma$ to a strategy $\sigma_2$ that agrees with
$\chi_2$.

We now describe the sequence of strategies. Each part of the construction will
be considered independently. 

\paragraph{\bf The clock.}
We start by defining the sequence of strategies that occurs in the two clocks.
For each \emph{clock bit-string} $\C \in \{0, 1\}^n$, we define a sequence of
strategies $\kappa^{\C}_1$, $\kappa^{\C}_2$, \dots, $\kappa^{\C}_{\length(\C)}$.
Greedy all-switches strategy improvement switches through each of these
strategies in turn, and then switches from $\kappa^{\C}_{\length(\C)}$ to
$\kappa^{\C+1}_1$, where $\C+1$ denotes the bit-string that results by adding
$1$ to the integer represented by $\C$. The sequence begins in the first
iteration after the valuation of $s^j$ is larger than the valuation of $r^j$.
We will first present the building blocks of this strategy, and the combine the
building blocks into the full sequence.

We begin by considering the vertices $t^j_l$ in the deceleration lane. Recall that
these states switch, in sequence,  from $r^j$ to $t^j_{l-1}$. This is formalised
in the following definition. For each $m \ge 1$, each $l$ in the range $1 \le l
\le \lale$, and each $j \in \{0, 1\}$,  we define:
\begin{align*}
\rho_m(t^j_0) &= \begin{cases} 
s^j & \text{if $m = 1$,} \\
r^j & \text{if $m > 1$.}
\end{cases} \\
\rho_{m}(t^j_{l}) &= \begin{cases}
s^j & \text{if $m = 1$,} \\
r^j & \text{if $m > 1$ and $m \le l+1$,} \\
t^j_{l-1} & \text{if $m > l+1$.} 
\end{cases} 
\end{align*}

We now move on to consider the vertices $d^j_i$, which represent the bits in the
counter. We begin by defining a sequence of strategies for the bits that are
$0$. Recall that these vertices switch to the states $a^j_{i}$ along the
deceleration lane until they run out of edges, at which point they switch to the
vertex $e^j_i$. This is formalised in the following definition. For each $i$ in
the range $1 \le i \le n$, each $m \ge 1$, and each $j \in \{0, 1\}$, we define:
\begin{equation*}
\rho_{m}(d^j_i) = \begin{cases}
s^j & \text{if $m = 1$,} \\
r^j & \text{if $m = 2$,} \\
a^j_{\laletwo + 2i} & \text{if $m = 3$,} \\
a^j_{m-3} & \text{if $4 \le m \le \laletwo + 2i + 3$,} \\
e^j_{i} & \text{if $m > \laletwo + 2i + 3$.} 
\end{cases}
\end{equation*}
Note that the first three iterations are special, because the edge to $a^j_{1}$
only becomes switchable in the third iteration. The edge to $r^j$
and the edge to $a^j_{\laletwo + 2i}$ prevent the edge to $e^j_i$ being switched
before this occurs.

We now give a full strategy definition for the vertices $d^j_i$. The bits that
are $0$ follow the strategy that we just defined, and the bits that are $1$
always choose the edge to $e^j_i$. For each bit-string $\C \in \{0, 1\}^n$, each
$i$ in the range $1 \le i \le n$, each $m \ge 1$, and each $j \in \{0, 1\}$, we
define:
\begin{equation*}
\rho^\C_m(d^j_i) = \begin{cases}
\rho_m(d^j_i) & \text{if $\C_i = 0$,} \\
e^j_i & \text{if $\C_i = 1$.}
\end{cases}
\end{equation*}

Finally, we consider the other vertices in the clock. To define strategies for
these vertices, we must first define some notation. For each $i$ in the range $1
\le i \le n$, we define $\nexbit(\C, i)$ to be a partial function that gives the
index of the first $1$ that appears higher than index $i$: that is, the smallest
index $j > i$ such that $\C_j = 1$. We now define the strategies. These
strategies all depend on the current clock bit-string $\C$, and have no
dependence on how far the deceleration lane has switched, so the parameter $m$ is
ignored. For each bit-string $\C \in \{0, 1\}^n$, each $m \ge 1$, each $i$ in
the range $1 \le i \le n$, and each $j \in \{0, 1\}$, we define:
\begin{align*}
\rho^\C_m(g^j_i) &= \begin{cases}
k^j_i & \text{if $\C_i = 0$,} \\
f^j_i & \text{if $\C_i = 1$.}
\end{cases} \\
\rho^\C_m(k^j_i) &= \begin{cases}
g^j_{\nexbit(\C, i)} & \text{if $\nexbit(\C, i)$ is defined,} \\
x & \text{otherwise.}
\end{cases} \\
\rho^\C_m(r^j) &= \begin{cases}
g^j_{\nexbit(\C, 0)} & \text{if $\nexbit(\C, 0)$ is defined,} \\
x & \text{otherwise.}
\end{cases} \\
\rho^\C_m(s^j) &= \begin{cases}
f^j_{\nexbit(\C, 0)} & \text{if $\nexbit(\C, 0)$ is defined,} \\
x & \text{otherwise.}
\end{cases}
\end{align*}

When the clock transitions between two clock bit-strings, there is a single
iteration in which the strategies defined above are not followed. This occurs
one iteration after the vertex $d^j_{\lsz(\C)}$ switches to $e^j_{\lsz(\C)}$. In
this iteration, the vertices $g^j_{\lsz(\C)}$ and $s^j$ switch to
$f^j_{\lsz{\C}}$, while every other vertex continues to use the strategies that
were defined above. We now define a special \emph{reset strategy} that captures
this. For each bit-string $\C \in \{0, 1\}^n$, and every vertex $v$ in either of
the two clocks we define:
\begin{equation*}
\rho^\C_{\text{Reset}}(v) = \begin{cases}
f^j_{\lsz(\C)} & \text{if $v = g^j_{\lsz(\C)}$ or $v = s^j$,} \\
\rho^\C_{\length(\C)}(v) & \text{otherwise.}
\end{cases}
\end{equation*}

We can now combine the strategies defined above in order to define the full
sequence of strategies that are used in the clocks. In the first $\length(\C)
-1$ iterations, we follow the sequence defined by the strategies
$\rho^{\C}_m(v)$, and in the final iteration we use the strategy
$\rho^{\C}_{\text{Reset}}(v)$. Formally, for each bit-string $\C \in \{0,
1\}^n$, each $m$ in the range $1 \le m \le \length(\C)$, and every vertex $v$ in
either of the two clocks, we define:
\begin{equation*}
\kappa^{\C}_m(v) = \begin{cases}
\rho^{\C}_m(v) & \text{if $m \le \length(\C) -1$,}\\
\rho^{\C}_{\text{Reset}}(v) & \text{if $m = \length(\C)$.}
\end{cases}
\end{equation*}
Friedmann showed the following lemma.

\begin{lem}[\cite{F11}]
\label{lem:fri}
Let $\C \in \{0, 1\}^n$. If we start all-switches strategy improvement at
$\kappa^{\C}_1$ for clock~$j$, then it will proceed by switching through the
sequence $\kappa^\C_1$, $\kappa^\C_2$, $\dots$, $\kappa^\C_{\length(\C)}, \kappa^{\C + 1}_1$.
\end{lem}

\paragraph{\bf The circuits.}

For each bit-string $B \in \{0, 1\}^n$, we give a sequence of strategies
$\sigma^B_1$, $\sigma^B_2$, $\dots$, which describes the sequence of strategies
that occurs when $B$ is the input of the circuit. The sequence is indexed from the
point at which the circuit's clock advances to the next bit-string. That is,
$\sigma^B_1$ occurs one iteration after the valuation of $s^j$ exceeds the
valuation of $r^j$. 

Recall that all of the gates with the same depth are evaluated in the same
iteration. We can now make this more precise: each gate $i$ will be evaluated in
the strategy $\sigma^B_{d(i) + 2}$. After this iteration, there will then be two
cases based on whether the gate evaluates to $1$ or $0$. To deal with this, we
require the following notation. For each bit-string $B$ and each gate $i$, we
define $\eval(B, i)$ to be $1$ if gate $i$ outputs true on input $B$, and $0$
otherwise.

\paragraph{\bf Or gates.}

Before the gate is evaluated, the state $o^j_i$ chooses the edge to $r^j$. Once
the gate has been evaluated, there are four possibilities. If both input gates
evaluate to false, then the state $o^j_i$ continues to use the edge to $r^j$. If
one of the two inputs is true, then $o^j_i$ will switch to the corresponding
input state. The case where both  inputs are true is the most complicated.
Obviously, $o^j_i$ will switch to one of the two input states, and in fact, it
switches to the one with the highest valuation. Since the overall correctness of
our construction does not care which successor is chosen in this case, we simply
define $\ornext(i,B,m)$ to be the successor with the highest valuation in step
$m$ of the sequence for bit-string $B$.

We can now formally define the sequence of strategies used by an \org-gate. For
every gate $i \in \org$, every pair of bit-strings $B \in \{0, 1\}^n$, and every
$m \ge 1$ we define:
\begin{equation}
\label{eqn:odef}
\sigma^{B}_m(o^j_i) = \begin{cases}
s^j_i & \text{if $m = 1$,} \\
r^j_i & \text{if $m > 1$ and $m \le d(i) + 2$,} \\
r^j_i & \text{if $m > d(i) + 2$ and $\eval(B, \inp_1(i)) = 0$ and 
$\eval(B, \inp_2(i)) = 0$,} \\
\inputstate(i,j,1) & \text{if $m > d(i) + 2$ and $\eval(B, \inp_1(i)) = 1$ and 
$\eval(B, \inp_2(i)) = 0$,} \\
\inputstate(i,j,2) & \text{if $m > d(i) + 2$ and $\eval(B, \inp_1(i)) = 0$ and 
$\eval(B, \inp_2(i)) = 1$,} \\
\ornext(i,B,m) & \text{if $m > d(i) + 2$ and $\eval(B, \inp_1(i)) = 1$ and $\eval(B, \inp_2(i)) = 1$.} 
\end{cases}
\end{equation}

\paragraph{\bf Not gates.}
There are two components of the \notg-gate gadget: the modified deceleration
lane and the state $d^j_i$. We begin by considering the modified deceleration
lane. 

We first define a strategy for the case where the gate evaluates to false. In
this case, the input gate evaluates to true, which causes the modified
deceleration lane to continue switching after iteration $d(i) + 2$. We formalise
this in the following definition, which is almost identical to the definition
given for the deceleration lane used in the clock. For each $i \in \notg$, each
$l$ in the range $1 \le l \le \lale$ with $l \ne d(i)$, each $j \in \{0, 1\}$,
and each $m \ge 1$, we define: 
\begin{align}
\label{eqn:tnotdef}
\sigma_m(t^j_{i, 0}) &= \begin{cases}
s^j & \text{if $m = 1$,} \\
r^j & \text{if $m > 1$.}
\end{cases} \\
\sigma_m(t^j_{i,l}) &= \begin{cases}
s^j & \text{if $m = 1$,} \\
r^j & \text{if $m > 1$ and $m \le l+1$,} \\
t^j_{i, l-1} & \text{if $m > l+1$.} 
\end{cases} 
\end{align}

On the other hand, if the gate evaluates to false, then the deceleration lane
stops switching. This is formalised in the following definition, which uses the
previous definition to give the actual strategy used by the modified
deceleration lane. For each $i \in \notg$, each $B \in \{0, 1\}^n$, each $l$ in
the range $0 \le l \le \lale$ with $l \ne d(i)$, each $j \in \{0, 1\}$, and each
$m \ge 1$, we define: 
\begin{equation*}
\sigma^{B}_m(t^j_{i,l}) = \begin{cases}
\sigma_m(t^j_{i,l}) & \text{if $l \le d(i)+1$, or $l > d(i) + 1$ and $\eval(B,
\inp(i)) = 1$,} \\
s^j & \text{if $l > d(i)+1$ and $m = 1$ and $\eval(B, \inp(i)) = 0$,} \\
r^j & \text{if $l > d(i)+1$ and $m > 1$ and $\eval(B, \inp(i)) = 0$.} 
\end{cases}
\end{equation*}

We now turn out attention to the state $d^j_i$, where we again begin by
considering the case where the gate evaluates to false. In this case, the state
$d^j_i$ continues switching to the modified deceleration lane. This is
formalised in the following definition, which is almost identical to the
definition given for the corresponding states in the clock. For all $i \in \notg
\cup \inputg$, all $B \in \{0, 1\}^n$, for all $m \ge 1$, and all $j \in \{0,
1\}$, we define:
\begin{equation*}
\sigma_m(d^j_i) = \begin{cases}
s^j & \text{if $m = 1$,} \\
r^j & \text{if $m = 2$,} \\
a^j_{i, \lale} & \text{if $m = 3$,} \\
a^j_{i, m-3} & \text{if $4 \le m \le \lale+3$,} \\
e^j_i & \text{if $\lale + 3 < m$.}
\end{cases}
\end{equation*}

On the other hand, if the gate evaluates to true, then after iteration $d(i) +
2$, the state $d^j_i$ switches to $e^j_i$. This is formalised in the following
definition, where the previous definition is used in order to give the actual
sequence of strategies for the state $d^j_i$. For all $B \in \{0,1\}^n$, all $i
\in \notg$, all $j \in \{0, 1\}$, and all $m \ge 1$ we define:
\begin{equation}
\label{eqn:notd}
\sigma^B_m(d^j_i) = \begin{cases}
\sigma_m(d^j_i) & \text{if $m \le d(i) + 2$, or $m > d(i) + 2$ and $\eval(B, \inp(i)) = 1$,} \\
e^j_i & \text{if $m > d(i)+2$ and $\eval(B, \inp(i)) = 0$.} 
\end{cases}
\end{equation}

\paragraph{\bf Input/output gates.} We now describe the sequence of strategies
used in the input/output gates. These strategies are almost identical to the
strategies that would be used in a \notg-gates with depth $d(C)+1$, but with a
few key differences. Firstly, whereas the \notg-gates used edges to $r^j$ and
$s^j$, these have instead been replaced with the edges to $y^j$ and $z^j$ from
the circuit movers. Secondly, the circuit movers cause a one iteration delay at
the start of the sequence. Note, however, that despite this delay, the
input/output gates are still evaluated on iteration $d(C) + 3$.

We begin by giving the strategies for the modified deceleration lane used in the
input/output gates. For each $i \in \notg$, each $l$ in the range $1 \le l \le
\lale$ with $l \ne d(C)$, each $j \in \{0, 1\}$, and each $m \ge 1$, we define: 
\begin{align*}
\sigma_m(t^j_{i,l}) &= \begin{cases}
z^j & \text{if $m = 2$,} \\
y^j & \text{if $m = 1$ or $m > 1$ and $m \le l+2$,} \\
t^j_{i, l-1} & \text{if $m > l+2$.} 
\end{cases} \\
\sigma_m(t^j_{i, 0}) &= \begin{cases}
z^j & \text{if $m = 2$,} \\
y^j & \text{if $m = 1$ or $m > 2$.}
\end{cases}
\end{align*}
Then, for each $i \in \notg$, each $B \in \{0, 1\}^n$, each $l$ in the range $0
\le l \le \lale$ with $l \ne d(C)$, each $j \in \{0, 1\}$, and each $m \ge 1$,
we define: 
\begin{equation*}
\sigma^{B}_m(t^j_{i,l}) = \begin{cases}
\sigma_m(t^j_{i,l}) & \text{if $l < d(C) + 3$, or $l \ge d(C)+3$ and $B_i = 0$,} \\
z^j & \text{if $l > d(C)+ 3$ and $m = 2$ and $B_i = 1$,} \\
y^j & \text{if $l > d(C)+ 3$ and either $m = 1$ or $m > 2$ and $B_i = 1$.} 
\end{cases}
\end{equation*}
Finally, we give the strategy for the state $d^j_i$. We reuse the
strategy $\sigma_{m-1}$ from the \notg-gate definitions, but with a one
iteration delay. For all $B \in \{0,1\}^n$, all $i \in \inputg$, all $j \in \{0,
1\}$, and all $m
> 1$ we define:
\begin{equation}
\label{def:inputd}
\sigma^B_m(d^j_i) = \begin{cases}
\sigma_{m-1}(d^j_i) & \text{if $1 < m \le d(C) + 3$, or $m > d(C) + 3$ and $B_i = 0$,} \\
e^j_i & \text{if $m > d(i)+3$ and $B_i = 1$.} 
\end{cases}
\end{equation}

\paragraph{\bf The circuit mover states.}
Finally, we describe the sequence of strategies used in the states that move the
input/output gates between the circuits. These strategies do not depend on the
current input bit-string to the circuit. Instead, they depend on the state of
both of the clocks, and are parameterized by the value of the delay function that we defined
earlier.

Formally, for every $m \ge 1$, every $i \in \inputg$, and every clock-bit string $\C \in
\{0, 1\}^n$ we define:
\begin{align}
\label{eqn:zj}
\sigma^{\C}_m(z^j) &= \begin{cases}
s^j & \text{if $m = 1$,} \\
r^j & \text{if $m > 1$.} \\
\end{cases} \\
\label{eqn:yj}
\sigma^{\C}_m(y^j) &= \begin{cases}
r^{1-j} & \text{if $m = 1$ or $m \ge \delay(j,\C) + 1$} \\
r^j & \text{if $m > 1$ and $m < \delay(j,\C) + 1$.} \\
\end{cases} \\
\label{eqn:pj}
\sigma^{\C}_m(p^j_i) &= \begin{cases}
p^{j}_{i,1} & \text{if $m = 1$ or $m \ge \delay(j,\C) + 1$,} \\
o^j_{\inp(i)} & \text{if $1 < m \le \delay(j,\C) + 1$.} \\
\end{cases} \\
\label{eqn:hj}
\sigma^{\C}_m(h^j_{i,0}) &= \begin{cases}
h^{j}_{i,2} & \text{if $m = 1$ or $m \ge \delay(j,\C) + 1$,} \\
h^j_{i,1} & \text{if $1 < m < \delay(j,\C) + 1$.} \\
\end{cases} 
\end{align}

\paragraph{\bf Putting it all together.}

We can now define a combined sequence of strategies for the entire construction.
We will defined a sequence of strategies 
$\chi^{B,\C,j}_1$, 
$\chi^{B,\C,j}_2$, \dots, which describes a computation in circuit $j$ under the
following conditions:
\begin{itemize}
\item The clock for circuit $j$ currently stores $\C$ in its binary counter.
\item The input to circuit $j$ is $B$. 
\end{itemize}

Before stating the strategies, we first define some necessary notation. For
every clock bit-string $\C \in \{0, 1\}$, and every $j \in \{0, 1\}$ we define:
\begin{equation*}
\OC(\C, j) = \begin{cases}
\C-1 & \text{if $j = 0$,} \\
\C & \text{if $j = 1$.} 
\end{cases}
\end{equation*}
This gives the bit-string used in the \emph{other clock,} when circuit $j$ is
computing. Since clock~$0$ is ahead of clock~$1$, we have that $\OC(\C, 0)$ is
the bit-string before $\C$, while $\OC(\C, 1)$ is the same as $\C$.

We can now define the sequence.
For each bit-string $B \in \{0, 1\}^n$, each bit-string $\C \in \{0, 1\}^n$,
each $m \ge 1$, and every vertex $v$ we define: 
\begin{align*}
\chi^{B,\C,j}_m(v) &= \begin{cases}
\kappa^{\C}_m(v) & \text{if $v$ is in clock~$j$,} \\
\kappa^{\OC(\C, j)}_{m + \delay(1-j, \OC(\C, j))}(v) & \text{if $v$ is in
clock~$1-j$,} \\
\sigma^{B}_m(v)  & \text{if $v$ is in a $\notg$ or $\org$ gate in circuit~$j$,} \\
\sigma^{F(B)}_m(v) & \text{if $v$ is in an input/output gate in circuit~$j$,} \\
\sigma^{B}_{m+\delay(1-j, \OC(\C, j))}(v) & \text{if $v$ is an input/output gate in
circuit~$1-j$,} \\
\sigma^{\C}_m(v)  & \text{if $v$ is a circuit mover state in circuit~$j$,} \\
\sigma^{\OC(\C, j)}_{m+\delay(1-j, \OC(\C, j))}(v) & \text{if $v$ is a circuit
mover state in circuit~$1-j$.}
\end{cases} \\
\end{align*}
The first two cases of this definition deal with the clocks: the clock in
circuit $j$ follows the sequence for bit-string $\C$, while the clock in circuit
$1-j$ continues to follow the sequence for bit-string $\OC(\C, j)$. Observe
that the clock for circuit $1-j$ has already been running for $\delay(1-j,
\OC(\C, j))$ iterations, so the strategies for this clock start on iteration $1 +
\delay(1-j, \OC(\C, j))$. The next two cases deal with the gate gadgets in
circuit $j$: the \notg and \org gates follow the sequence for bit-string $B$,
and then the input/output gates for circuit $j$, which are in output mode, store
$F(B)$. The next case deals with the input/output gates in circuit $1-j$ are in
input mode, and so follow the strategy for bit-string $B$. The final two cases
deal with the circuit mover states, which follow the strategies for the clock
bit-string used in their respective clocks. Observe that no strategy is
specified for the gate gadgets in circuit $1-j$, because the strategy chosen
here is irrelevant. 

For technical convenience, we define:
\begin{align*}
\chi^{B, \C, 0}_{\delay(0,\C)} &= \chi^{F(B), \C, 1}_1 \\
\chi^{B, \C, 1}_{\delay(1,\C)} &= \chi^{F(B), \C+1, 0}_1 
\end{align*} 
Using this definition, we can now state the main technical claim of the paper. 
\begin{lem}
\label{lem:main}
Let $B \in \{0, 1\}^n$ be a bit-string, let $C \in \{0, 1\}^n$ be a
bit-string such that $C \ne (1,1,\dots,1)$, and let $j \in \{0, 1\}$. 
If greedy all-switches strategy improvement is applied to $\chi^{B, \C,
0}_1$, then it will pass through the sequence:
\begin{equation*}
\chi^{B, \C, j}_1, \chi^{B, \C, j}_2, \dots, \chi^{B, \C, j}_{\delay(j,\C)}.
\end{equation*}
\end{lem}
\noindent Unfortunately, the proof of this lemma is quite long, and the vast
majority of it is presented in the appendix. In Section~\ref{sec:proof}, we give
an overview of the proof, and describe how each of the individual appendices fit
into the overall proof.

\paragraph{\bf Best responses.}
Recall that, for each strategy considered, strategy improvement computes a
best-response for the opponent. Now that we have defined the sequence of
strategies, we can also define the best-responses to these strategies. For each
strategy $\chi^{B, \C, j}_i$, we define a strategy $\mu^{B,\C,j}_i \in \sodd$
that is a best-response to $\chi^{B, \C, j}_i$. We will later prove that these
strategies are indeed best-responses.

We begin by considering the vertices $e^j_i$ for each \notg-gate $i$. Recall
that these vertices only pick the edge to $h^j_{i,0}$ in the case where they are
forced to by $d^j_i$ selecting the edge to $e^j_i$. As defined above, this only
occurs in the case where the \notg-gate evaluates to true. Formally, for each
bit-string $B \in \{0, 1\}^n$, each bit-string $\C \in \{0, 1\}^n$, each $m \ge
1$, and every $i \in \notg$ we define:
\begin{equation*}
\mu^{B,\C,j}_m(e^j_i) = \begin{cases}
h^j_{i,0} & \text{if $m > d(i)+2$ and $\eval(B, \inp(i)) = 0$,} \\
d^j_i & \text{otherwise.} 
\end{cases}
\end{equation*}

For the input/output gadgets in circuit $1-j$, which will provide the input to
circuit $j$, the situation is the same. The vertex $e^{1-j}_i$ chooses the edge
to $h^{1-j}_{i, 0}$ if and only if $B_i$ is $1$. Formally, for each bit-string
$B \in \{0, 1\}^n$, each bit-string $\C \in \{0, 1\}^n$, each $m \ge 1$, and
every vertex $i \in \inputg$ we define: 
\begin{equation*}
\mu^{B,\C,j}_m(e^{1-j}_i) = \begin{cases}
h^{1-j}_{i,0} & \text{if $B_i = 1$,} \\
d^{1-j}_i & \text{if $B_i = 0$.} 
\end{cases} 
\end{equation*}

For the input/output gadgets in circuit $1-j$, the situation is the largely the
same as for a \notg-gate with depth $d(C) + 1$, and the edge chosen depends on
$F(B)_i$. However, one difference is that we do not define a best-response for
the case where $m = 1$, because the input/output gadget does not reset until the
second iteration, and our proof does not depend on the best response chosen in
iteration one. Formally, for each bit-string $B \in \{0, 1\}^n$, each bit-string
$\C \in \{0, 1\}^n$, each $m > 1$, and every vertex $i \in \inputg$ we define:
\begin{equation*}
\mu^{B,\C,j}_m(e^j_i) = \begin{cases}
%d^j_i & \text{if $m = 1$ and $B_i = 0$,} \\
%h^j_{i,0} & \text{if $m = 1$ and $B_i = 1$,} \\
h^j_{i,0} & \text{if $m > d(i)+3$ and $F(B)_i = 1$,} \\
d^j_i & \text{otherwise.} 
\end{cases} 
\end{equation*}

Finally, we define the best responses for the vertices $q^j$ as follows.  For
each bit-string $B \in \{0, 1\}^n$, each bit-string $\C \in \{0, 1\}^n$, each
$m$ in the range $1 \le m \le \delay(j, \C) -1$, and every vertex $i \in
\inputg$ we define: 
\begin{align*}
\mu^{B,\C,j}_m(q^j_{i,0}) &= \begin{cases}
e^j_{i} & \text{if $m = 1$, } \\
q^j_{i,1} & \text{if $m > 1$. }
\end{cases} \\
\mu^{B,\C,j}_m(q^{1-j}_{i,0}) &= \begin{cases}
 e^{1-j}_i.  
\end{cases}
\end{align*}

\section{The Proof}
\label{sec:proof}

In this section we give the proof for Lemma~\ref{lem:main}. Let $B, \C \in \{0,
1\}^n$ be two bit-strings, let  $j \in \{0, 1\}$, and let $m$ be in the range $1
\le m \le \delay(j, \C) - 1$. We must show that greedy all-switches strategy
improvement switches $\chi^{B, \C, j}_m$ to $\chi^{B, \C, j}_{m+1}$.

Let $\sigma$ be a strategy that agrees with $\chi^{B, \C, j}_m$. Since we are
using the all-switches switching rule, we can consider each vertex $v$
independently, and must show that the most appealing outgoing edge at $v$ is the
one specified by $\chi^{B, \C, j}_{m+1}$ (in our construction there will always
be exactly one most appealing edge, so we do not care how ties are broken by
the switching rule). Hence, the majority of the proof boils down to calculating the
valuation of each outgoing edge of $v$, and then comparing these valuations.

To compare the valuation of two outgoing edges $(v, u)$ and $(v, w)$, we usually
use the following technique. First we consider the two paths $\pi_1$ and $\pi_2$
that start at $u$ and $w$, respectively, and follow $\sigma$ and $\br(\sigma)$.
Then we find the first vertex $v'$ that is contained in both paths. Since the
$\sqsubseteq$ relation only cares about the maximum difference between the two
paths, all priorities that are visited after $v'$ are irrelevant, since they
appear in both $\val^{\sigma}(u)$ and $\val^{\sigma}(w)$. On the other hand,
since each priority is assigned to at most one vertex, all of the priorities
visited by $p_1$ before reaching $v'$ are contained in $\val^{\sigma}(u)$ and not
contained in $\val^{\sigma}(w)$, and all of the priorities visited by $p_2$
before reaching $v'$ are contained in $\val^{\sigma}(w)$ and not contained in
$\val^{\sigma}(u)$. So it suffices to find the largest priority on the prefixes
of $p_1$ before $v'$ and the prefix of $p_2$ before $v'$. The parity of this
priority then determines whether $\val^{\sigma}(u) \sqsubseteq \val^{\sigma}(w)$
according to the rules laid out in the definition of $\sqsubseteq$.

We now give an outline of the proof.
\begin{itemize}
\item The fact that the two clocks switch through their respective strategies
follows from Lemma~\ref{lem:fri}.

\item The difference in valuation between the states $r^j$ and $s^j$ of the
clock are the driving force of the construction. In Appendix~\ref{app:clock}, we
give two lemmas that formalize this difference.

\item Next, in Appendix~\ref{app:br}, we prove that the best-response strategies
defined in Section~\ref{sec:strategies} are in fact the best responses. That is,
we show that $\mu^{B,\C,j}_m$ is a best response to every strategy $\sigma$ that
agrees with $\chi^{B, \C, j}_{m}$.

\item In Appendix~\ref{app:outputs}, we give two key lemmas that describe the
valuations of the output stats $o^j_i$. These lemmas show three important
properties. Firstly, if $m \le d(i) + 2$, then the valuation of $o^j_i$ is low,
so there is no incentive to switch to $o^j_i$ before gate $i$ is evaluated.
Secondly, if $m > d(i) + 2$ and the gate evaluates to $0$, then the valuation of
$o^j_i$ remains low. Finally, if $m > d(i) + 2$ and the gate evaluates to $1$,
then the valuation of $o^j_i$ is high. These final two properties allow the
gates with depth strictly greater than $i$ to compute their outputs correctly.

\item The rest of the proof consists of proving that all vertices switch to the
correct outgoing edge. The states $o^j_i$ in the \org gate gadgets are dealt
with in Appendix~\ref{app:org}. The states $t^j_{i, l}$ in the \notg gates are
dealt with in Appendix~\ref{app:nott}. The states $d^j_{i}$ in the \notg gates
are dealt with in Appendix~\ref{app:notd}. The states $z^j$ and $z^{1-j}$ are
dealt with in Appendix~\ref{app:z}, and the states $y^j$ and $y^{1-j}$ are dealt
with in Appendix~\ref{app:y}. The states $p^j$ and $p^{1-j}$ are dealt with in
Appendix~\ref{app:p} and the states $h^j_{i, 0}$ are dealt with in
Appendix~\ref{app:h}. Finally, the states in the \inputg gates, which behave in
a largely identical way to the \notg gates, are dealt with in
Appendix~\ref{app:input}.

\end{itemize}
All of the above combines to provide a proof for Lemma~\ref{lem:main}.

Having shown this Lemma, we can now give the reduction from the problem $\bitswitch$ to
the problem $\edgeswitch$. Given a circuit iteration instance $(F, B, z)$, we produce the
parity game $G$ corresponding to $F$, we use $\chi^{B, 1, 0}_2$ as the initial
strategy, and if $i \in \inputg$ is the input/output gate corresponding to index
$z$, then we will monitor whether the edge from $d^0_i$ to $e^0_i$ is ever
switched by greedy all-switches strategy improvement. We therefore produce the
instance $\edgeswitch(G, (d^0_i, e^0_i), \chi^{B, 1, 0}_2)$. We must take care
to ensure that $\chi^{B, 1, 0}_2$ is a terminating strategy, which is proved in
the following Lemma.

\begin{lem}
We have that $\chi^{B, 1, 0}_2$ is a terminating strategy.
\end{lem}
\begin{proof}
Firstly, since we use the same strategies as Friedmann in the clock, we do not
need to prove that theses portions of the strategies are terminating, because
this has already been shown by Friedmann. In particular, this implies that all
paths starting at $s^j$ and $r^j$ for $j \in \{0, 1\}$ will eventually arrive at
the sink $x$. Therefore, it is sufficient to show that the first component of
$\val_{\text{VJ}}^{\chi^{B, 1, 0}_2}(v)$ is $1$ for every vertex $v$ in the
circuits.

Observe that, in the strategy $\chi^{B, 1, 0}_2$, we have that the only possible
cycles that player Odd can form in the best response are two-vertex cycles of
the form $d^j_i$ and $e^j_i$, but these cycles have an even priority, so the
best response can not choose them. In particular, it is not possible to form a
cycle that passes through the input/output gadgets in both circuits, because the
path that starts at $o^0_i$ for every input/output gate $i$ must eventually
arrive at $s^j$. Thus, we have that all paths starting at all vertices in the
circuits that follow $\chi^{B, 1, 0}_2$ and its best response will eventually
arrive at this sink~$x$. 
\end{proof}

Now all that remains is to argue that $\edgeswitch(G, (d^0_i, e^0_i), \chi^{B,
1, 0}_2)$ is true if and only if $\bitswitch(F, B, z)$ is true. To do this, we
simply observe that the sequence of strategies used in Lemma~\ref{lem:main} only
ever specify that $d^0_i$ must be switched to $e^0_z$ in the case where there is
some even $j$ such that $F^j(B)_z = 1$. At all other times, the vertex $d^0_i$
chooses an edge other that $e^j_i$. Hence, the reduction is correct, and we have
shown Theorem~\ref{thm:edgeswitch}.

\section{Other Algorithms}
\label{sec:other}

\paragraph{\bf Other strategy improvement algorithms.}
As we mentioned in the introduction, Theorem~\ref{thm:edgeswitch} implies several
results about other algorithms. In particular, discounted games and
simple-stochastic games both have natural strategy improvement algorithms given
by Puri~\cite{puri95}, and Condon~\cite{condon93}, respectively. Friedmann
showed that if you take a one-sink parity game, and apply the natural reduction
from parity games to either discounted or simple stochastic games, then the
greedy variants of Puri's and Condon's algorithms will switch exactly the same
edges as the algorithm of V\"oge and Jurdzi\'nski~\cite[Corollory 9.10 and Lemma
9.12]{F11}. Hence, Theorem~\ref{thm:edgeswitch} also implies the discounted
and simple-stochastic cases of Corollary~\ref{cor:edgeswitch}.

One case that was missed by Friedmann was mean-payoff games. There is a natural
strategy improvement algorithm~\cite{FV97} for mean-payoff games that adopts the
well-known \emph{gain-bias} formulation from average-reward
MDPs~\cite{puterman94}. In this algorithm, the valuations has two components:
the \emph{gain} of a vertex gives the long-term average-reward that can be
obtained from that vertex under the current strategy, and the \emph{bias}
measures the short term deviation from the long-term average.

We argue that, if we apply the standard reduction from parity games to
mean-payoff games, and then set the reward of $x$ to $0$, then the gain-bias
algorithm for mean-payoff games switches exactly the same edges as the algorithm
of V\"oge and Jurdzi\'nski. The standard reduction from parity games to
mean-payoff games~\cite{puri95,jurdzinski98} replaces each priority $p$ with the
weight $(-m)^{p}$, where $m$ denotes the number of vertices in the parity game.
By setting the weight of $x$ to $0$, we ensure that the long-term average reward
from each state is $0$. Previous work has observed~\cite{FS14}, that if the gain
is $0$ at every vertex, then the bias represents the \emph{total reward} that
can be obtained from each state. It is not difficult to prove that, after the
standard reduction has been applied, the total reward that can be obtained from
a vertex $v$ is larger than the total reward from a vertex $u$ if and only if
$\val^{\sigma}(u) \sqsubset \val^{\sigma}(v)$ in the original parity game. This
is because the rewards assigned by the standard reduction grow quickly enough so
that only the largest priority visited matters. Hence, we also have the
mean-payoff case of Corollary~\ref{cor:edgeswitch}.

Bj\"orklund and Vorobyov have also devised a strategy improvement algorithm for
mean-payoff games. Their algorithm involves adding an extra \emph{sink} vertex,
and then adding edges from every vertex of the maximizing player to the sink.
Their valuations are also the total reward obtained before reaching the sink. We
cannot show a similar result for their algorithm, but we can show a result for a
\emph{variant} of their algorithm that only gives additional edges to a subset
of the vertices of the maximizing player. To do this, we do the same reduction
as we did for the gain-bias algorithm, and then we only add an edge from $x$ to
the new sink added the Bj\"orklund and Vorobyov. The same reasoning as above
then implies that the Bj\"orklund-Vorobyov algorithm will make the same switches
as the V\"oge-Jurdzi\'nski algorithm.

\paragraph{\bf Unique sink orientations.}

As mentioned in the introduction, there is a relationship between strategy
improvement algorithms and sink-finding algorithms for unique sink orientations.
Our result already implies a similar lower bound for sink-finding algorithms
applied to unique sink orientations. However, since the vertices in our parity
game have more than two outgoing edges, these results only hold for unique sink
orientations of \emph{grids}. The more commonly studied model is unique sink
orientations of \emph{hypercubes}, which correspond to \emph{binary} parity
games, where each vertex has at most two outgoing edges. We argue that our
construction can be formulated as a binary parity game.

Friedmann has already shown that his construction can be formulated as a binary
parity game~\cite{F11}, so we already have that the clocks can be transformed
so that they are binary. Furthermore, since our \notg-gates and our
\inputg-gates are taken directly from Friedmann's bit gadget, we can apply
Friedmann's reduction to make these binary. In particular, note that all of the
extra states that we add to the input/output gate are binary, so these states do
not need any modification.

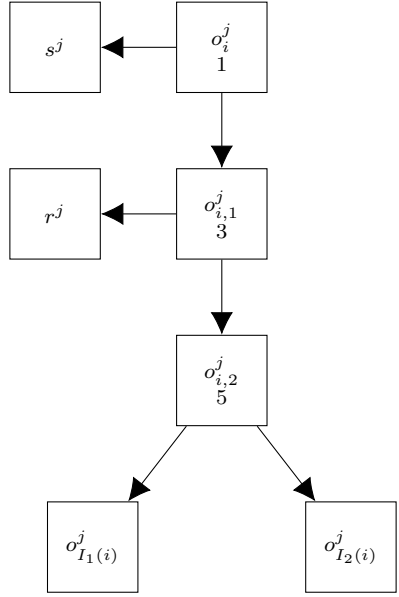
\begin{figure}
%%%%%%%%%%%%%%%%%%%%%%%%%%%%%%%%%%%%%%%%%%%%%%%%%%%%%%%%%%%%%%%%%%%%%%%%%%%%
% OR GATE 
%%%%%%%%%%%%%%%%%%%%%%%%%%%%%%%%%%%%%%%%%%%%%%%%%%%%%%%%%%%%%%%%%%%%%%%%%%%%
\begin{center}
\begin{tikzpicture}[node distance=1cm,font=\scriptsize]
\node [even] (o) {$o^j_{i,2}$ \\ $5$};
\node [even] (i1) [below left=1cm and 0.5cm of o] {$o^j_{\inp_1(i)}$};
\node [even] (i2) [below right=1cm and 0.5cm of o] {$o^j_{\inp_2(i)}$};

\node [even] (o_2) [above=1cm of o] {$o^j_{i,1}$ \\ $3$};
\node [even] (o_1) [above=1cm of o_2] {$o^j_{i}$ \\ $1$};

\node [even] (r) [left=1cm of o_2] {$r^j$};
\node [even] (s) [left=1cm of o_1] {$s^j$};

\path[->]
	(o_2) edge (r)
    (o_2) edge (o)
	(o_1) edge (s)
    (o_1) edge (o_2)
	(o) edge (i1)
	(o) edge (i2)
	;
\end{tikzpicture}
\end{center}
\caption{The binary \org gate.}
\label{fig:ormodified}
\end{figure}

The only remaining part of the construction is the \org-gate, which has four
outgoing edges. We replace the existing gadget with a modified
gadget, shown in Figure~\ref{fig:ormodified}.

\begin{center}
\begin{tabular}{l|l|l|l|l}
Vertex & Conditions  & Edges  & Priority & Player\\
\hline
$o^j_i$ & $j \in \{0, 1\}$, $i \in \org$ & $s^j$, $o^j_{i,1}$
& $\pp(4, i, 0, j, 1)$ & Even \\
$o^j_{i,1}$ & $j \in \{0, 1\}$, $i \in \org$ & $r^j$, $o^j_{i,2}$& $\pp(4, i, 1,
j, 1)$ & Even \\
$o^j_{i,2}$ & $j \in \{0, 1\}$, $i \in \org$ & $\inputstate(i,j,1)$,
$\inputstate(i,j,2)$& $\pp(4, i, 2, j, 1)$ & Even
\end{tabular}
\end{center}

This gadget replaces the single vertex of the original \org-gate, with three
binary vertices. The only significant difference that this gadget makes to the
construction is that now it can take up to two strategy improvement iterations
for the \org-gate to compute its output. This is because, we may have to wait
for $o^j_{i,2}$ to switch before $o^j_{i, 1}$ can switch. The vertex $o^j_{i}$
always chooses the edge $o^j_{i, 1}$ during the computation, because the
valuation of $r^j$ is larger than the valuation of $s^j$.

To deal with this, we can redesign the construction so that each \notg-gate $i$
is computed on iteration $2i$ rather than iteration $i$, and each \org-gate is
computed before iteration $2i$. This involved making the following changes:
\begin{itemize}
\item The length of the deceleration lane in the two clocks must be extended by
$2k$, to account for the $2k$ extra iterations it takes for the circuits to
compute ($k$ extra iterations for circuit $0$ and $k$ extra iterations for
circuit $1$). Moreover, the delays for both of the clocks must be increased by
$k$.
\item For the same reason, the length of the modified deceleration lanes in the \notg and \inputg
gates must be increased by $2k$.  
\item Finally, the edge to $o^j_{\inputstate(i, j)}$ must be moved from $t_{i,
d(i)}$ to $t_{i, 2 d(i)}$.
\end{itemize}
Once these changes have been made, we have produced a binary parity game. 

One final thing we must be aware of is that we only get a unique-sink
orientation if there is never a tie between the valuation of two vertices. This,
however, always holds in a one-sink game where every vertex has a distinct
priority, because all paths necessarily contain a distinct set of
priorities, which prevents ties in the $\sqsubseteq$ ordering. Therefore, we
have the \PSPACE-completeness result for the \textsc{BottomAntipodal} algorithm
claimed in Corollary~\ref{cor:uso}.

\section{The optimal strategy result}
\label{sec:optstrat}

We are also able to prove a result about the complexity of determining which
optimal strategy is found by the V\"oge-Jurdzi\'nski algorithm. However, we
cannot formulate this in the context of a one-sink game, because any result of
this nature must exploit \emph{ties in valuations.} In a one-sink game, since
every vertex has a different priority, no two paths can have the same set of
priorities, so ties are not possible. Hence, for a one-sink game, there will be
a unique optimal strategy, and so the complexity of finding it can be no harder
than solving the parity game itself, and this problem is not \PSPACE-complete
unless \PSPACE = \UP $\cap$ \coUP.

On the other hand, ties in valuations are possible in the original
V\"oge-Jurdzi\'nski algorithm. This is because the first component of their
valuation is not necessarily $1$, and so the second component does not
necessarily contain every priority along the relevant path (recall that
priorities smaller than the first component are not included in the second
component). These facts mean that it is possible to construct parity games that
have multiple optimal strategies under the V\"oge-Jurdzi\'nski valuation.

\paragraph{\bf Our modified construction.}
We will use a slight modification of our construction to show that computing the
optimal strategy found by the V\"oge-Jurdzi\'nski algorithm is \PSPACE-complete.
The key difference is the addition of a \emph{third} clock with $n+1$ bits,
which will be indexed by $2$. We remove a single edge from this clock: the edge
from $e^2_{n+1}$ to $h^2_{n+1}$ is removed. 

Recall that in the clock construction, the odd vertices $e^j_{i}$ do not use the
edge to $h^j_i$ unless they are forced to by the vertex $d^j_i$ selecting the
edge to $e^j_i$. Hence, until the $n+1$th bit is flipped, the third clock
behaves like any other clock. When the $n+1$th bit is flipped, after $2^n$
iterations have taken place, a new cycle is formed with a very large even
priority.

We also modify the edges that leave $d^1_z$. For each edge $e = (d^1_z, u)$ we
do the following:
\begin{enumerate}
\item We delete $e$.
\item We introduce a new vertex $v_u$ owned by player Even. This vertex is
assigned an insignificant priority that, in particular, is much smaller than the
priorities assigned to $e^2_{n+1}$ and $d^2_{n+1}$.
\item We add the edges $(d^1_z, v_u)$, $(v_u, u)$, and $(v_u, f^2_{n+1})$.
\end{enumerate}
The following table summarises the extra clock that we add to the construction,
and the new outgoing edges from $d^1_z$. For ease of notation, we define $U = \{
y^1, z^1, e^1_z\} \cup \{a^1_{z,l} \; : \; 1 \le l \le \lale\}$ to be the
original outgoing edges from $d^1_z$ that will now be replaced. Moreover, we
assume that each vertex $u$ is represented by a number in the range $1 \le u \le
|U|$, which will be used as part of the priority for $v_u$.

\begin{small}
\begin{center}
\begin{tabular}{l|l|l|l|l}
Vertex & Conditions  & Edges  & Priority & Player \\
\hline
$t_0^2$ & & $r^2$, $s^2$ & $\pp(2, 0, 2k+4n+4, 2, 0)$ & Even \\
$t_l^2$ & $1 \le l \le \lale$ & $r^2$, $s^2$, $t^2_{l-1}$ &
$\pp(2, 0, l, 2, 1)$ & Even \\
$a_l^2$ & $1 \le l \le \lale$ & $t^2_l$ & $\pp(2, 0, l
+1, 2, 0)$ & Even \\
\hline
$d^2_i$ & $1 \le i \le n$ & $e^2_i$, $s^2$, $r^2$, & $\pp(1, i, 0, 2, 1)$ & Even  \\
& & $a^2_l $ for $1 \le l \le \laletwo + 2i$  & &\\
$e^2_i$ & $1 \le i \le n$ & $d_i$ & $\pp(1, i, 1, 2, 0)$ &
Odd \\
$g^2_i$ & $1 \le i \le n$ & $f^2_i$ & $\pp(1, i, 2, 2, 1)$ &
Even \\
$k^2_i$ & $1 \le i \le n$ & $x$, $g^2_l$, for $i < l \le
n$ & $\pp(8, i, 0, 2, 1)$ & Even \\
$f^2_i$ & $1 \le i \le n$ & $e^2_i$ & $\pp(8, i, 1, 2, 1)$ &
Even \\
$h^2_i$ & $1 \le i \le n$ & $k^2_i$ & $\pp(8, i, 2, 2, 0)$ &
Even \\
\hline
$s^2$ & & $x$, $f^2_l$ for $1 \le l \le n$ & $\pp(7, 0, 0,
2, 0)$ & Even \\
$r^2$ & & $x$, $g^2_l$ for $1 \le l \le n$ & $\pp(7, 0, 1,
2, 0)$ & Even \\ \hline
$v_u$ & $u \in U$ & $u$ & $\pp(0, 0, 0, u, 0)$ & Even \\
$d^1_z$ & & $v_u$ for all $u \in U$ & $\pp(4,i,0,j,1)$ & Even \\
\end{tabular}
\end{center}
\end{small}

\paragraph{\bf \PSPACE-completeness.} We now argue how this modified construction
provides a \PSPACE-hardness proof for the optimal strategy decision problem.
Before the $n+1$th bit of the third clock flips, the edge $(v_u, f^2_{n+1})$ is
never switchable due to the large odd priority assigned to $f^2_{n+1}$, so this
modification does not affect the computation of $F^{2^n}(B)$. On the other hand,
once the $n+1$th bit of the third clock flips, all edges of the form 
$(v_u, f^2_{n+1})$ immediately become switchable, because the first component of
the valuation of $f^2_{n+1}$ is now a large even priority, and not $1$. So all
of these edges will be switched simultaneously. 

The key thing to note is that, since the priorities assigned to the vertices
$v_u$ are insignificant, they do not appear in the second component of the
valuation, and so vertex $d^1_z$ is now indifferent between all of its outgoing
edges. Moreover, the vertices $v_u$ never switch away from 
$f^2_{n+1}$ for the following reasons:
\begin{itemize}
\item The only even cycle that can be forced by player Even is the one that uses
$d^2_{n+1}$ and $e^2_{n+1}$. So, these vertices must select a strategy that
reaches this cycle eventually. 
\item All priorities used in the circuits are smaller than the priority of the
cycle between $d^2_{n+1}$ and $e^2_{n+1}$. So, the second component of the
valuation function is irrelevant, and the only way of improving the strategy
would be to find a shorter path to the cycle.
\item The vertices $v_u$ are the only vertices that have edges to the third
clock, so the only way a vertex $v_u$ could reach the third clock would be to
travel through both circuits to reach $d^1_z$, and then use a different vertex
$v_{u'}$,  but this would be a much longer path, and therefore this would have a
lower valuation.
\end{itemize}
Hence, the vertices $v_u$ will never switch away from $f^2_{n+1}$.

Observe that, after $2^n$ iterations, the input/output gadgets in circuit $1-j$
store the value of $F^{2^n}(B)$, and therefore $d^1_z$ chooses the edge to
$e^j_z$ if and only if the $z$th bit of $F^{2^n}(B)$ is $1$. The above argument
implies that $d^1_z$ does not switch again, so in the optimal strategy found by
the algorithm, the vertex $d^1_z$ chooses the edge to $e^j_z$ if and only if the
$z$th bit of $F^{2^n}(B)$ is $1$. Thus, we have that computing the optimal
strategy found by the V\"oge-Jurdzi\'nski strategy improvement algorithm is
\PSPACE-complete, as claimed in Theorem~\ref{thm:optstrat}.

\paragraph{\bf Other games.}
We also get similar results for the gain-bias algorithm for mean-payoff games,
and the standard strategy improvement algorithms for discounted and simple
stochastic games. For the most part, we can still rely on the proof of Friedmann
for these results. This is because, although we do not have a one-sink game, the
game behaves as a one-sink game until the $n+1$th bit in the third clock is
flipped. An easy way to see this is to reinstate the edge between $e^2_{n+1}$
and $h^2_{n+1}$ to create a one-sink game, and observe that, since the edge is
not used in the best response until the $n+1$th bit is flipped, it cannot affect
the sequence of strategies visited by strategy improvement. Once the $n+1$th bit
has flipped, we only care about making $d^1_z$ indifferent between its outgoing
edges, and in this section we explain how this is achieved.

For mean-payoff games, we use the same reduction as we did in
Section~\ref{sec:other} to our altered construction. After doing this, we set
the weight of the vertices $v_u$ to $0$ to ensure that $d^1_z$ will be exactly
indifferent between all of its outgoing edges once these vertices switch to
$f^2_{n+1}$. This gives the result for the gain-bias algorithm in mean-payoff
games.

For discounted games, once the standard reduction from mean-payoff to discounted
games has been applied, the proof of Friedmann already implies that the
discounted game algorithm makes the same decisions as the V\"oge-Jurdzi\'nski
algorithm for the vertices other than $d^1_z$. The only worry is that the
discount factor may make the vertex $d^1_z$ not indifferent between some of its
outgoing edges. However, it is enough to note that all paths from $d^1_z$ to
$f^2_{n+1}$ have length $2$, and therefore the vertex will be indifferent no
matter what discount factor is chosen. This gives the result for the standard
strategy improvement algorithm for discounted games.

Finally, after applying the standard reduction from discounted to
simple-stochastic games, the proof of Friedmann can be applied to argue that the
valuations in the simple stochastic game are related to the valuations in the
discounted game by a linear transformation. Hence, $d^1_z$ will still be
indifferent between its outgoing edges after the $n+1$th bit is flipped. This
gives the result for the standard strategy improvement algorithm for simple
stochastic games. Thus, we have the claimed results from
Corollary~\ref{cor:optstrat}.

\section{Open problems}
\label{sec:conc}

Strategy improvement generalizes policy iteration which solves mean-payoff and
discounted-payoff Markov decision processes~\cite{puterman94}. The exponential
lower bounds for greedy all-switches have been extended to MDPs. Fearnley showed
that the second player in Friedmann's construction~\cite{F11} can be
simulated by a probabilistic action, and used this to show an exponential lower
bound for the all-switches variant of policy iteration of average-reward
MDPs~\cite{F10}.  This technique cannot be applied to the construction in this
paper, because we use additional Odd player states (in particular the vertices
$q^j_{i,1}$) that cannot be translated in this way. Can our \PSPACE-hardness
results be extended to \allswitches for MDPs?

Also, there are other pivoting algorithms for parity games that deserve
attention. It has been shown that Lemke's algorithm and the Cottle-Dantzig
algorithm for the P-matrix linear complementarity problem (LCP) can be applied
to parity, mean-payoff, and discounted games~\cite{js08,fjs10}. It would be
interesting to come up with similar \PSPACE-completeness results for these
algorithms, which would also then apply to the more general P-matrix LCP
problem.
%
%We will use a technique that simulates one of the players in a parity game
%with random transitions in an MDP; this technique was used by Fearnley to
%extend Friedmann's exponential-time parity game examples for \allswitches to
%MDPs~\cite{F10}.

%Extend to MDPs? Detailed explanation of how to do it (or not?).

\newpage
\bibliographystyle{abbrv}
%plainnat.bst         
%abbrvnat.bst         
%unsrtnat.bst
\bibliography{references}

\newpage
\appendix

%\section{Priorities}

%Priority reference sheet:
%\begin{itemize}
%\item 0 - sink.
%\item 1 - small priorities in clock-bit gadgets.
%\item 2 - deceleration lane in the clock.
%\item 3 - mover gadget.
%\item 4 - small priorities in gate gadgets.
%\item 5 - deceleration lane for \notg gates.
%\item 6 - large priorities in gate gadgets.
%\item 7 - s and r
%\item 8 - large priorities in the clock-bit gadgets.
%\end{itemize}

%\section{Friedmann's Construction}

%\begin{itemize}
%\item We have $n$ bits, where $n$ is obtained from the circuit iteration
%instance.
%\item To simplify the description of the construction, we changed the name of
%the vertex $c$ to $t_0$.
%\item The priorities we give here are identical in order to the priorities that
%Friedmann specifies. We just needed more space.
%\end{itemize}

\newpage

\section{Facts about the clock}
\label{app:clock}

In this section we prove two important lemmas about the clocks. The first lemma
shows an important property about the difference in valuation between $r^j$ and
$s^j$ for the clock used by circuit $j$. The second lemma considers the
difference in valuation \emph{across} the two clocks, by  comparing the
valuations of $r^j$, $s^j$, $r^{1-j}$, and $s^{1-j}$.

\begin{lem}
\label{lem:clockrs}
Let $\sigma$ be a strategy that agrees with $\kappa^\C_m$ for
some $m$ in the range $1 \le m \le \length(\C)$, some clock-value $\C
\in \{0, 1\}^n$, and some $j \in \{0, 1\}$. We have:
\begin{enumerate}
\item \label{itm:rsone} If $m = \length(\C) -1$ then $\val^{\sigma}(r^j)
\sqsubset \val^{\sigma}(s^j)$.
\item \label{itm:rstwo} If $m < \length(\C) - 1$ then $\val^{\sigma}(s^j)
\sqsubset \val^{\sigma}(r^j)$.
\end{enumerate}
In both cases, we have that $\maxdiff^{\sigma}(s^j, r^j) \ge \pp(7, 0, 0, 0,
0)$. 
\end{lem}
\begin{proof}
We begin with the first case. In this case, by definition, we have that the path
that starts at $s^j$ and follows $\sigma$ moves to $f^j_i$ for some $i$, whereas
the path that starts at $r^j$ and follows $\sigma$ moves to $g^j_{i'}$ for some
$i' \ne i$. There are two possibilities.
\begin{enumerate}
\item If $i > i'$, then since $i$ is the least significant zero in $\C$, we must
have that $i' = \nexbit(\C, 0) = 1$. Hence, the path that starts at $g^j_{i'}$
passes through the bit gadgets for all bits strictly smaller than $i$ before
eventually arriving at $g^j_{\nexbit(\C, i)}$ (or $x$ if $\nexbit(\C, i)$ is not
defined). In particular, since the path does \emph{not} pass through $h^j_i$ the
largest priority on the path is strictly smaller than $\pp(8, i, 2, j, 0)$.

On the other hand, the path that starts at $f^j_i$ eventually arrives at
$g^j_{\nexbit(\C, i)}$ (or $x$ if $\nexbit(\C, i)$ is not defined) and it
\emph{does} pass through $h^j_i$. The largest priority on this path is  $\pp(8,
i, 2, j, 0)$, and since the priority is even, we can conclude that
$\val^{\sigma}(r^j) \sqsubset \val^{\sigma}(s^j)$.

\item If $i < i'$, then since $i'$ is the least significant one in $\C$, we must
have that $i = 1$. Hence, the path that starts at $f^j_i$ eventually moves to
$k^j_i$ and then directly to $g^j_{i'}$. The largest priority on this path is
$\pp(8, i', 2, j, 0)$, and since this is even, we can conclude that
$\val^{\sigma}(r^j) \sqsubset \val^{\sigma}(s^j)$.
\end{enumerate}

We now move on to the second case. In this case, by definition, we have that the
path that starts at $s^j$ moves to $f^j_i$ for some $i$, whereas the path that
starts at $r^j$ moves to $g^j_i$ and then to $f^j_i$. Since the priority on
$g^j_i$ is strictly smaller than $\pp(7, 0, 1, j, 0)$, we have that
$\maxdiff^{\sigma}(r^j, s^j) = \pp(7, 0, 1, j, 0)$, which the priority assigned
to $r^j$. Since this priority is even, we have that $\val^{\sigma}(s^j)
\sqsubset \val^{\sigma}(r^j)$, as required.

Observe that in all cases considered above, we have shown that
$\maxdiff^{\sigma}(r^j, s^j) \ge \pp(7, 0, 0, 0, 0)$ as required. Hence, we have
completed the proof of this lemma. 
\end{proof}

\begin{lem}
\label{lem:crossclock}
Let $\sigma$ be a strategy that agrees with $\chi^{B, \C, j}_m$ for some $m \ge
1$, some bit-strings $B, \C \in \{0, 1\}^n$, and some $j \in \{0, 1\}$. We have:
\begin{enumerate}
\item \label{itm:ccone} If $m = \delay(j, \C) - 1$, then $\val^{\sigma}(r^{1-j})
\sqsubset \val^{\sigma}(r^{j}) \sqsubset \val^{\sigma}(s^{1-j})$ and
\begin{align*}
\maxdiff^\sigma(r^{1-j}, s^j) \ge \pp(7, 0, 0, 0, 0)
\end{align*}
and
$\maxdiff^\sigma(r^{1-j}, r^j) \ge \pp(7, 0, 0, 0, 0)$.
\item \label{itm:cctwo} If $m < \delay(j, \C) - 1$, then
$\val^{\sigma}(r^{1-j}) \sqsubset \val^{\sigma}(s^j) \sqsubset \val^{\sigma}(r^{j})$ and
\begin{align*}
\maxdiff^\sigma(r^{1-j}, s^{j}) \ge \pp(7, 0, 0,
0, 0)
\end{align*}
and $\maxdiff^\sigma(s^{j}, r^{1-j}) \ge \pp(7, 0, 0, 0, 0)$.
\end{enumerate}
\end{lem}
\begin{proof}
We begin with the second claim. The fact that $\val^{\sigma}(s^j) \sqsubset
\val^{\sigma}(r^j)$ follows from part~\ref{itm:rstwo} of
Lemma~\ref{lem:clockrs}, so it is sufficient to show that
$\val^{\sigma}(r^{1-j}) \sqsubset \val^{\sigma}(s^j)$. There are two cases to
consider, based on whether $j = 0$ or $j = 1$.
\begin{enumerate}
\item If $j = 0$, then clock $j$ uses bit-string $\C$, and clock $1-j$ uses
bit-string $\C - 1$. Observe that the clock strategies specify that the path
starting at $s^{j}$ visits $h^j_i$ if and only if $\C_i = 1$. Similarly, the
path starting at $r^{1-j}$ visits $h^{1-j}_i$ if and only if $(\C - 1)_i = 1$.
If $i'$ is the index of the least significant $1$ in $\C$, then we have that the
path that starts at $s^j$ visits $h^j_{i'}$ and $k^j_{i'}$, and the path that
starts at $s^{1-j}$ does not visit these vertices. Moreover, these two paths are
the same after this point. Hence, we have that $\maxdiff(s^j, r^{1-j})$ is
$\pp(8, i', 2, j, 0)$, and since this priority is even, we can conclude that
$\val^{\sigma}(r^{1-j}) \sqsubset \val^{\sigma}(s^j)$.

\item If $j = 1$, then both clocks use bit-string $\C$. Hence, the paths
starting at $s^j$ and $r^{1-j}$ use the same path through their respective
clocks. So, if $i'$ is the index of the most significant $1$ in $\C$, then we
have that 
$\pp(8, i', 2, 0, 0)$ is the largest priority on the path starting at $r^{1-j} =
r^0$, and 
$\pp(8, i', 2, 1, 0)$ is the largest priority on the path starting at $s^{j} =
s^1$. Thus, $\maxdiff^{\sigma}(s^j, r^{1-j}) = \pp(8, i', 2, 1, 0)$, and since
this priority is even and contained in $\val^{\sigma}(s^j)$ we can conclude that
$\val^{\sigma}(r^{1-j}) \sqsubset \val^{\sigma}(s^j)$.
\end{enumerate}

We now move on to the first claim. Here the same reasoning as we gave for the
second case can be used to prove that $\val^{\sigma}(r^{1-j}) \sqsubset
\val^{\sigma}(s^j)$, and therefore Lemma~\ref{lem:clockrs} implies that
$\val^{\sigma}(r^{1-j}) \sqsubset \val^{\sigma}(r^j)$. What remains is to prove
that 
$\val^{\sigma}(r^j) \sqsubset \val^{\sigma}(s^{1-j})$. Again there are two cases
to consider.
\begin{enumerate}
\item If $j = 0$, then clock $j$ uses bit-string $\C$, and clock $1-j$ is about
to transition from bit-string $\C-1$ to bit-string $\C$. In fact, the path from
$s^{1-j}$ is already the path for bit-string $\C$, so the proof from item~$2$
above can be reused.
\item If $j = 1$, then clock $j$ uses bit-string $\C$, and clock $1-j$ is about
to transition from bit-string $\C$ to bit-string $\C+1$. In fact, the path from
$s^{1-j}$ is already the path for bit-string $\C+1$, so the proof from item~$1$
above can be reused.
\end{enumerate}
Finally, we observe that all of the maximum difference priorities used in the
proof are strictly larger than $\pp(7, 0, 0, 0, 0)$, which completes the proof.
\end{proof}

\section{Best responses}
\label{app:br}

In this section, we prove that the best responses defined in
Section~\ref{sec:strategies} are indeed best responses to $\chi^{B, \C, j}_m$.
There are two types of odd vertices used in the construction: the vertices
$e^j_i$ used in the \notg and \inputg gates, and the vertices $q^j_{i, 0}$ used
in the \inputg gates. We begin by proving a general lemma concerning the
vertices $e^j_i$ used in the \notg and \inputg gates.

\begin{lem}
\label{lem:nocycle}
Let $\sigma $ be a strategy that agrees with $\chi^{B, \C, j}_m$ for some
bit-strings $B, \C \in \{0, 1\}^n$, some $m$ in the range $1 \le m \le \delay(j,
\C) - 1$,  and some $j \in \{0, 1\}$. For every $i \in \notg \cup \inputg$, and
every $l \in \{0, 1\}$, we have that if $\sigma(d^l_{i}) = e^l_i$, then
$\br(\sigma)(e^l_i) \ne d^l_i$.
\end{lem}
\begin{proof}
Note that if player Odd uses the edge from $e^l_i$ to $d^l_i$, then this would
create a cycle with largest priority $\pp(1, i, 1, j, 0)$, which is even. Since
the game is a one-sink game, and since the initial strategy is terminating, we
have that Odd can eventually reach the odd cycle at $x^j$ from the vertices
$r^j$, $r^{1-j}$, $s^j$, and $s^{1-j}$. Furthermore, Odd can reach one of these
four vertices by moving to $h^l_i$. Since the odd cycle at $x^j$ has priority
smaller than $\pp(1, i, 1, j, 0)$, we can conclude that $\br(\sigma)(e^j_i) \ne
d^l_i$. 
\end{proof}

We now proceed to prove individual lemma for each of the vertices that belong to
player Odd. Each type of Odd vertex will be considered in a different
subsection.

\subsection{The vertices $q^l_{i, 0}$} 

We now consider the vertices $q^l_{i,
0}$ for $i \in \inputg$. The first lemma considers the case where $l = j$, and
the second lemma considers the case where $l = 1-j$.

\begin{lem}
\label{lem:brqj}
Let $\sigma $ be a strategy that agrees with $\chi^{B, \C, j}_m$ for some
bit-strings $B, \C \in \{0, 1\}^n$, some $m$ in the range $1 \le m \le \delay(j,
\C) - 1$,  and some $j \in \{0, 1\}$. 
For every $i \in \inputg$, we have that $\br(\sigma)(q^{j}_{i,0}) = \mu^{B,
\C, j}_m(q^j_{i,0})$.
\end{lem}
\begin{proof}
There are two cases to consider.
\begin{itemize}
\item If $m = 1$, then then we must show that the edge to $e^j_{i}$ is chosen by
Odd in the best response.
Consider a strategy $\tau$
where
$\tau(e^j_i) = h^j_{i,0}$, and 
$\tau(q^j_{i,0}) = e^j_i$. When $\tau$ is played against $\sigma$, the path that
starts at $e^j_i$ eventually arrives at $r^{1-j}$, and the largest priority on this path is
strictly smaller than 
$\pp(6, d(C) + 2, 0, j, 0)$.
On the other hand, taking the edge to $q^j_{i, 1}$ leads directly to $r^{1-j}$
while visiting the priority 
$\pp(6, d(C) + 2, 0, j, 0)$. Since this priority is even, we can conclude that
Odd would prefer to play $\tau$ than to use the edge from $q^j_{i, 0}$ to
$q^j_{i, 1}$ in his best response. Therefore, we must have that
$\br(\sigma)(q^j_{i, 0}) = e^j_i$, as required.

\item If $m > 1$, then we must show that the edge to $q^{j}_{i,1}$ is the least
appealing edge at $q^{j}_{i, 0}$. Observe that the path that starts at $e^{j}_i$
and follows $\sigma$ eventually arrives at $r^j$, and every priority on this
path that is strictly smaller than $\pp(7, 0, 0, 0, 0)$. On the other hand, the
path that starts at $q^j_{i, 1}$ moves directly to $r^{1-j}$. Hence, we can
apply Lemma~\ref{lem:crossclock} (both parts) to argue that
$\val^{\sigma}(q^{j}_{i, 1}) \sqsubset \val^{\sigma}(e^j_i)$, as required. \qedhere
\end{itemize} 
\end{proof}

\begin{lem}
Let $\sigma $ be a strategy that agrees with $\chi^{B, \C, j}_m$ for some
bit-strings $B, \C \in \{0, 1\}^n$, some $m$ in the range $1 \le m \le \delay(j,
\C) - 1$,  and some $j \in \{0, 1\}$. 
For every $i \in \inputg$, we have that $\br(\sigma)(q^{1-j}_{i,0}) = \mu^{B,
\C, j}_m(q^{1-j}_{i,0}$.
\end{lem}
\begin{proof}
We must show that the edge to $e^{1-j}_i$ is the least
appealing edge at $q^{1-j}_{i, 0}$. Observe that the path that starts at
$e^{1-j}_i$ and follows $\sigma$ will eventually arrive at either $r^j$ or
$r^{1-j}$. In either case, the largest priority on this path will be strictly
smaller than 
$\pp(6, d(C) + 2, 0, j, 0)$. On the other hand, the path that starts at
$q^{1-j}_{i, 1}$ moves directly to $r^j$, and the largest priority on this path
is 
$\pp(6, d(C) + 2, 0, j, 0)$. Since this priority is even, we can conclude that
$\val^{\sigma}(e^{1-j}_i) \sqsubset \val^{\sigma}(q^{1-j}_{i, 0})$, as required.
\end{proof}

\subsection{The vertices $e^l_i$ in \notg gates}

The following lemma considers the vertices $e^l_i$ for $l = j$ and $i \in
\notg$. We do not need to prove a lemma for the case where $l = 1-j$ and $i \in
\notg$, because these vertices are in the non-computing circuit, and we do not
specify strategies for the \notg and \org gadgets in the non-computing circuits.

\begin{lem}
\label{lem:brnot}
Let $\sigma$ be a strategy that agrees with $\chi^{B, \C, j}_m$ for some
bit-strings $B, \C \in \{0, 1\}^n$, some $m$ in the range $1 \le m \le \delay(j,
\C) - 1$,  and some $j \in \{0, 1\}$. 
For every $i \in \notg$, we have that $\br(\sigma)(e^{j}_{i}) = \mu^{B,
\C, j}_m(e^j_i)$.
\end{lem}
\begin{proof}
There are three cases to consider.
\begin{enumerate}
\item If $m = 1$ then, the path that starts at $d^j_i$ and follows $\sigma$
moves directly to $s^j$. On the other hand, the path that starts at
$h^j_i$ moves directly to $r^j$.
All of the priorities on these paths are strictly
smaller than $\pp(7, 0, 0, 0, 0)$, so we can apply Lemma~\ref{lem:clockrs}
part~\ref{itm:rstwo} to argue that $\val^{\sigma}(d^j_i) \sqsubset
\val^{\sigma}(d^j_i)$, so therefore we have $\br(\sigma)(e^j_i) = d^j_i$.

\item If $m > 1$ and either $m \le d(i) + 2$, or $m > d(i) + 2$ and $\eval(B,
\inp(i)) = 1$, then observe that the path that starts at $d^j_i$ and follows
$\sigma$ eventually arrives at $r^j$, and that the largest priority on this path
is strictly smaller than $\pp(6,i,1,j,0)$. On the other hand, the path that
starts at $h^j_i$ and follows $\sigma$ moved directly to $r^j$, and the largest
priority on this path is $\pp(6,i,1,j,0)$. Since this priority is even, we can
conclude that $\val^{\sigma}(d^j_i) \sqsubset \val^{\sigma}(e^j_i)$, so
therefore we have $\br(\sigma)(e^j_i) = d^j_i$.

\item If $m > d(i) + 2$ and $\eval(B, \inp(i)) = 0$, then we have
$\sigma(d^j_i) = e^j_i$, so we can apply Lemma~\ref{lem:nocycle} to prove that
$\br(\sigma)(e^j_i) = h^j_i$.
\qedhere
\end{enumerate} 
\end{proof}

\subsection{The vertices $e^j_i$ in input/output gates}

The following lemmas consider the vertices $e^l_i$ when $i \in \inputg$. The
first lemma considers the case where $l = j$, and the second lemma considers the
case where $l = 1-j$.

\begin{lem}
\label{lem:br1}
Let $\sigma$ be a strategy that agrees with $\chi^{B, \C, j}_m$ for some
bit-strings $B, \C \in \{0, 1\}^n$, some $m$ in the range $1 \le m \le \delay(j,
\C) - 1$,  and some $j \in \{0, 1\}$. 
For every $i \in \inputg$, we have that $\br(\sigma)(e^{j}_{i}) = \mu^{B,
\C, j}_m(e^j_i)$.
\end{lem}
\begin{proof}
There are a number of cases to consider.
\begin{enumerate}
%\item If $m = 1$ and $B_i = 0$, then the path that starts at $d^j_i$ eventually
%reaches $r^{1-j}$, and the largest priority on this path is strictly smaller
%than $\pp(6, 0, 1, j, 0)$. On the other hand, the path that starts at $h^j_{i,
%0}$ moves to $h^j_{i, 2}$ and then to $r^{1-j}$, and the largest priority on
%this path is $\pp(6, 0, 1, j, 0)$. Since this priority is even, we can conclude
%that $\val^{\sigma}(d^j_i) \sqsubset \val^{\sigma}(h^j_{i,0})$, and therefore we
%have $\br(\sigma)(e^j_i) = d^j_i$.

%\item If $m = 1$, and $B_i = 1$, then we have $\sigma(d^j_i) = e^j_i$, and so we
%can apply Lemma~\ref{lem:nocycle} to argue that $\br(\sigma)(e^j_i) =
%h^j_{i,0}$.

\item If $m > 1$ and $F^2(B)_i = 0$ and either $m \le d(i)+3$, or $m > d(i) +3$
and $F^2(B)_i = 0$, then the path that starts at $d^j_i$ eventually arrives at
$r^j$, and the largest priority on this path is strictly smaller than 
$\pp(6,d(C)+1,1,j,0)$. On the other hand, the path that starts at $h^j_{i, 0}$
and follows $\sigma$ passes through $h^j_{i, 1}$ and then arrives at $r^j$. The
largest priority on this path is 
$\pp(6,d(C)+1,1,j,0)$, and since this priority is even, we can conclude that
$\val^{\sigma}(d^j_i) \sqsubset \val^{\sigma}(h^j_{i, 0})$. Therefore, we have
that 
$\br(\sigma)(e^j_i) = d^j_{i}$.

\item If $m > 1$ and $m > d(i) + 3$ and $F^2(B)_i = 1$, then $\sigma(d^j_i) =
e^j_i$, and so we can apply Lemma~\ref{lem:nocycle} to argue that
$\br(\sigma)(e^j_i) = h^j_{i,0}$.
\qedhere
\end{enumerate} 
\end{proof}

\begin{lem}
\label{lem:br2}
Let $\sigma$ be a strategy that agrees with $\chi^{B, \C, j}_m$ for some
bit-strings $B, \C \in \{0, 1\}^n$, some $m$ in the range $1 \le m \le \delay(j,
\C) - 1$,  and some $j \in \{0, 1\}$. 
For every $i \in \inputg$, we have that $\br(\sigma)(e^{1-j}_{i}) = \mu^{B,
\C, j}_m(e^{1-j}_i)$.
\end{lem}
\begin{proof}
There are a number of cases to consider.
\begin{enumerate}
\item If $m = 1$ and $B_i = 0$, then the path that starts at $d^{1-j}_i$ and
follows $\sigma$ will eventually reach $r^{1-j}$, and the largest priority on
this path is strictly smaller than $\pp(6,d(C)+1,1,j,0)$. On the other hand, the path that starts at
$h^{1-j}_{i, 0}$ moves to $h^{1-j}_{i, 1}$ and then arrives at $r^{1-j}$, and
the largest priority on this path is $\pp(6,d(C)+1,1,j,0)$. Since this priority
is even, we have that $\val^{\sigma}(d^{1-j}_i) \sqsubset
\val^{\sigma}(h^{1-j}_{i,0})$ and therefore $\br(\sigma)(e^{1-j}_i) =
d^{1-j}_{i}$.

\item $m > 1$ and $B_i = 0$, then the path that starts at $d^{1-j}_i$ moves
to $r^j$, and the largest priority on this path is strictly smaller than $\pp(6,
0, 1, j, 0)$. On the other hand, the path that starts at $h^{1-j}_{i, 0}$ moves
to $h^{1-j}_{i, 2}$ and then arrives at $r^{j}$, and the largest priority on
this path is $\pp(6, 0, 1, j, 0)$. Since this priority is even, we have that
$\val^{\sigma}(d^{1-j}_i) \sqsubset \val^{\sigma}(h^{1-j}_{i,0})$ and therefore
$\br(\sigma)(e^{1-j}_i) = d^{1-j}_{i}$.

\item If $B_i = 1$, then we have $\sigma(d^{1-j}_i) = e^{1-j}_i$, and we can
apply Lemma~\ref{lem:nocycle} to argue that $\br(\sigma)(e^{1-j}_i) =
h^{1-j}_{i,0}$.
\qedhere
\end{enumerate} 
\end{proof}

\subsection{Gate outputs}
\label{app:outputs}

In this section we give two key lemmas that describe the valuation of the output
states $o^j_i$. The first lemma considers the case where $m = 1$ or $m = 2$, and
the second lemma considers the case where $m \ge 3$.

\begin{lem}
\label{lem:oval1}
Let $\sigma$ be a strategy that agrees with $\chi^{B, \C, j}_m$ for some $\C, B
\in \{0, 1\}^n$, some $j \in \{0, 1\}^n$, and for $m = 1$ or $m = 2$. For every
gate $i$, we have $\val^{\sigma}(o^j_i) \sqsubset \val^{\sigma}(r^j)$.
\end{lem}
\begin{proof}
There are four cases to consider
\begin{enumerate}
\item We begin by showing the claim for an input/output gate $i$ from circuit $1-j$ in
the case where $m = 1$. Note that
Lemma~\ref{lem:br2} implies that $\br(\sigma)(e^{1-j}_i) = d^{1-j}_i$.
Observe that by definition, the path that starts at $d^j_i$ and follows $\sigma$
will trace a path through the gadgets for circuit $1-j$, and will eventually
reach $r^{1-j}$. Furthermore, the largest priority possible that can be seen
along this path is $\pp(6, d(C)+1, 1, j, 0) < \pp(7, 0, 0, 0, 0)$. Hence, we can
apply Lemma~\ref{lem:crossclock} to argue that $\val^{\sigma}(o^{1-j}_i)
\sqsubset \val^{\sigma}(r^j)$.

\item Now we consider an input/output gate $i$ from circuit $1-j$ in the case where $m =
2$. Again, Lemma~\ref{lem:br2} implies that 
$\br(\sigma)(e^{1-j}_i) = d^{1-j}_i$, but in this case 
since $m = 2$, we have that the vertices $y^{1-j}$ and $p^{1-j}_i$ have both switched to $r^j$. Hence, 
the path that starts at $o^{1-j}_i$ 
will eventually arrive at $r^j$.
It can be verified that, whatever path is taken from $o^{1-j}_{i}$ to $r^j$, the largest priority 
along this path is $\pp(6, i, 0, j, 1)$ on the vertex $o^{1-j}_{i}$. Since this
priority is odd, we have that $\val^{\sigma}(o^{1-j}_i) \sqsubset
\val^{\sigma}(r^j)$.

\item Next we consider the case where $i$ is a \org-gate. If $m = 1$ then we
have that $\sigma(o^j_i) = s^j$, and if $m = 2$ then $\sigma(o^j_i) = r^j$. In
both cases,  we can use Lemma~\ref{lem:clockrs} and the fact that the priority
assigned to $o^j_i$ is odd, to conclude that $\val^{\sigma}(o^{j}_i) \sqsubset
\val^{\sigma}(r^j)$.

\item Finally, we consider the case where $i$ is a \notg-gate. We can apply
Lemma~\ref{lem:brnot} to argue that $\br(\sigma)(e^j_i) = d^j_i$. If $m = 1$
then we have that $\sigma(d^j_i) = s^j$, and if $m = 2$ then $\sigma(d^j_i) =
r^j$. In both cases, the  highest priority on the path from $o^j_i$ to either
$s^j$ or $r^j$ is the odd priority from $o^j_i$, so we can use this fact, along
with Lemma~\ref{lem:clockrs} to conclude that $\val^{\sigma}(o^{j}_i) \sqsubset
\val^{\sigma}(r^j)$.
\qedhere
\end{enumerate} 
\end{proof}

\begin{lem}
\label{lem:oval}
Let $\sigma$ be a strategy that agrees with $\chi^{B, \C, j}_m$ for some $\C, B
\in \{0, 1\}^n$, some $j \in \{0, 1\}^n$, and some $m$ in the range $3 \le m \le
\delay(j, \C$. For every gate $i$, we have:
\begin{enumerate}
\item \label{itm:nineone} If $m \le d(i) + 2$, then $\val^{\sigma}(o^j_i) \sqsubset
\val^{\sigma}(r^j)$.
\item \label{itm:ninetwo} If $m > d(i) + 2$ and $\eval(B, i) = 0$, then 
$\val^{\sigma}(o^j_i) \sqsubset \val^{\sigma}(r^j)$.
\item \label{itm:ninethree} If $m > d(i) + 2$ and $\eval(B, i) = 1$, then
$\val^{\sigma}(r^j)  \sqsubset \val^{\sigma}(o^j_i)$ and:
\begin{equation*}
\pp(6,0,0,0,0) \le \maxdiff^{\sigma}(r^j, o^j_i) \le \pp(6, i, 1, j, 0).
\end{equation*}
\end{enumerate}
\end{lem}
\begin{proof}
We will prove this claim by induction over the depth of the gates. For the base
case we consider an input/output gate $i$ from circuit $1-j$, which provide the input
values for circuit $j$. Since we consider these gates to have depth $0$, we
always have $m > d(i) + 2$, so there are two cases to prove based on whether
$B_i$ is zero or one. First, observe that since $m \ge 3$, the circuit mover
gadgets attached to the input/output gadget for bit $i$ in circuit $1-j$ have
$\sigma(y^{1-j}) = r^{j}$. Since $\delay(1-j,\C) \ge d(C) + 3$, the definition
given in~\eqref{def:inputd} implies that the strategy at $d^{1-j}_i$ is
determined by $B_i$. So we have the following two cases.
\begin{itemize}
\item If $B_i = 0$, then $\sigma(d^{1-j}_i) = a^{1-j}_{i,l}$ for some $l$. By
definition, in the strategy $\sigma$, all paths from $a^{1-j}_{i,l}$ eventually
arrive at $r^j$, and the maximum priority on any of these paths is smaller than
$\pp(6,0,0,j,1)$. Hence, the largest priority on the path from $o^{1-j}_i$ to
$r^j$ is the priority $\pp(6,0,0,j,1)$ on the vertex $o^{1-j}_i$, and since this
is an odd priority, we have $\val^{\sigma}(o^{1-j}_i) \sqsubset
\val^{\sigma}(r^{j})$.
\item If $B_i = 1$, then $\sigma(d^{1-j}_i) = e^{1-j}_i$ and
Lemma~\ref{lem:br2} then implies $\br(\sigma)(e^{1-j}_i) = h^{1-j}_i$. So, the
path that starts at $o^{1-j}_i$ and follows $\sigma$ passes through $e^{1-j}_i$,
$h^{1-j}_{i,0}$, $h^{1-j}_{i,2}$, and then arrives at $r^j$. The largest
priority on this path is $\pp(6,0,1,j,0) > \pp(6, 0, 0, j, 1)$ on the state
$h^{1-j}_{i,2}$, so we have $\val^{\sigma}(r^j) \sqsubset
\val^{\sigma}(o^{1-j}_i)$
\end{itemize}
Hence, the base case of the induction has been shown. The inductive step will be
split into two cases, based on whether $i$ is a \notg-gate or an \org-gate.

Suppose that the inductive hypothesis holds for all gates $i$ with $d(i) < k$,
and let $i$ be a \org-gate with $d(i) = k$. We must prove three cases.
\begin{itemize} \item The first two cases use the same proof. If $m \le d(i)+2$,
or if $m > d(i) + 2$ and $\eval(B, i) = 0$, then by definition we have
$\sigma(o^j_i) = r^j$. Since the priority assigned to $o^j_i$ is odd, we have
$\val^{\sigma}(o^j_i) \sqsubset \val^{\sigma}(r^j)$, as required.

\item If $m > d(i) + 2$ and $\eval(B, i) = 1$, then by definition we have that
$\sigma(o^j_i) = \inputstate(i, j, l)$ for some gate $l$ with $l \in \{1, 2\}$,
and we know that $\eval(B, \inp_l(i)) = 1$. Hence, we can apply
the inductive hypothesis to argue that $\maxdiff^{\sigma}(r^j,
\inputstate(i,j,l))$ is even,
and it satisfies:
\begin{equation*}
\pp(6,0,0,0,0) \le \maxdiff^{\sigma}(r^j, \inputstate(i,j,l)) \le \pp(6, l, 1, j, 0).
\end{equation*}
Since the priority assigned to $o^j_i$ is smaller than $\pp(6,0,0,0,0)$, we have
that the same two properties apply to $\maxdiff^{\sigma}(r^j, o^j_i)$. Hence,
$\val^{\sigma}(r^j)  \sqsubset \val^{\sigma}(o^j_i)$, and the required bounds on
$\maxdiff^{\sigma}(r^j, o^j_i)$ hold because $\pp(6, l, 1, j, 0) < \pp(6, i, 1,
j, 0)$.
\end{itemize}

Now suppose that the inductive hypothesis holds for all gates $i$ with $d(i) <
k$, and let $i$ be a \notg-gate with $d(i) = k$. We must prove three cases.
\begin{itemize}
\item
If $m \le d(i) + 2$, then by definition, the path that starts at $o^j_i$ and follows
$\sigma$ passes through $e^j_i$, and Lemma~\ref{lem:brnot} implies that it then
moves to $d^j_i$, a vertex of the form $a^j_{i,l}$, a number of vertices
of the form $t^j_{i,l}$, before finally arriving at $r^j$. It can easily be
verified that the largest priority on this path is $\pp(6, i, 0, j, 1)$ from the
vertex $o^j_i$. So, we have $\maxdiff^{\sigma}(r^j, o^j_i) = \pp(6, i, 0, j,
1)$, and since this priority is odd, we have $\val^{\sigma}(o^j_i) \sqsubset
\val^{\sigma}(r^j)$, as required.

\item 
If $m > d(i) + 2$ and $\eval(B, i) = 0$, then we must have $\eval(B,
\inp(i)) = 1$. Hence, by definition, the path that starts at $o^j_i$ and follows
$\sigma$ first passes through $e^j_i$, and then Lemma~\ref{lem:brnot} implies that
it then passes through $d^j_i$, and then some vertex of the form $a^j_{i,l}$,
followed by a number of vertices of the form $t^j_{i,l}$, before finally
arriving at $t^j_{i, d(i)}$ and then moving to $\inputstate(i,j)$. The largest
priority on this path is $\pp(6, i, 0, j, 1)$ from the vertex $o^j_i$. By the
inductive hypothesis, we have $\maxdiff^{\sigma}(r^j, \inputstate(i,j)) < \pp(6,
\inp(i), 1, j, 0)$, and since $\inp(i) < i$ we have that $\pp(6, i, 0, j, 1) >
\pp(6, \inp(i), 1, j, 0)$. Hence, we have that $\maxdiff^{\sigma}(r^j,
o^j_{\inp(i)}) = \pp(6, i, 0, j, 1)$, and since this is odd, we have that
$\val^{\sigma}(o^j_i) \sqsubset \val^{\sigma}(r^j)$, as required.

\item If $m > d(i) + 2$ and $\eval(B, i) = 1$, then by definition we have
that the path that starts at $o^j_i$ and follows $\sigma$ passes through
$e^j_i$, and then Lemma~\ref{lem:brnot} implies that it passes through $h^j_i$, and
then reaches $r^j$. The largest priority on this path is $\pp(6,i,1,j,0)$ on the
vertex $h^j_i$. Hence we have $\maxdiff^{\sigma}(r^j, o^j_i) \le \pp(6, i, 1, j,
0)$, and since this priority is even, we have that $\val^{\sigma}(r^j)
\sqsubset \val^{\sigma}(o^j_i)$. Hence, we have shown both of the required
properties for this case.
\end{itemize}
Now that we have shown the two versions of the inductive hypothesis, we have
completed the proof. 
\end{proof}

\section{Or gates}
\label{app:org}

The following pair of lemmas show that the states $o^j_i$ in the \org gate
gadgets correctly switch to the outgoing edge specified by
$\chi^{B,\C,j}_{m+1}$. The first lemma considers the case where $1 \le m <
\delay(j, \C) - 1$, and the second lemma considers the case where $m = \delay(j,
\C) - 1$.

\begin{lem}
Let $\sigma$ be a strategy that agrees with $\chi^{B, \C, j}_m$ for some $\C, B
\in \{0, 1\}^n$, some $j \in \{0, 1\}^n$, and some $m$ in the range $1 \le m < 
\delay(j, \C) - 1$. For each \org-gate $i$, greedy all-switches strategy
improvement will switch $o^j_i$ to $\chi^{B,\C,j}_{m+1}(o^j_i)$.
\end{lem}
\begin{proof}
In this proof we will show the that the most appealing edge is the one that is
specified in Equation~\eqref{eqn:odef}. This boils down to a case analysis.

First suppose that $m < d(i)+2$. We must prove that the edge to $r^j$ is the
most appealing edge at $o^j_i$. By Lemma~\ref{lem:clockrs}, we have that
$\val^{\sigma}(s^j) \sqsubset \val^{\sigma}(r^j)$. Furthermore, since
$d(\inp_1(i)) = d(\inp_2(i)) = d(i) - 1$, part~\ref{itm:nineone} of Lemma~\ref{lem:oval} implies
that $\val^{\sigma}(\inputstate(i,j,l)) \sqsubset \val^{\sigma}(r^j)$ for
$l \in \{1, 2\}$. Hence, we have that $r^j$ is the most appealing edge at
$o^j_i$, as required.

Now suppose that $d(i) + 2 \le  m \le \delay(j, \C) -1$. There are three cases to
consider.
\begin{itemize}
\item If both input gates are false, then Lemma~\ref{lem:clockrs} and
part~\ref{itm:ninetwo}  of
Lemma~\ref{lem:oval} imply that $r^j$ will continue to be the most appealing
edge at $o^j_i$, as required.

\item If $\inp_l(i)$ is true and $\inp_{1-l}(i)$ is false, for some $l \in \{1,
2\}$, then part~\ref{itm:ninethree}  of Lemma~\ref{lem:oval} implies that
$\val^{\sigma}(r^j) \sqsubset \val^{\sigma}(\inputstate(i,j,l))$, and
$\val^{\sigma}(\inputstate(i,j,1-l)) \sqsubset \val^{\sigma}(r^j)$.
Therefore, the most appealing edge at $o^j_i$ is the one to
$\inputstate(i,j,l)$, as required.

\item Finally,  if both input gates are true, then Lemma~\ref{lem:oval} implies
that the highest appeal edge at $o^j_i$ is either $\inputstate(i,j,1)$ or
$\inputstate(i,j,2)$. Since $\ornext(i,B,m)$ is defined to be the successor
with highest appeal, we have that the highest appeal edge at $o^j_i$ is the one
to $o^j_{I_{\ornext(i)}(i,B,m)}$, as required. 
\end{itemize}
This completes the proof that greedy all-switches strategy improvement will
switch $o^j_i$ to $\chi^{B,\C,j}_{m+1}(o^j_i)$. 
\end{proof}

\begin{lem}
Let $\sigma$ be a strategy that agrees with $\chi^{B, \C, j}_m$ for some $\C, B
\in \{0, 1\}^n$, some $j \in \{0, 1\}^n$, and for $m = \delay(j, \C)-1$. For
each \org-gate $i$, greedy all-switches strategy improvement will switch
$o^{1-j}_i$
to $\chi^{B,\C,j}_{m+1}(o^{1-j}_i)$.
\end{lem}
\begin{proof}
We must show that the edge to $s^{1-j}$ is the most appealing edge at
$o^{1-j}_i$. It can be verified that all paths starting at
$\inputstate(i,j,1)$ and $\inputstate(i,j,2)$ either reach
$r^{1-j}$, $s^{1-j}$, or $r^j$. Furthermore, the largest possible priority on
these paths is strictly smaller than $\pp(7, 0, 0, 0, 0)$. Hence, we can apply
part~\ref{itm:rsone} of Lemma~\ref{lem:clockrs} and part~\ref{itm:ccone} of
Lemma~\ref{lem:crossclock} to conclude that the edge to $s^{1-j}$ is the most
appealing outgoing edge at $o^{1-j}_i$. 
\end{proof}

\section{The states $t^j_{i,l}$ in \notg gates}
\label{app:nott}

In this section we show that the states $t^j_{i,l}$ in the \notg gate gadgets
correctly switch to the outgoing edge specified by $\chi^{B,\C,j}_{m+1}$. The
first two lemmas consider the case where $1 \le m < \delay(j, \C) - 1$, and the
third lemma considers the case where $m = \delay(j, \C) - 1$. The first lemma
deals with the case where $l < d(i)$, while the second lemma deals with the case
where $l > d(i)$. Observe that there is no need to deal with the case where $l =
d(i)$, because $t^j_{i, d(i)}$ only has one outgoing edge.

\begin{lem}
\label{lem:nottsmall}
Let $\sigma$ be a strategy that agrees with $\chi^{B, \C, j}_m$ for some $\C, B
\in \{0, 1\}^n$, some $j \in \{0, 1\}^n$, and some $m$ in the range $1 \le m <
\delay(j, \C)-1$. For each \notg-gate $i$, and each $l$ int the range $1 \le l <
d(i)$, greedy all-switches strategy improvement will switch $t^j_{i,l}$ to
$\chi^{B,\C,j}_{m+1}(t^j_{i,l})$.
\end{lem}
\begin{proof}
Lemma~\ref{lem:clockrs} implies that the state $t^j_{i,l}$ will not switch to
$s^j$. In the rest of this proof we consider the two remaining outgoing edges at
this state. We have two cases to consider.
\begin{enumerate}
\item
\label{itm:tone}
 We first deal with the case where $m < l + 1$, where we must show that
the edge to $r^j$ is the highest appeal edge at $t^j_{i,l}$. Let $v =
\sigma(t^j_{i,l})$ be the successor of $t^j_{i,l}$ according to $\sigma$ (if $m
= 1$ then $v = s^j$, and otherwise $v = r^j$). Since $m \le l$, the definition in
Equation~\eqref{eqn:tnotdef} we have that $\sigma(t^j_{i,l-1}) = v$. Since the
priority assigned to $t^j_{i,l-1}$ is odd, we therefore have that
$\val^{\sigma}(t^j_{i,l-1}) \sqsubset \val^{\sigma}(v)$. Since we already know
from Lemma~\ref{lem:clockrs} that $\val^{\sigma}(s^j) \sqsubset
\val^{\sigma}(r^j)$, we have therefore proved that the edge to $r^j$ is the most
appealing edge at $t^j_{i,l}$.
\item 
\label{itm:ttwo}
Now we deal with the case where $m \ge l + 1$, where we must show that the most
appealing edge at $t^j_{i,l}$ is the one to $t^j_{i,l-1}$. Since $m > l$, the
definition in Equation~\eqref{eqn:tnotdef} implies that the path that starts at
$t^j_{i,l-1}$ and follows $\sigma$ will pass through $t^j_{i,l'}$ for all $l'$
in the range $0 \le l' < l$ before arriving at $r^j$. The highest priority on
this path is $\pp(5, i, 2k + 4n + 4, j, 0)$ on the vertex $t^j_{i,0}$. Since
this priority is even we have that $\val^{\sigma}(r^j) \sqsubset
\val^{\sigma}(t^j_{i,l-1})$, and therefore the edge to $t^j_{i,l-1}$ is the most
appealing edge at $t^j_{i,l}$.
\qedhere
\end{enumerate} 
\end{proof}

\begin{lem}
\label{lem:nottlarge}
Let $\sigma$ be a strategy that agrees with $\chi^{B, \C, j}_m$ for some $\C, B
\in \{0, 1\}^n$, some $j \in \{0, 1\}^n$, and some $m$ in the range $1 \le m <
\delay(j,\C) - 1$. For each \notg-gate $i$, and each $l > d(i)$, greedy
all-switches strategy improvement will switch $t^j_{i,l}$ to
$\chi^{B,\C,j}_{m+1}(t^j_{i,l})$.
\end{lem}
\begin{proof}
Lemma~\ref{lem:clockrs} implies that the state $t^j_{i,l}$ will not switch to
$s^j$. In the rest of this proof we consider the two remaining outgoing edges at
this state.

\begin{enumerate}
\item
First we deal with the case where $m < l + 1$, where we must show that the edge
to $r^j$ is the most appealing edge at $t^j_{i,l}$. When $l > d(i) + 1$, the
proof of this fact is identical to Item~\ref{itm:tone} in the proof of
Lemma~\ref{lem:nottsmall}. For $l = d(i) + 1$, we invoke Lemma~\ref{lem:oval} to
argue that, since $m < l + 1 = d(i) + 2$, we have that $\val^{\sigma}(o^j_i)
\sqsubset \val^{\sigma}(r^j)$. Since the priority assigned to $t^j_{i, d(i)}$ is
odd, we therefore also have that $\val^{\sigma}(t^j_{i,d(i)}) \sqsubset
\val^{\sigma}(r^j)$. Therefore, the edge to $r^j$ is the most appealing edge at
$t^j_{i, d(i) + 1}$.

\item
Now we deal with the case where $m \ge l + 1$, and where $\eval(B, \inp(i)) =
0$. Here we must show that the edge to $r^j$ is the most appealing edge at
$t^j_{i, l}$. For each $l > d(i) + 1$, the proof is identical to the proof of
the first case in the proof of this lemma. For $l = d(i) + 1$, we invoke
Lemma~\ref{lem:oval} to argue that, since $\eval(B, \inp(i)) = 0$ we must
have $\val^{\sigma}(o^j_{\inp(i)}) \sqsubset \val^{\sigma}(r^j)$, and therefore
the edge to $r^j$ is the most appealing edge at $t^j_{i, l}$.

\item
Finally, we deal with the case where $m \ge l + 1$ and where $\eval(B,
\inp(i)) = 1$. Here we must show that edge to $t^{j}_{i,l-1}$ is the most
appealing edge at $t^j_{i,l}$. From the definition given in
Equation~\eqref{eqn:tnotdef}, we have that the path that starts at
$t^{j}_{i,l-1}$ and follows $\sigma$ will pass through $t^{j}_{i,l'}$ for each
$l'$ in the range $d(i) \le l' < l$ before moving to $o^j_{\inp(i)}$. Since $m
\ge l + 1 > d(i)$, we have that $m \ge d(i) + 2$, and therefore
Lemma~\ref{lem:oval} implies that $\val^{\sigma}(r^j)  \sqsubset
\val^{\sigma}(o^j_{\inp(i)})$ and that $\maxdiff^{\sigma}(r^j, o^j_{\inp(i)})
\ge \pp(6,0,0,0,0)$. All priorities on the path from $t^j_{i,l-1}$ to
$o^j_{\inp(i)}$ are smaller than $\pp(6,0,0,0,0)$, so we can conclude that
$\val^{\sigma}(r^j)  \sqsubset \val^{\sigma}(t^j_{i,l-1})$, as required.
\qedhere
\end{enumerate} 
\end{proof}

\begin{lem}
Let $\sigma$ be a strategy that agrees with $\chi^{B, \C, j}_m$ for some $\C, B
\in \{0, 1\}^n$, some $j \in \{0, 1\}^n$, and for $m = \delay(j, \C)-1$. For
each \notg-gate $i$, and each $l$ int the range $1 \le l < \lale$, greedy
all-switches strategy improvement will switch $t^{1-j}_{i,l}$ to
$\chi^{B,\C,j}_{m+1}(t^{1-j}_{i,l})$.
\end{lem}
\begin{proof}
We must show that the edge to $s^{1-j}$ is the most appealing edge at
$t^{1-j}_{i,l}$. It can be verified that all paths starting at $t^{1-j}_{i,l-1}$
reach one of $r^{1-j}$, $s^{1-j}$, or $r^j$. Furthermore, the largest possible
priority on these paths is strictly smaller than $\pp(7, 0, 0, 0, 0)$. Hence, we
can apply part~\ref{itm:rsone} of Lemma~\ref{lem:clockrs} and
part~\ref{itm:ccone} of Lemma~\ref{lem:crossclock} to conclude that the edge to
$s^{1-j}$ is the most appealing outgoing edge at $t^{1-j}_{i,l}$. 
\end{proof}

\section{The state $d^j_i$ in \notg gates}
\label{app:notd}

In this section we show that the states $d^j_{i}$ in the \notg gate gadgets
correctly switch to the outgoing edge specified by $\chi^{B,\C,j}_{m+1}$. The
first lemma considers the case where $m = 1$, the second lemma considers the
case where $m = 2$, the third lemma considers the case where $3 \le m < d(i) +
2$, the fourth lemma considers the case where $d(i) + 2 \le m < \delay(j, \C) -
1$ and the gate outputs 0, the fifth lemma considers the case where $d(i) + 2
\le m < \delay(j, \C)$ and the gate outputs 1, and the final lemma considers the
case where $m = \delay(j, \C) - 1$.

\begin{lem}
\label{lem:notm1}
Let $\sigma$ be a strategy that agrees with $\chi^{B, \C, j}_m$ for some $\C, B
\in \{0, 1\}^n$, some $j \in \{0, 1\}^n$, and where $m=1$. For each
\notg-gate $i$, greedy all-switches strategy improvement will switch $d^j_i$ to
$\chi^{B,\C,j}_{m+1}(d^j_i)$.
\end{lem}
\begin{proof}
According to the definition given in Equation~\eqref{eqn:notd}, we must show
that the edge to $r^j$ is the most appealing edge at $d^j_i$. We do so by a case
analysis.
\begin{enumerate}
\item First we consider the vertex $s^j$. Here we can apply part~\ref{itm:rstwo}
of Lemma~\ref{lem:clockrs} to argue that $\val^{\sigma}(s^j) \sqsubset
\val^{\sigma}(r^j)$.

\item Next we consider a vertex $a^j_{i, l}$ with $l \ne d(i)$. Here the definition given in
Equation~\eqref{eqn:tnotdef} implies that the path that starts at $a^j_{i,l}$
and follows $\sigma$ passes through $t^j_{i,l}$ and then arrives at $s^j$.
The largest priority on this path is $\pp(5, i, l+1, j, 0)$ on the vertex
$a^j_{i,l}$. However, since this priority is smaller than $\pp(7, 0, 0, 0, 0)$,
we can apply part~\ref{itm:rstwo} of Lemma~\ref{lem:clockrs} to prove that
$\val^{\sigma}(a^j_{i,l}) \sqsubset \val^{\sigma}(r^j)$.

\item Next we consider a vertex $a^j_{i, l}$ with $l = d(i)$. Here we can apply
Lemma~\ref{lem:oval1} to argue that $\val^{\sigma}(o^j_{\inp(i)}) \sqsubset
\val^{\sigma}(r^j)$. Furthermore, the largest priority on the path from $a^j_{i,
l}$ to $o^j_i$ is the odd priority on $t^j_{i, d(i)}$. Hence, we can conclude
that $\val^{\sigma}(a^j_{i, d(i)}) \sqsubset \val^{\sigma}(r^j)$.

\item Finally, we consider the vertex $e^j_i$. Lemma~\ref{lem:brnot} implies that
$\br(\sigma)(e^j_i) = d^j_i$. Hence, the path that starts at $e^j_i$ moves to
$d^j_i$ and then to $s^j$. The highest priority on this path is $\pp(4, i ,1, j,
0) < \pp(7, 0, 0, 0, 0)$. Therefore, we can apply Lemma~\ref{lem:clockrs} to
show that $\val^{\sigma}(e^j_i) \sqsubset \val^{\sigma}(r^j)$.
\end{enumerate}
Hence, we have shown that the edge to $r^j$ is the most appealing outgoing edge
at $d^j_i$, so greedy all-switches strategy improvement will switch $d^j_i$ to
$r^j$. 
\end{proof}

\begin{lem}
\label{lem:notm2}
Let $\sigma$ be a strategy that agrees with $\chi^{B, \C, j}_m$ for some $\C, B
\in \{0, 1\}^n$, some $j \in \{0, 1\}^n$, and where $m=2$. For each
\notg-gate $i$, greedy all-switches strategy improvement will switch $d^j_i$ to
$\chi^{B,\C,j}_{m+1}(d^j_i)$.
\end{lem}
\begin{proof}
According to the definition given in Equation~\eqref{eqn:notd}, we must show
that the edge to $a^j_{i, \lale}$ is the most appealing edge at $d^j_i$. We do
so by a case analysis.
\begin{enumerate}
\item First we consider the vertex $r^j$. Observe that the path that starts at
$a^j_{i, \lale}$ and follows $\sigma$ visits $t^j_{i, \lale}$ and then arrives
at $r^j$. The highest priority on this path is the even priority assigned to
$a^j_{i, \lale}$, so therefore we can conclude that $\val^{\sigma}(r^j)
\sqsubset \val^{\sigma}(a^j_{i, \lale})$.

\item Next we consider the vertex $s^j$. Here we can apply
Lemma~\ref{lem:clockrs} to argue that $\val^{\sigma}(s^j) \sqsubset
\val^{\sigma}(r^j)$, and we have already shown that 
$\val^{\sigma}(r^j) \sqsubset \val^{\sigma}(a^j_{i, \lale})$.

\item Next we consider a vertex $a^j_{i, l}$ with $l \ne d(i)$ and $l \ne
a^j_{i, \lale}$. The path that starts at $a^j_{i,l}$
and follows $\sigma$ passes through $t^j_{i,l}$ and then arrives at $r^j$.
The largest priority on this path is 
the even priority assigned to $a^j_{i, l}$. However, since $l < \lale$, we have
that this priority is smaller than the even priority assigned to $a^j_{i,
\lale}$. Therefore, we have 
$\val^{\sigma}(a^j_{i, l}) \sqsubset \val^{\sigma}(a^j_{i, \lale})$.

\item Next we consider the vertex $a^j_{i, d(i)}$. Here we can apply
Lemma~\ref{lem:oval1} to argue that $\val^{\sigma}(o^j_{\inp(i)}) \sqsubset
\val^{\sigma}(r^j)$. Furthermore, the largest priority on the path from $a^j_{i,
l}$ to $o^j_i$ is the odd priority on $t^j_{i, d(i)}$. Since the highest
priority on the path from $a^j_{i, \lale}$ to $r^j$ is even, we we can conclude
that $\val^{\sigma}(a^j_{i, d(i)}) \sqsubset \val^{\sigma}(a^j_{i, \lale})$.

\item Finally, we consider the vertex $e^j_i$. Lemma~\ref{lem:brnot} implies that
$\br(\sigma)(e^j_i) = d^j_i$. Hence, the path that starts at $e^j_i$ moves to
$d^j_i$ and then to $r^j$. The highest priority on this path is $\pp(4, i ,1, j,
0)$ on the vertex $e^j_i$. However, this is smaller than the largest priority on
the path from $a^j_{i, \lale}$ to $r^j$, so we can conclude that
$\val^{\sigma}(e^j_i) \sqsubset \val^{\sigma}(a^j_{i, \lale})$.
\qedhere
\end{enumerate} 
\end{proof}

\begin{lem}
\label{lem:dbase}
Let $\sigma$ be a strategy that agrees with $\chi^{B, \C, j}_m$ for some $\C, B
\in \{0, 1\}^n$, some $j \in \{0, 1\}^n$, and some $m$ in the range $3 \le m <
d(i) + 2$. For each \notg-gate $i$, greedy all-switches strategy improvement
will switch $d^j_i$ to $\chi^{B,\C,j}_{m+1}(d^j_i)$.
\end{lem}
\begin{proof}
Lemma~\ref{lem:clockrs} implies that the state $d^j_i$ will not switch to
$s^j$. In the rest of this proof we consider the other outgoing edges at
this state.

Since we are in the case where $m < d(i) + 2$ the definition from
Equation~\eqref{eqn:notd} specifies that greedy all-switches strategy
improvement must switch to $a^j_{i,m-2}$. Hence must argue that the edge to
$a^j_{i,m-2}$ is the most appealing edge at $d^j_i$, and we will start by
considering the appeal of this edge. The definition in
Equation~\eqref{eqn:tnotdef} implies that the path that starts at $a^j_{i,m-2}$
passes through $t^j_{i,l}$ for all $l \le m-2$ before arriving at $r^j$. The
highest priority on this path is $\pp(5, i, 2k+4n+4, j, 0)$ on the vertex
$t^j_0$, and the second highest priority on this path is $\pp(5, i , m-1, j, 0)$
on the vertex $a^j_{i,m-2}$. We will now show that all other edges are less
appealing.
\begin{enumerate}
\item First we consider the vertex $r^j$. Since $\pp(5, i, 2k+4n+4, j, 0)$ is
even, we immediately get that $\val^{\sigma}(r^j) \sqsubset
\val^{\sigma}(a^j_{i,m-2})$.
\item 
\label{itm:two}
Next we consider a vertex $a^j_{i,l}$ with $l < m-2$. The path that
starts at this vertex and follows $\sigma$ passes through $t^j_{i,l'}$  for all
$l' \le l$ before arriving at $r^j$. The highest priority on this path is
$\pp(5, i, 2k+4n+4, j, 0)$ on the vertex $t^j_0$, and the second highest
priority on this path is $\pp(5, i , l+1, j, 0)$ on the vertex $a^j_{i,l}$.
Hence, we have that $\maxdiff^{\sigma}(a^j_{i,m-2}, a^j_{i, l})$ is $\pp(5, i ,
m-1, j, 0)$, and since this is even, we can conclude that
$\val^{\sigma}(a^j_{i,l}) \sqsubset \val^{\sigma}(a^j_{i,m-2})$.
\item 
\label{itm:three}
Next we consider a vertex $a^j_{i,l}$ with $l > m-2$  and with $l \ne
d(i)$. The path that starts at this vertex and follows $\sigma$ passes through
$t^j_{i,l}$ and then moves immediately to $r^j$. The highest priority on this
path is $\pp(5, i , l+1, j, 0)$ on the vertex $a^j_{i,l}$. Since $l+1 < 2k +
4n + 4$ we have that $\maxdiff^{\sigma}(a^j_{i,m-2}, a^j_{i, l})$ is $\pp(5, i ,
2k+4n+4, j, 0)$, and since this priority is even, we can conclude that
$\val^{\sigma}(a^j_{i,l}) \sqsubset \val^{\sigma}(a^j_{i,m-2})$.
\item 
\label{itm:four}
Next we consider the vertex $a^j_{i, d(i)}$. The path that starts at this
vertex passes through $t^j_{i, d(i)}$, and then moves to $\inputstate(i, j)$.
Since $m < d(i) + 2$, we have that $m \le d(\inp(i)) + 2$, and therefore the
first case of Lemma~\ref{lem:oval} tells us that
$\val^{\sigma}(\inputstate(i,j)) \sqsubset \val^{\sigma}(r^j)$. Hence we can
conclude that $\val^{\sigma}(a^j_{i, d(i)}) \sqsubset
\val^{\sigma}(a^j_{i,m-2})$.
\item 
\label{itm:five}
Finally, we consider the vertex $e^j_i$. By Lemma~\ref{lem:brnot}, the path
that starts at $e^j_i$ and follows $\sigma$ passes through $d^j_i$. If $m=3$, it
then moves to $a^j_{i, \lale}$, and if $m >3$ it then moves to $a^j_{i, m-3}$.
In either case, since the priority assigned to $e^j_i$ is smaller than the
priorities assigned to the vertices $a^j_l$ and $t^j_l$, we can reuse the
arguments made above to conclude that $\val^{\sigma}(e^j_i) \sqsubset
\val^{\sigma}(a^j_{i,m-2})$.
\end{enumerate}
Therefore, we have shown that the edge to $a^j_{i,m-2}$ is the most appealing
outgoing edge at
$d^j_i$, and so this edge will be switched by greedy all-switches strategy
improvement. 
\end{proof}

\begin{lem}
\label{lem:notd1}
Let $\sigma$ be a strategy that agrees with $\chi^{B, \C, j}_m$ for some $\C, B
\in \{0, 1\}^n$, some $j \in \{0, 1\}^n$, and some $m$ in the range $d(i) + 2
\le m < \delay(j, \C) - 1$. For each \notg-gate $\eval(B, \inp(i)) = 1$
then greedy all-switches strategy improvement will switch $d^j_i$ to
$\chi^{B,\C,j}_{m+1}(d^j_i)$.
\end{lem}
\begin{proof}
Lemma~\ref{lem:clockrs} implies that the state $d^j_i$ will not switch to
$s^j$. In the rest of this proof we consider the other outgoing edges at
this state.

Since we are in the case where $m \ge d(i) + 2$ and $\eval(B, \inp(i)) = 1$,
the definition from Equation~\eqref{eqn:notd} specifies that greedy all-switches
strategy improvement must switch to $a^j_{i,m-2}$. Hence must argue that the
edge to $a^j_{i,m-2}$ is the most appealing edge at $d^j_i$, and we will start
by considering the appeal of this edge. The definition in
Equation~\eqref{eqn:tnotdef} implies that the path that starts at $a^j_{i,m-2}$
passes through $t^j_{i,l}$ for all $l$ in the range $d(i) \le l \le m-2$ before
arriving at $\inputstate(i, j)$. Since $m \ge d(i) + 2$ we have $m > d(\inp(i) +
2$, and so Lemma~\ref{lem:oval} implies that $\maxdiff^{\sigma}(r^j,
\inputstate(i, j)) \ge \pp(6, 0, 0, 0, 0)$. We now consider the other outgoing
edges from $d^j_i$.
\begin{enumerate}
\item  First we consider $r^j$, where Lemma~\ref{lem:oval} immediately gives
that $\val^{\sigma}(r^j) \sqsubset \val^{\sigma}(a^j_{i,m-2})$.
\item Next we consider a vertex $a^j_{i, l}$ with $l < d(i)$. Using the same
reasoning as Item~\ref{itm:two} in the proof of Lemma~\ref{lem:dbase}, we can
conclude that the highest priority on the path from $a^j_{i, l}$ to $r^j$ is
$\pp(5, i, 2k+4n+4, j, 0) < \pp(6, 0, 0, 0, 0)$ and so therefore
$\val^{\sigma}(a^j_{i,l}) \sqsubset \val^{\sigma}(a^j_{i,m-2})$.
\item Next we consider a vertex $a^j_{i, l}$ with $l$ in the range $d(i) \le l <
m -2$. The path that starts at $a^j_{i, l}$ and follows $\sigma$ passes through
$t^j_{i, l'}$ for all $l'$ in the range $d(i) \le l' \le l$ and then arrives at
$\inputstate(i,j)$. The highest
priority on this path is $\pp(5, i, l+1, j, 0)$. On the other hand, the largest
priority on the path from $a^j_{i, m-2}$ to $\inputstate(i,j)$ is 
$\pp(5, i, m-1, j, 0)$. Since this priority is even, we can conclude that
$\val^{\sigma}(a^j_{i,l}) \sqsubset \val^{\sigma}(a^j_{i,m-2})$.
\item Next we consider a vertex $a^j_{i, l}$ with $l > m - 2$. Using the same
reasoning as Item~\ref{itm:three} in the proof of Lemma~\ref{lem:dbase} we can
conclude that the highest priority on the path from $a^j_{i, l}$ to $r^j$ is
$\pp(5, i , l+1, j, 0) < \pp(6, 0, 0, 0, 0)$ and so therefore
$\val^{\sigma}(a^j_{i,l}) \sqsubset \val^{\sigma}(a^j_{i,m-2})$.
\item Finally, we consider the vertex $e^j_i$, where we can use the same
reasoning as Item~\ref{itm:five} in the proof of Lemma~\ref{lem:dbase} to conclude
that $\val^{\sigma}(e^j_{i}) \sqsubset \val^{\sigma}(a^j_{i,m-2})$.
\end{enumerate}
Therefore, we have shown that the edge to $a^j_{i,m-2}$ is the most appealing
outgoing edge at $d^j_i$, and so this edge will be switched by greedy
all-switches strategy improvement. 
\end{proof}

\begin{lem}
\label{lem:notd0}
Let $\sigma$ be a strategy that agrees with $\chi^{B, \C, j}_m$ for some $\C, B
\in \{0, 1\}^n$, some $j \in \{0, 1\}^n$, and some $m$ in the range $d(i) + 2
\le m < \delay(j, \C) - 1$. For each \notg-gate $\eval(B, \inp(i)) = 0$
then greedy all-switches strategy improvement will switch $d^j_i$ to
$\chi^{B,\C,j}_{m+1}(d^j_i)$.
\end{lem}
\begin{proof}
Lemma~\ref{lem:clockrs} implies that the state $d^j_i$ will not switch to
$s^j$. In the rest of this proof we consider the other outgoing edges at
this state.

Since $m \ge d(i) + 2$ and $\eval(B, \inp(i) = 0)$, the definition in
Equation~\eqref{eqn:notd} specifies that the edge to $e^j_i$ is the most
appealing edge at $d^j_i$. To prove this, we first show that all other edges are
less appealing than $a^j_{i, d(i) -1}$, and we will then later show that $e^j_i$
is more appealing than $a^j_{i, d(i) -1}$. The definition in
Equation~\eqref{eqn:tnotdef} implies that the path that starts at
$a^j_{i,d(i)-1}$ passes through $t^j_{i,l}$ for all $l$ in the range $0 \le l
\le d(i) - 1$ before arriving at $r^j$. The largest priority on this path is
$\pp(5, i, d(i), j 0)$ on the state $a^j_{i, d(i)-1}$. We now consider the other
outgoing edges.
\begin{enumerate}
\item First we consider the vertex $r^j$. Since $\pp(5, i, 2k+4n+4, j, 0)$ is
even, we immediately get that $\val^{\sigma}(r^j) \sqsubset
\val^{\sigma}(a^j_{i,m-2})$.
\item Next we consider a vertex $a^j_{i, l}$ with $l < d(i) -1$. Here we can use
the same reasoning as we used in Item~\ref{itm:two} in the proof of
Lemma~\ref{lem:dbase} to argue that $\val^{\sigma}(a^j_{i,l}) \sqsubset
\val^{\sigma}(a^j_{i,d(i)-1})$.
\item Next we consider a vertex $a^j_{i, l}$ with $l > d(i)$. Here we can use
the same reasoning as we used in Item~\ref{itm:three} in the proof of
Lemma~\ref{lem:dbase} to argue that $\val^{\sigma}(a^j_{i,l}) \sqsubset
\val^{\sigma}(a^j_{i,d(i)-1})$.
\item Finally, we consider the vertex $a^j_{i,d(i)}$. Here we can use the
same reasoning as we used in Item~\ref{itm:four} in the proof of
Lemma~\ref{lem:dbase} to argue that $\val^{\sigma}(a^j_{i,d(i)}) \sqsubset
\val^{\sigma}(a^j_{i,d(i)-1})$, although this time we will use case two of
Lemma~\ref{lem:oval}.
\end{enumerate}
So, we have shown that every edge other than the one to $e^j_i$ is less
appealing than the edge to $a^j_{i, d(i)-1}$.

Now we will show that the edge to $e^j_i$ is more appealing than the edge to
$a^j_{i, d(i)-1}$. There are two cases to consider.
\begin{itemize}
\item If $m = d(i) + 2$, then $\sigma(d^j_i) = a^j_{i, d(i)-1}$. Since
Lemma~\ref{lem:brnot} implies that $\br(\sigma)(e^j_i) = d^j_i$, we have that the
path that starts at $e^j_i$ and follows $\sigma$ passes through $d^j_i$ and then
arrives at $a^j_{i, d(i)-1}$. The largest priority on this path is $\pp(4, i, 1,
j, 0)$, and since this is even, we can conclude that $\val^{\sigma}(a^j_{i, d(i)
- 1})
\sqsubset \val^{\sigma}(e^j_i)$.
\item If $m > d(i) + 2$, then $\sigma(d^j_i) = e^j_i$. In this case
Lemma~\ref{lem:brnot} implies that $\br(\sigma)(e^j_i) = h^j_i$. The path that
starts at $e^j_i$ and follows $\sigma$ passes through $h^j_i$ and then moves
directly to $r^j_i$. The largest priority on this path is $\pp(6, i, 1, j 0)$,
which is bigger than $\pp(5, i, d(i), j 0)$. Therefore,
$\maxdiff^{\sigma}(e^j_i, a^j_{i, d(i) - 1} = \pp(6, i, 1, j 0)$, and since this
is even, we can conclude that $\val^{\sigma}(a^j_{i, d(i) - 1}) \sqsubset
\val^{\sigma}(e^j_i)$.
\end{itemize}
Hence, we have shown that the edge to $e^j_i$ is the most appealing edge at
$d^j_i$. 
\end{proof}

\begin{lem}
Let $\sigma$ be a strategy that agrees with $\chi^{B, \C, j}_m$ for some $\C, B
\in \{0, 1\}^n$, some $j \in \{0, 1\}^n$, and for $m = \delay(j, \C)-1$. For
each \notg-gate $i$ greedy
all-switches strategy improvement will switch $d^{1-j}_{i}$ to
$\chi^{B,\C,j}_{m+1}(d^{1-j}_{i})$.
\end{lem}
\begin{proof}
We must show that the edge to $s^{1-j}$ is the most appealing edge at
$d^{1-j}_{i}$. It can be verified that all paths starting at the vertices
$a^{1-j}_{i,l}$ and $e^j_i$ reach one of $r^{1-j}$, $s^{1-j}$, or $r^j$.
Furthermore, the largest possible priority on all of these paths is strictly
smaller than $\pp(7, 0, 0, 0, 0)$. Hence, we can apply part~\ref{itm:rsone} of
Lemma~\ref{lem:clockrs} and part~\ref{itm:ccone} of Lemma~\ref{lem:crossclock}
to conclude that the edge to $s^{1-j}$ is the most appealing outgoing edge at
$d^{1-j}_{i}$. 
\end{proof}

\section{The vertices $z^l$}
\label{app:z}

In this section we show that the states $z^l$ correctly switch to the outgoing
edge specified by $\chi^{B,\C,j}_{m+1}$. The first lemma considers the case
where $l = j$, while the second lemma considers the case where $l = 1-j$.

\begin{lem}
Let $\sigma$ be a strategy that agrees with $\chi^{B, \C, j}_m$ for some $\C, B
\in \{0, 1\}^n$, some $j \in \{0, 1\}^n$, and $m$ in the range $1 \le m \le
\delay(j, \C) - 1$. The greedy all-switches strategy improvement algorithm will switch
$z^{j}$ to $\chi^{B,\C,j}_{m+1}(z^{j})$.
\end{lem}
\begin{proof}
We must show that the edge to $r^j$ is the most appealing edge at $z^j$. This
follows immediately from part~\ref{itm:rstwo} of Lemma~\ref{lem:clockrs}. 
\end{proof}

\begin{lem}
Let $\sigma$ be a strategy that agrees with $\chi^{B, \C, j}_m$ for some $\C, B
\in \{0, 1\}^n$, some $j \in \{0, 1\}^n$, and some $m$ in the range $1 \le m \le
\delay(j, \C) - 1$. The greedy all-switches strategy improvement algorithm will switch
$z^{1-j}$ to $\chi^{B,\C,j}_{m+1}(z^{1-j})$.
\end{lem}
\begin{proof}
There are two cases to consider.
\begin{enumerate}
\item If $m < \delay(j, \C) - 1$, then we must show that the edge to $r^{1-j}$
is the most appealing edge at $z^{1-j}$. This follows immediately by applying
part~\ref{itm:rstwo} of Lemma~\ref{lem:clockrs}.
\item If $m = \delay(j, \C) - 1$, then since $\delay(j, \C) + \delay(1-j, \C) =
\length(\C)$, we must show that the edge to $s^{1-j}$ is the most appealing edge
at $z^{1-j}$. This follows immediately from part~\ref{itm:rsone} of
Lemma~\ref{lem:clockrs}.
\qedhere
\end{enumerate} 
\end{proof}

\section{The vertices $y^l$}
\label{app:y}

In this section we show that the states $y^l$ correctly switch to the outgoing
edge specified by $\chi^{B,\C,j}_{m+1}$. The first lemma considers the case
where $l = j$, while the second lemma considers the case where $l = 1-j$.

\begin{lem}
Let $\sigma$ be a strategy that agrees with $\chi^{B, \C, j}_m$ for some $\C, B
\in \{0, 1\}^n$, some $j \in \{0, 1\}^n$, and some $m$ in the range $1 \le m \le
\delay(j, \C) - 1$. The greedy all-switches strategy improvement algorithm will switch
$y^j$ to $\chi^{B,\C,j}_{m+1}(y^j)$.
\end{lem}
\begin{proof}
The definition of $\chi^{B,\C,j}_{m+1}(y^{j}_i)$ specifies that the edge chosen
at $y^{j}$ is defined by $\sigma_{m+\delay(j, \C)}(y^{j})$, which is given in
Equation~\eqref{eqn:yj}.  According to this definition, we must show that the
edge to $r^j$ is the most appealing outgoing edge at $y^j$. This follows
immediately from parts~\ref{itm:ccone} and~\ref{itm:cctwo} of
Lemma~\ref{lem:crossclock}. 
\end{proof}

\begin{lem}
Let $\sigma$ be a strategy that agrees with $\chi^{B, \C, j}_m$ for some $\C, B
\in \{0, 1\}^n$, some $j \in \{0, 1\}^n$, and some $m$ in the range $1 \le m \le
\delay(j, \C) - 1$. The greedy all-switches strategy improvement algorithm will
switch $y^{1-j}$ to $\chi^{B,\C,j}_{m+1}(y^{1-j})$.
\end{lem}
\begin{proof}
According to Equation~\eqref{eqn:yj}, we must show that the edge to $r^j$ is the
most appealing edge at $y^{1-j}$. This follows immediately  from
Lemma~\ref{lem:crossclock}. 
\end{proof}

\section{The vertices $p^l_i$}
\label{app:p}

In this section we show that the vertices $p^l_i$ correctly switch to the
outgoing edge specified by $\chi^{B,\C,j}_{m+1}$. The first lemma considers the
case where $l = j$, while the second lemma considers the case where $l = 1-j$.

\begin{lem}
Let $\sigma$ be a strategy that agrees with $\chi^{B, \C, j}_m$ for some $\C, B
\in \{0, 1\}^n$, some $j \in \{0, 1\}^n$, and some $m$ in the range $1 \le m \le
\delay(j, \C) - 1$. For each $i \in \inputg$, the greedy all-switches strategy
improvement algorithm will switch $p^j_{i}$ to $\chi^{B,\C,j}_{m+1}(p^j_i)$.
\end{lem}
\begin{proof}
Since $m$ is in the range $1 \le m \le \delay(j, \C) - 1$, the definition given
in Equation~\eqref{eqn:pj} specifies that $o^j_{\inp(i)}$ should be the most
appealing outgoing edge at $p^j_i$. There are two cases to consider.
\begin{enumerate}
\item 
If $m = 1$, then observe that the path that starts at $o^j_{\inp(i)}$ and
follows $\sigma$ will eventually reach $s^j$, no matter whether $\inp(i)$ is a
\notg-gate or an \org-gate. Furthermore, the largest priority on this path is
strictly smaller than $\pp(7, 0, 0, 0, 0)$. On the other hand, the path that
starts at $p^j_{i,1}$ moves immediately to $r^{1-j}$, and the largest priority
on this path is strictly smaller than $\pp(7, 0, 0, 0, 0)$. Hence, we can apply
part~\ref{itm:cctwo} of Lemma~\ref{lem:crossclock} to argue that
$\val^{\sigma}(p^{j}_{i,1}) \sqsubset \val^{\sigma}(o^j_{\inp(i)})$, as
required.
\item If $m \ge 1$, then then observe that the path that starts at
$o^j_{\inp(i)}$ and follows $\sigma$ will eventually reach $r^j$, and again the
largest priority on this path is strictly smaller than $\pp(7, 0, 0, 0, 0)$.
Hence, we can apply part~\ref{itm:cctwo} of Lemma~\ref{lem:crossclock} to argue
that $\val^{\sigma}(p^{j}_{i,1}) \sqsubset \val^{\sigma}(o^j_{\inp(i)})$, as
required.
\qedhere
\end{enumerate} 
\end{proof}

\begin{lem}
Let $\sigma$ be a strategy that agrees with $\chi^{B, \C, j}_m$ for some $\C, B
\in \{0, 1\}^n$, some $j \in \{0, 1\}^n$, and some $m$ in the range $1 \le m \le
\delay(j, \C) - 1$. For each $i \in \inputg$, the greedy all-switches strategy
improvement algorithm will switch $p^{1-j}_{i}$ to
$\chi^{B,\C,j}_{m+1}(p^{1-j}_i)$.
\end{lem}
\begin{proof}
The definition given in Equation~\eqref{eqn:pj} specifies that the edge to
$p^{1-j}_{i}$ should be the most appealing edge at $p^{1-j}_i$. The path that
starts at $o^{1-j}_{\inp(i)}$ and follows $\sigma$ must eventually arrive at
either $s^{1-j}$ or $r^{1-j}$. In particular, observe that $r^j$ cannot be
reached due to the vertices $q^j_{i, 0}$, which by Lemma~\ref{lem:brqj} selects
the edge towards $q^j_{i, 1}$. On the other hand, the path that starts at
$p^{1-j}_{i,1}$ moves directly to $r^j$. Moreover the largest priorities on both
of these paths are strictly smaller than $\pp(7, 0, 0, 0, 0)$. Hence, we can
apply Lemmas~\ref{lem:clockrs} and~\ref{lem:crossclock} to argue that
$\val^{\sigma}(o^j_{\inp(i)}) \sqsubset \val^{\sigma}(r^j)$. 
\end{proof}

\section{The vertices $h^l_{i,0}$}
\label{app:h}

In this section we show that the vertices $h^l_{i,0}$ correctly switch to the
outgoing edge specified by $\chi^{B,\C,j}_{m+1}$. The first lemma considers the
case where $l = j$, while the second lemma considers the case where $l = 1-j$.

\begin{lem}
Let $\sigma$ be a strategy that agrees with $\chi^{B, \C, j}_m$ for some $\C, B
\in \{0, 1\}^n$, some $j \in \{0, 1\}^n$, and some $m$ in the range $1 \le m \le
\delay(j, \C) - 1$. For each $i \in \inputg$, the greedy all-switches strategy
improvement algorithm will switch $h^j_{i, 0}$ to $\chi^{B,\C,j}_{m+1}(h^j_{i,
0})$.
\end{lem}
\begin{proof}
We must show that the most appealing edge at $h^{j}_{i,0}$ should be
$h^j_{i,1}$. Observe that both $h^j_{i,1}$ and $h^j_{i,2}$ move directly to
$r^j$ and $r^{1-j}$, respectively. Moreover, the priorities assigned to these
vertices are strictly smaller than $\pp(7, 0, 0, 0, 0)$. Since $h^j_{i,1}$ moves
to $r^{j}$, we can apply part~\ref{itm:cctwo} of Lemma~\ref{lem:crossclock} to
prove that $h^j_{i,1}$ is the most appealing edge at $h^j_{i, 0}$. 
\end{proof}

\begin{lem}
Let $\sigma$ be a strategy that agrees with $\chi^{B, \C, j}_m$ for some $\C, B
\in \{0, 1\}^n$, some $j \in \{0, 1\}^n$, and some $m$ in the range $1 \le m \le
\delay(j, \C) - 1$. The greedy all-switches strategy improvement algorithm will switch
$h^{1-j}_{i, 0}$ to $\chi^{B,\C,j}_{m+1}(h^{1-j}_{i,0})$.
\end{lem}
\begin{proof}
We must show that the most appealing edge at $h^{j}_{i,0}$ should be
$h^j_{i,1}$. Observe that both $h^{1-j}_{i,1}$ and $h^{1-j}_{i,2}$ move directly
to $r^{1-j}$ and $r^{j}$, respectively. Moreover, the priorities assigned to
these vertices are strictly smaller than $\pp(7, 0, 0, 0, 0)$. We will use these
fact in order to apply Lemma~\ref{lem:crossclock} in the following case
analysis. According to Equation~\eqref{eqn:hj}, there are two cases to consider.
Since $h^j_{i,1}$ moves to $r^{j}$, we can apply part~\ref{itm:cctwo} of
Lemma~\ref{lem:crossclock} to prove that $h^j_{i,1}$ is the most appealing edge
at $h^j_{i, 0}$. 
\end{proof}

\section{Input/output gates}
\label{app:input}

In this section we show that the vertices in the input/output gadgets  correctly
switch to the outgoing edge specified by $\chi^{B,\C,j}_{m+1}$. The first two
lemmas deal with the case where the input/output gadget for circuit $j$ resets.
Note that this occurs one iteration later than the rest of the vertices in
circuit $j$, which is why we prove separate lemmas for this case. Otherwise, the
input/output gadgets behave as if they are \notg gates, so the proofs that we have
already given for the \notg gates can be applied with only minor changes. These is
formalized in the final two lemmas of this section. The first of these lemmas
considers the input/output gadgets in circuit $j$ and the second considers the
input/output gadgets in circuit $1-j$.

\begin{lem}
\label{lem:tinput1}
Let $\sigma$ be a strategy that agrees with $\chi^{B, \C, j}_m$ for some $\C, B
\in \{0, 1\}^n$, some $j \in \{0, 1\}^n$, and for $m = 1$. For
each \inputg-gate $i$, and each $l$ int the range $1 \le l < \lale$, greedy
all-switches strategy improvement will switch $t^{j}_{i,l}$ to
$\chi^{B,\C,j}_{m+1}(t^{j}_{i,l})$.
\end{lem}
\begin{proof}
We must show that the edge to $z^j$ is the most appealing edge at $t^j_{i, l}$.
All paths that start at $t^j_{i, l}$ and follow $\sigma$ will eventually arrive
at $r^{1-j}$, either via $y^j$, or via $p^j_i$. On the other hand, the path that
starts at $z^j$ moves directly to $s^j$. Therefore, part~\ref{itm:cctwo} of
Lemma~\ref{lem:crossclock} implies that the edge to $z^j$ is the most appealing
edge at $t^j_{i, l}$. 
\end{proof}

\begin{lem}
Let $\sigma$ be a strategy that agrees with $\chi^{B, \C, j}_m$ for some $\C, B
\in \{0, 1\}^n$, some $j \in \{0, 1\}^n$, and for $m = 1$. For
each \inputg-gate $i$ greedy
all-switches strategy improvement will switch $d^{j}_{i}$ to
$\chi^{B,\C,j}_{m+1}(d^{j}_{i})$.
\end{lem}
\begin{proof}
This proof uses the same argument as the proof of Lemma~\ref{lem:tinput1},
because all paths that start at $d^j_i$ will eventually arrive at $r^{1-j}$.
\qedhere
\end{proof}

\begin{lem}
\label{lem:inpj}
Let $\sigma$ be a strategy that agrees with $\chi^{B, \C, j}_m$ for some $\C, B
\in \{0, 1\}^n$, some $j \in \{0, 1\}^n$, and for $m$ in the range $2 \le m \le
\delay(j, \C) - 1$. For each \inputg-gate $i$, and each $l$ int the range $1 \le
l < \lale$, greedy all-switches strategy improvement will switch $t^{j}_{i,l}$
to $\chi^{B,\C,j}_{m+1}(t^{j}_{i,l})$ and $d^{j}_{i}$ to
$\chi^{B,\C,j}_{m+1}(d^{j}_{i})$.
\end{lem}
\begin{proof}
Since $m \ge 2$, we have that $\sigma(y^j) = r^j$ and $\sigma(z^j) = s^j$.
Hence, both  $t^{j}_{i,l}$ and $d^j_i$ behave in exactly the same way as the
states $t^{j}_{i',l}$ and $d^j_{i',l}$ for $i' \in \notg$, with the exception
that these states one step behind the \notg-gate vertices, due to the delay
introduced by $y^j$. Note, however, that this is account for by placing the edge
to $p^j_i$ on $t^j_{i, d(C)}$, rather than $t^j_{i, d(C)+1}$, as would be
expected for a \notg-gate with depth $d(C) + 1$. Therefore, to prove this lemma,
we can use exactly the reasoning as gave for
Lemmas~\ref{lem:nottsmall},~\ref{lem:nottlarge},~\ref{lem:notm1},~\ref{lem:notm2},~\ref{lem:dbase},~\ref{lem:notd0},
and~\ref{lem:notd1}. This is because all of the reasoning used there is done
relative to $r^j$ and $s^j$, and since $y^j$ and $z^j$ have insignificant
priorities, none of this reasoning changes. 
\end{proof}

\begin{lem}
Let $\sigma$ be a strategy that agrees with $\chi^{B, \C, j}_m$ for some $\C, B
\in \{0, 1\}^n$, some $j \in \{0, 1\}^n$, and for $m$ in the range $2 \le m < 
\delay(j, \C) - 1$. For each \inputg-gate $i$, and each $l$ int the range $1 \le
l < \lale$, greedy all-switches strategy improvement will switch $t^{1-j}_{i,l}$
to $\chi^{B,\C,j}_{m+1}(t^{1-j}_{i,l})$ and $d^{1-j}_{i}$ to
$\chi^{B,\C,j}_{m+1}(d^{1-j}_{i})$.
\end{lem}
\begin{proof}
Much like the proof of Lemma~\ref{lem:inpj}, this claim can be proved using
essentially the same reasoning given as was given for
Lemmas~\ref{lem:nottsmall},~\ref{lem:nottlarge},~\ref{lem:notm1},~\ref{lem:notm2},~\ref{lem:dbase},~\ref{lem:notd0},
and~\ref{lem:notd1}, because an \inputg-gate behaves exactly like a \notg-gate.

In particular, note that since we have defined $\chi^{B, \C, 0}_{\delay(0,\C)} =
\chi^{B, \C, 1}_1$ and $\chi^{B, \C, 1}_{\delay(1,\C)} = \chi^{B, \C+1,
0}_1$, the gate gadgets in circuit $1-j$ continue to have well-defined
strategies, so the $d^{1-j}_i$ will continue to act like a \notg-gate between
iterations $1$ and $2$.

The one point that we must pay attention to is that, in the transition between
$\chi^{B,\C,j}_{1}$ and $\chi^{B,\C,j}_{2}$, the state $y^{1-j}$ switches from
$r^{1-j}$ to $r^{j}$. Note, however, that both $h^j_{i, 0}$ and $p^j_i$ both
switch to vertex that eventually leads to $r^j$ at exactly the same time, so all
paths that exit the gadget will switch from $r^{1-j}$ to $r^j$. Since almost all
of the reasoning in the above lemmas is done relative to $r^j$, the relative
orders over the edge appeals cannot change.

The only arguments that must be changed are the ones that depend on
Lemma~\ref{lem:oval}. Here, the fact that the priority $\pp(5, i, 2k+4n+5, j,
0)$ is assigned to $p^{1-j}_{i,1}$ is sufficient to ensure that
$\val^{\sigma}(r^j) \sqsubset \val^{\sigma}(t^{1-j}_{i, d(C)})$, so the
deceleration lane will continue switching. On the other hand, since $\pp(5, i,
2k+4n+5, j, 0) < \pp(7, 0, 0, 0, 0)$, this priority is not large enough to cause
$d^j_i$ to switch away from $e^j_i$, if $B_i = 1$. 
\end{proof}

\end{document}